\documentclass[fleqn,10pt]{wlscirep}
\usepackage{setspace}
\usepackage[utf8]{inputenc}
\usepackage[T1]{fontenc}
\usepackage{bm}
\title{Oxide spin-orbitronics: spin-charge interconversion and topological spin textures}

\author[1,2]{Felix Trier}
\author[3,4]{Paul No\"el}
\author[5]{Joo-Von Kim}
\author[3]{Jean-Philippe Attan\'e}
\author[3]{Laurent Vila}
\author[1,*]{Manuel Bibes}
\affil[1]{Unit\'e Mixte de Physique, CNRS, Thales, Universit\'e Paris-Saclay, Palaiseau, France}
\affil[2]{Department of Energy Conversion and Storage, Technical University of Denmark, Fysikvej, Building 310, 2800 Kgs. Lyngby, Denmark}
\affil[3]{Université Grenoble Alpes, CEA, CNRS, INP-G, Spintec, Grenoble, France}
\affil[4]{Dept. of Materials, ETH Z{\"u}rich, Hönggerbergring 64, 8093 Z{\"u}rich, Switzerland}
\affil[5]{Centre de Nanosciences et de Nanotechnologies, CNRS, Universit\'e Paris-Saclay, Palaiseau, France}

\affil[*]{e-mail: manuel.bibes@cnrs-thales.fr}

\begin{abstract}
Quantum oxide materials possess a vast range of properties stemming from the interplay between the lattice, charge, spin and orbital degrees of freedom, in which electron correlations often play an important role. Historically, the spin-orbit coupling was rarely a dominant energy scale in oxides. It however recently came to the forefront, unleashing various exotic phenomena connected with real and reciprocal-space topology that may be harnessed in spintronics. In this article, we review the recent advances in the new field of oxide spin-orbitronics with a special focus on spin-charge interconversion from the direct and inverse spin Hall and Edelstein effects, and on the generation and observation of topological spin textures such as skyrmions. We highlight the control of spin-orbit-driven effects by ferroelectricity and give perspectives for the field.
\end{abstract}

\begin{document}

\flushbottom
\maketitle

\thispagestyle{empty}

\section*{Introduction}
The coupling between electron spin and charge by the spin-orbit interaction lies at the core of multiple phenomena in condensed matter physics. The presence of a large spin-orbit interaction in materials where the inversion symmetry is broken enables a plethora of intriguing solid-state phenomena such as spin-momentum locking, spin-orbit torque, and topologically nontrivial spin textures arising from antisymmetric magnetic exchange interaction. Collectively, this broad and rapidly emerging area of research is known as \textit{spin-orbitronics} \cite{soumyanarayanan_emergent_2016}. The advantage of spin-orbitronics over conventional spintronic concepts, where ferromagnetic elements have traditionally been employed to generate and detect spin-currents, notably lies in the efficient spin-charge interconversion through the spin-orbit coupling, which can be entirely achieved without the need for magnetic materials. The spin-orbit interaction is also key to the existence of topological spin textures such as skyrmions which have been proposed as nanoscale magnetic bits for information storage\cite{sampaioNucleationStabilityCurrentinduced2013}. Although spin-orbitronics was firstly considered in metals and semiconductors, it is now quickly entering the field of oxides heterostructures~\cite{bibes_ultrathin_2011} with a number of unique characteristics.

Oxide interfaces have been at the centre of attention in recent years, due to the intimate coupling between the spin, charge and lattice degrees of freedom occurring at such interfaces that consequently serve as a basis for a variety of emergent phenomena \cite{hwang_emergent_2012}. In particular, oxide interfaces provide an exciting test-bed for studying and harnessing the spin-orbit interaction through the Rashba effect \cite{gariglio_spinorbit_2019} and Dzyaloshinskii–Moriya interaction, and in this way represent a versatile platform to generate, control and detect spin currents or magnetic textures. Moreover, oxide interfaces can nowadays be grown with exquisite precision \cite{ramesh_creating_2019} and ultrasharp interfaces, which has opened the door for their use into spin-orbitronics\cite{varignonNewSpinOxide2018}. The significance of oxides for spin-orbitronics lies also in their broad range of functional material properties, which is often much richer than in traditional materials. One of these material properties, ferroelectricity, is for instance commonly found in many oxides but remains absent from most other families. Interestingly, ferroelectricity has in recent years entered the stage as a new degree of freedom to control spin-orbit-driven effects. In this context, recent important advances in the literature of oxide spin-orbitronics include the demonstration of a giant spin-charge conversion efficiency in SrTiO$_3$-based two-dimensional electron gases (2DEGs) \cite{vaz_mapping_2019}, its control by ferroelectricity\cite{noel_non-volatile_2020}, the observation of magnetic skyrmions tunable by ferroelectricity in ultrathin oxide heterostructures \cite{wang_ferroelectrically_2018}, and the realization of skyrmion embryos at room temperature in multiferroics \cite{chauleau_electric_2019}. These milestone achievements have consequently not gone unnoticed, and several major companies including Intel, IBM and Thales have inaugurated their own research activities on oxide spin-orbitronics.

Here, we will review recent advances in the emerging field of oxide spin-orbitronics. In the first part, we will cover spin-charge interconversion using oxide materials through the direct and inverse spin Hall and Edelstein effects (Section 1), realized in the prominent oxide platform SrTiO$_3$, as well as in ruthenates and iridates. As a concluding remark in that section, we will cover the ferroelectric control of spin-charge interconversion. Subsequently, we will discuss chiral magnetism and skyrmionic structures in oxide systems (Section 2). Herein, we will cover non-centrosymmetric magnetic oxides, the generation and observation of skyrmions and skyrmion bubbles in oxide heterostructures and lastly the electrical control of skyrmions.

\section*{Spin-charge interconversion with oxides}

\subsection*{SrTiO$_3$-based 2DEGs}

\subsubsection*{Origin of the electron gas}
The 2DEG based on the perovskite SrTiO$_3$ was first discovered by Ohtomo and Hwang~\cite{ohtomo_high-mobility_2004} at the LaAlO$_3$/SrTiO$_3$ (LAO/STO) (001) interface, for LAO thicknesses of at least four unit cells~\cite{stemmer_two-dimensional_2014, yu_polarity-induced_2014, bristowe_origin_2014, cantoni_electron_2012, li_very_2011}. The interest towards the LAO/STO system then grew rapidly. It was favoured on one hand by the prospects of developing oxide-based electronic applications \cite{hwang_emergent_2012,chakhalian_whither_2012}, from field-effect resistance control \cite{cen_nanoscale_2008, thiel_tunable_2006} to solar cells~\cite{assmann_oxide_2013}, and on the other hand, by unexpected properties of the LAO/STO interface, such as the low-temperature (co-)existence of superconductivity and magnetism~\cite{li_coexistence_2011, reyren_superconducting_2007, caviglia_electric_2008, brinkman_magnetic_2007} and the manifestation of an unconventional quantum Hall effect~\cite{trier_quantization_2016}.

\begin{figure}[h!]
    \centering\includegraphics[width=0.8\paperwidth]{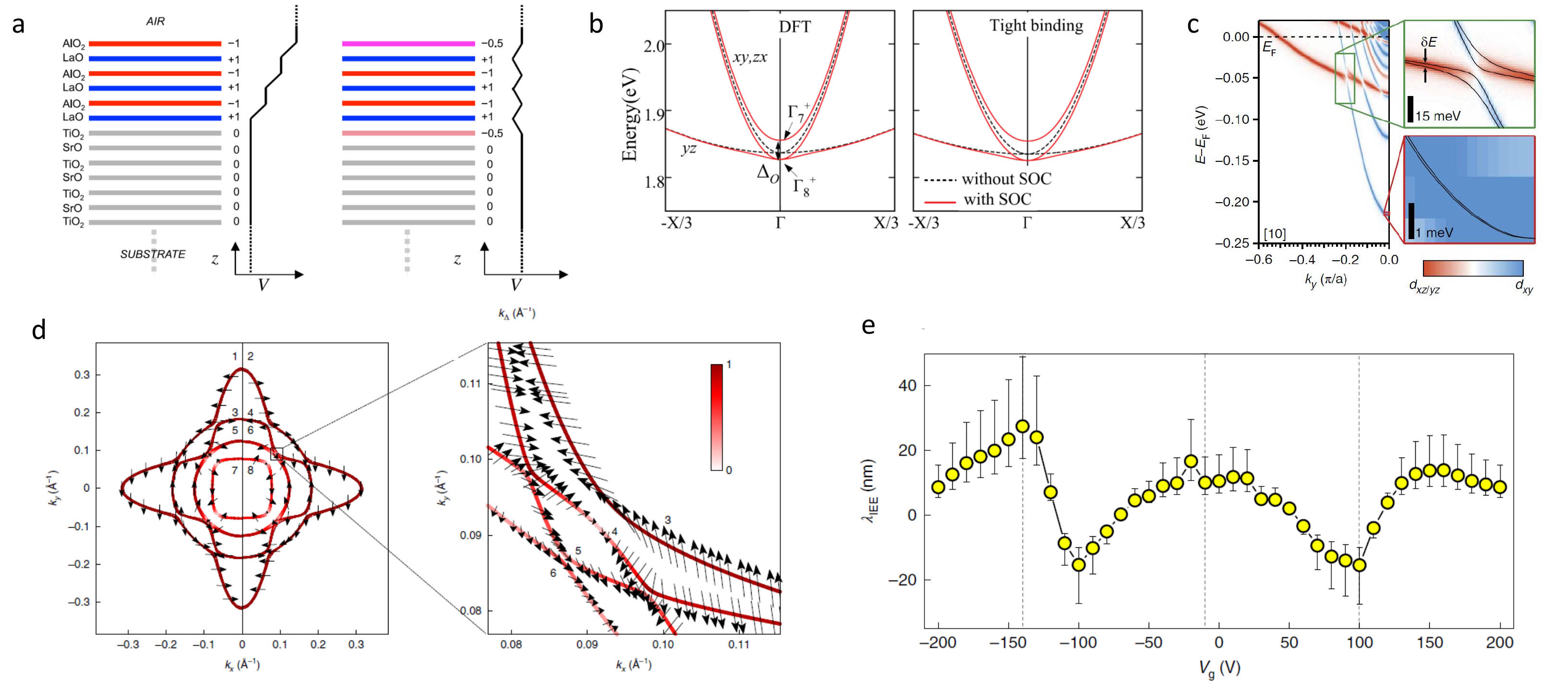}
    \caption{\textbf{SrTiO$_3$-based two-dimensional electron gas: origin and spin-dependent band structure}. (a) Sketch from Ref. \cite{bristowe_origin_2014} illustrating the polar catastrophe, based on atomic layers considered as charged planes, with the net charge (in units of electron/surface unit cell) given by ionic formal charges. The $z$ axis corresponds to the [001] direction. In the left case, the potential V diverges with increasing LAO thickness. In the right case, the electronic reconstruction leads to the appearance of an electron gas at the interface. 
    (b) Band structure of $t_{2g}$ orbitals in bulk STO, calculated in ref. \cite{zhong_theory_2013} by DFT and tight binding methods, exhibiting the heavy (\textit{yz}) and light (\textit{xy}, \textit{zx}) bands.
    (c) Electronic structure of the LAO/STO 2DEG along the [010] direction, from ref. \cite{king_quasiparticle_2014}. The magnified view reveals a weak Rashba-type spin splitting around the band bottom, which becomes enhanced by approximately one order of magnitude at the crossings of the light and heavy sub-bands.
    (d) Calculated Fermi surface and spin expectation values (direction given by the arrows) from ref. \cite{vaz_mapping_2019}, for a Fermi level near the band inversion region, and zoom on a zone with enhanced spin-splitting. The numbers denote the bands in energetically ascending order. The colour scale shows the absolute spin expectation value.
    (e) Spin to charge conversion efficiency (inverse Edelstein length $\lambda_{IEE}$) as a function of the back-gate voltage in a NiFe/Al/STO sample, measured through spin pumping ferromagnetic resonance experiments in Ref.~\cite{vaz_mapping_2019}. The conversion efficiency depends on the Fermi level position, and can be controlled both in amplitude and in sign.
}
\label{fig1}
\end{figure}

The initial explanation for the appearance of the 2DEG at the LAO/STO interface was based on the so-called polar catastrophe model~\cite{nakagawa_why_2006, chakhalian_whither_2012}. LAO can be viewed as consisting of alternating charged planes of LaO$^+$ and AlO$_2^{-}$ which result in a polar discontinuity appearing at the interface when deposited on top of the non-polar STO. When the diverging potential due to this polar discontinuity becomes large enough, an electronic reconstruction will spontaneously occur, whereupon half an electron per 2D unit cell is transferred from the LAO valence band at the surface to the STO conduction band at the interface (see Fig.~\ref{fig1}a).

Although this model explains why the 2DEG only appears for LAO thicknesses greater than four unit cells, it has been challenged by several observations. In particular, the polar catastrophe model fails to explain the appearance of a 2DEG when amorphous oxide overlayers are deposited on STO \cite{chen_metallic_2011, lee_creation_2012, liu_origin_2013} as well as the detection of Ti $3d$-like states below four unit cells of LAO by photoemission spectroscopy~\cite{sing_profiling_2009, takizawa_electronic_2011, rubano_spectral_2011, slooten_hard_2013}. Whereas cation intermixing seems to have been ruled out as the source of interface conductivity, these experimental observations rather point toward oxygen vacancies as the source of interfacial charge carriers, in particular in samples which have not been annealed in oxygen after deposition. Even though this open question has not yet been formally agreed upon in the research community, it appears that both the polar catastrophe as well as oxygen vacancies can contribute to 2DEG formation~\cite{yu_polarity-induced_2014}. At high deposition temperatures and oxygen pressures, electronic reconstruction should however become the dominant effect.  

In the following years, several other STO-based heterointerfaces were found to display metallic 2DEGs, for instance when STO is in contact with LaFeO$_3$ \cite{xu_reversible_2017},
KTaO$_3$ \cite{maznichenko_formation_2020,oja_d0_2012}, LaTiO$_3$ \cite{ohtomo_artificial_2002,ohtsuka_transport_2010}, LaGaO$_3$ \cite{perna_conducting_2010}, LaVO$_3$ \cite{hotta_polar_2007}, KNbO$_3$ \cite{oja_d0_2012}, NaNbO$_3$ \cite{oja_d0_2012}, GdTiO$_3$ \cite{moetakef_transport_2011}, LaVO$_3$ \cite{he_metal-insulator_2012}, NdGaO$_3$ \cite{annadi_electronic_2012},
PrAlO$_3$ \cite{annadi_electronic_2012,nazir_charge_2011}, 
NdGaO$_3$ \cite{annadi_electronic_2012,nazir_charge_2011}, 
NdAlO$_3$ \cite{nazir_charge_2011} or $\gamma$-Al$_2$O$_3$ \cite{chenHighmobilityTwodimensionalElectron2013}.

Interestingly, various studies showed that oxygen vacancies may provide mobile carriers and result in a 2DEG even if STO is not in hard contact with a material. A metallic gas can thus be obtained at the vacuum-cleaved surface of STO, independently of the STO bulk carrier density over more than seven decades \cite{santander-syro_two-dimensional_2011}. The carrier density of the gas appearing at a bare STO surface can be controlled through exposure of the surface to intense ultraviolet light \cite{meevasana_creation_2011}.

The use of epitaxially-grown oxide heterostructures and of bare surfaces might be relatively impractical for mass production, mostly because of contacting, deposition temperatures and reproducibility issues. A way to overcome this technical challenge was solved by the demonstration that 2DEGs can also be realised when a thin layer of an elementary reducing agent, such as Al, is deposited at the surface of STO \cite{rodel_universal_2016}. This breakthrough discovery represents an important step towards the development of low-cost oxide-based applications. Angle-resolved photo-emission spectroscopy (ARPES) and transport experiments have furthermore shown that the deposition of a ferromagnetic material above such a reducing layer preserves the 2DEG  \cite{vaz_tuning_2017, vaz_mapping_2019, noel_non-volatile_2020}.

\subsubsection*{Band structure and spin-charge interconversion}
With mobility values at low temperatures comparable or larger than that of silicon \cite{trier_electron_2018}, and the possibility to tune the transport properties using a back-gate \cite{thiel_tunable_2006, hurand_field-effect_2015}, STO-based 2DEGs are promising for electronic applications. For spintronics, the interest of STO-based 2DEGs lies in their peculiar band structure that gives rise to unparalleled high spin-charge conversion efficiencies \cite{lesne_highly_2016, vaz_mapping_2019, noel_non-volatile_2020}. This band structure had been studied both experimentally, using ARPES \cite{king_quasiparticle_2014, vaz_mapping_2019}, and theoretically \cite{zhong_theory_2013, khalsa_theory_2012, king_quasiparticle_2014}, in different systems (LAO/STO, metals/STO, vacuum/STO). In all cases, the interface band structure observed in ARPES is related to the bulk STO band structure, with 2DEG wave functions spreading across several STO atomic layers \cite{khalsa_theory_2012}. Bulk STO is a band insulator with unoccupied Ti 3$d$ $t_{2g}$ ($d_{yz}$,$d_{zx}$,$d_{xy}$) bands (in the absence of spin-orbit coupling). The $t_{2g}$ band structure calculated by DFT is shown in Fig. \ref{fig1}b, with a small energy dispersion for the heavy $d_{yz}$ band, and light $d_{xy}$ and $d_{zy}$ bands.

For STO-based heterointerfaces, the quantum well extends 200-300 meV below the Fermi level \cite{king_quasiparticle_2014, vaz_mapping_2019} (see Fig. \ref{fig1}c). The confinement in the 2DEG leads to the creation of sub-bands, so that the band structure is comprised of several light states having dominantly $d_{xy}$ orbital character, with corresponding concentric circular Fermi surfaces, and of heavy states deriving from $d_{xz/yz}$ orbitals, with Fermi surfaces ellipsoidal along the (010) and (100) directions. As the inversion symmetry is broken, the spin-orbit coupling (SOC) lifts the spin degeneracy of the 2DEG. Band inversion and orbital mixing lead to a spin splitting that is enhanced in certain \textit{k}-space points, and in particular at trivial and topologically non-trivial avoided band crossings \cite{vaz_mapping_2019} (see Fig. \ref{fig1}b,c). These features of the band structure have two major implications for spintronic applications. Firstly, the enhancement of the Rashba-like splitting can lead to highly spin-dependent transport properties. Secondly, these properties depend on the position of the Fermi level in the band structure, and can thus be tuned by electric fields from a gate. The possibility of modifying the magnitude of the Rashba SO interaction by using gate voltages was first explored in Refs. \cite{caviglia_tunable_2010, ben_shalom_tuning_2010}. Beyond this magnetoresistance control, the existence of tunable SOC effects in STO-based 2DEGs offers a new degree of freedom, which does not exist for the spin Hall effect in heavy metals, and thus provides an exciting way to control several spin-orbitronic effects.

\begin{figure}[h!]
    \centering\includegraphics[width=0.4\paperwidth]{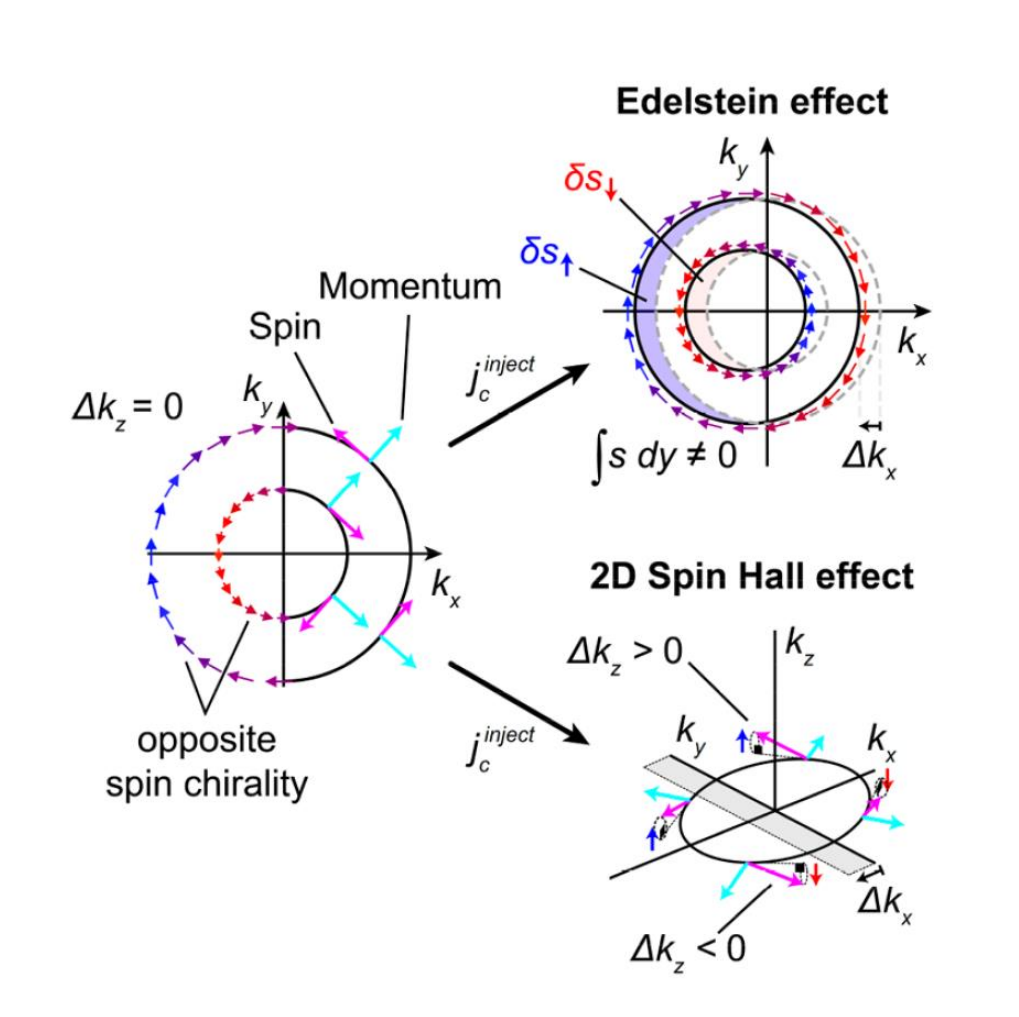}
    \caption{\textbf{Box1: Rashba-Edelstein effect and two-dimensional spin Hall effect}: The presence of finite spin-orbit coupling in a two-dimensional electron gas induces a spin splitting, which, in a simple free-electron picture with parabolic dispersion, results in two concentric circular Fermi contours with opposite spin chiralities. In these Fermi contours, the spins are locked perpendicularly to the momentum (left). This gives rise to two different spin-charge conversion possibilities. \textit{i)} The first is the direct Edelstein effect (top right), where a current injected along $x$ induces a shift of the Fermi contour, generating a spin accumulation $\delta s= \delta s\uparrow-\delta s \downarrow$ along $y$. This spin accumulation can diffuse towards a neighboring ferromagnetic layer, thus generating a spin current and eventually spin-orbit torques. The inverse Edelstein effect corresponds to the reciprocal mechanism: an incoming spin current generates a spin accumulation, spin-polarised along $y$, and thus a charge current along $x$. \textit{ii)} The second charge-to-spin conversion possibility is the direct two-dimensional spin Hall effect (bottom right). When the Fermi contours are displaced by $\Delta k$, the spins are not exactly perpendicular to $k$ and then start to precess around the local Rashba field, ultimately acquiring a finite component along +$z$ for $k_yY>0$ and –$z$ for $k_y<0$. The charge current applied along X thus results in the motion of electrons with opposite spins in opposite directions, i.e., in a pure spin current with a spin polarization along $z$. The reciprocal effect, the inverse 2D spin Hall effect, is also possible, a spin current being then converted into a charge current. Adapted from ref. \cite{trier_electric-field_2019}.}
\end{figure}

The possibility to tune the spin-to-charge conversion efficiency, i.e., the inverse Rashba-Edelstein effect (see Box 1), has been demonstrated using spin-pumping techniques \cite{lesne_highly_2016, vaz_mapping_2019}. The figure of merit of the conversion, the inverse Edelstein length ($\lambda_{IEE}$), is very large at low temperatures (\textit{T} $\approx$ 10 K), reaching $\lambda_{IEE}$ = 7 nm in LAO/STO \cite{lesne_highly_2016} and $\lambda_{IEE}$ = 20 nm in Al/STO \cite{vaz_mapping_2019}. For comparison, the product of the spin Hall angle by the spin diffusion length is equal to approximately 0.2 nm for Pt. These values of $\lambda_{IEE}$ in STO-based 2DEGs are higher than those observed in topological insulators or at Rashba interfaces with very high SOC, which has been attributed to the long carrier relaxation lifetime in STO \cite{shen_microscopic_2014, zhang_conversion_2016}, and to the spin-splitting enhancement due to orbital mixing at the vicinity of avoided band-crossing (cf. Fig. \ref{fig1}d) \cite{vaz_mapping_2019}. It is however important to note that the value of the inverse Edelstein length decreases to around 0.5 nm at room temperature \cite{chauleau_efficient_2016, vaz_mapping_2019}. As visible in Fig. \ref{fig1}e, both the sign of the conversion and its amplitude are found to vary spectacularly with the gate voltage. These variations are qualitatively close to what can be theoretically predicted from the band structure \cite{vaz_mapping_2019}. Additionally to the gate voltage, a control of the conversion might be obtained using strain \cite{sahin_strain_2019}. Beyond these spin-pumping experiments, attempts have also been made to detect the conversion using electrical spin injection from LSMO to the LAO/STO 2DEG \cite{telesio_study_2018}.

Whereas the spin-to-charge conversion could find applications in MESO devices (MESO stands for Magneto-Electric Spin-Orbit) \cite{manipatruni_scalable_2019}, the charge-to-spin conversion could be used to manipulate the magnetization by spin-orbit torques \cite{miron_perpendicular_2011, liu_spin-torque_2012}, leading to new reconfigurable spin-orbit torque memories and logic gates, benefiting to skyrmions or domain wall manipulation, and allowing the development of agile THz emitters and spin-wave logic architectures \cite{dieny_opportunities_2020, grollier_neuromorphic_2020}.

The charge-to-spin conversion has been observed in CoFeB/LAO/STO using the spin-torque ferromagnetic resonance (STFMR) technique~\cite{wang_room-temperature_2017} with a spin/charge current ratio at room temperature of 0.6 nm$^{-1}$. This ratio was found to decrease as the temperature was reduced, an observation explained by the authors by the suppression of inelastic tunneling at lower temperatures, which should effectively prevent the spin accumulation to access the ferromagnet. In LSMO/LAO/STO, planar Hall effect measurements also indicate that an in-plane current creates an effective in-plane field exerted on the magnetization, orthogonal to the current direction~\cite{yamanouchiCurrentinducedEffectiveMagnetic2020}.

The spin-charge interconversion can also be realised by the two-dimensional spin Hall effect (2D SHE, see Box 1)~\cite{sinova_universal_2004}, which leads to a pure spin current transverse to the applied current with an out-of plane spin polarization. The electrical generation and detection of this spin current through the direct and inverse 2D SHE have been demonstrated experimentally \cite{jinNonlocalSpinDiffusion2017} in LAO/STO nanoscale devices \cite{trier_electric-field_2019}. Moreover, both the spin diffusion length and conversion efficiency were demonstrated to be largely tunable by a backgate voltage\cite{trier_electric-field_2019}.

Beyond the spin-charge conversion, the symmetry breaking at the LAO/STO interface, combined with an externally applied magnetic field, lead to the appearance of non-reciprocal phenomena\cite{tokuraNonreciprocalResponsesNoncentrosymmetric2018} such as a large unidirectional magnetoresistance, often coined the bilinear magnetoresistance (BMR) \cite{dyrdal_spin-momentum-locking_2020, vaz_determining_2020, choe_gate-tunable_2019, he_observation_2018}, i.e., of a resistance component varying linearly with both the applied current and the applied field. As it is basically a consequence of the spin-orbit interaction, it has been shown that this likewise can be tuned using a backgate voltage~\cite{choe_gate-tunable_2019, vaz_determining_2020}. Importantly, the amplitude of the BMR can be used to extract the Rashba coefficient, with a good agreement with values deduced from weak antilocalization -- but over a broader range of carrier density -- and with theory~\cite{vaz_determining_2020}.

\subsection*{Other oxide-based two-dimensional electron gases and interfaces}

\subsubsection*{KTaO$_3$-based 2DEGs}
\begin{figure}
    \centering
    \includegraphics[width=0.3\paperwidth]{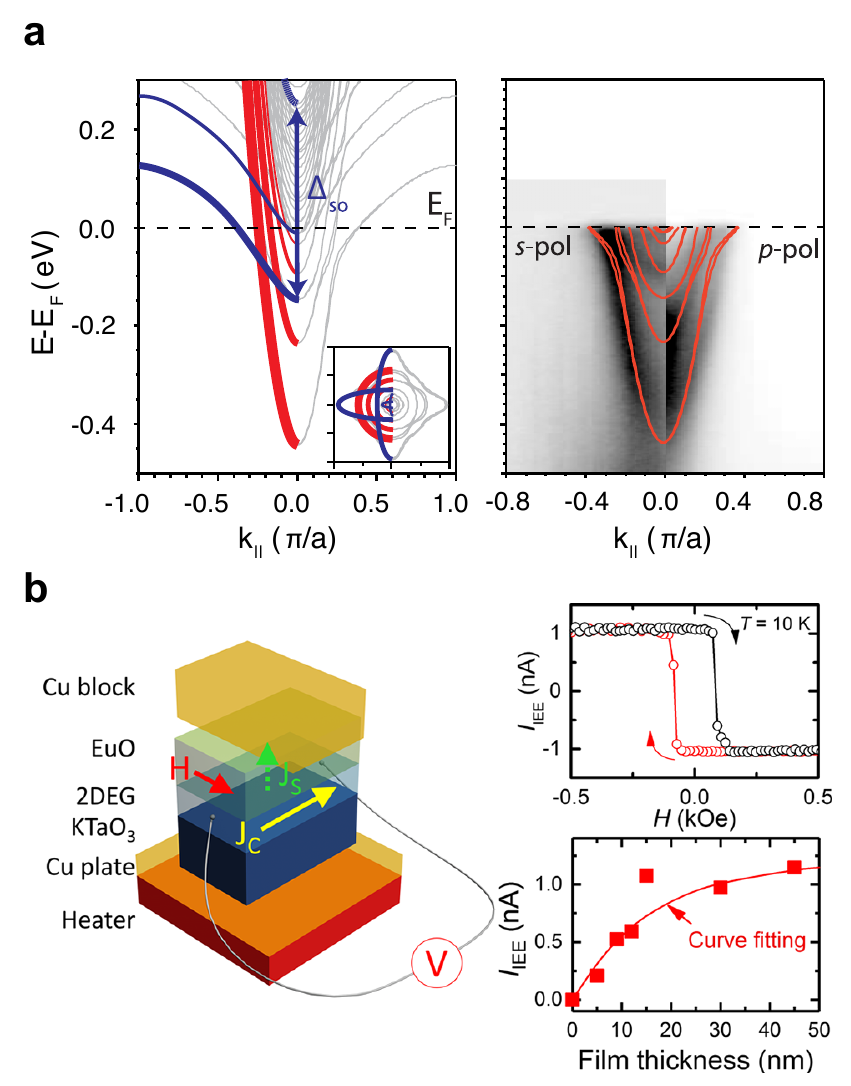}
    \caption{\textbf{Spin-charge conversion in KTaO$_3$ two-dimensional electron gases}. (a) Electronic structure of KTaO$_3$ two-dimensional electron gases, showing a relatively large band-splitting ($\Delta_{so}$) due to the spin-orbit coupling. Figure adapted from reference \cite{king_subband_2012}. (b) Spin-charge conversion at the EuO/KTaO$_3$ interface by the Inverse Edelstein Effect. Figure adapted from reference \cite{zhang_thermal_2019}.}
    \label{fig2}
\end{figure}

Recently, another perovskite oxide system, KTaO$_3$, has attracted the attention of the research community. Similarly to the case of SrTiO$_3$, when KTaO$_3$ is interfaced with other oxide thin films such as LaTiO$_3$ \cite{zou_latio_2015}, amorphous-LaAlO$_3$ \cite{zhang_highly_2017}, EuO \cite{zhang_high-mobility_2018} or LaVO$_3$ \cite{wadehra_planar_2020}, a 2DEG appears at the interface. However, in contrast with SrTiO$_3$, KTaO$_3$ is characterized by the 5\textit{d} heavy element tantalum, which results in a strong band-splitting due to the spin-orbit coupling of 400 meV \cite{nakamura_electric_2009, king_subband_2012, santander-syro_orbital_2012} (cf. Figure \ref{fig2}a). Due to this large spin-orbit coupling, the 2DEG has already been demonstrated to display clear signatures of weak anti-localization at low temperatures \cite{wadehra_planar_2020}. Moreover, when in proximity with the ferromagnetic EuO, the 2DEGs have also been demonstrated to display hysteretic magnetoresistance up to a temperature of 25 K \cite{zhang_high-mobility_2018}. Based on these ingredients, KTaO$_3$-based 2DEGs hold a great potential as a robust platform for spin/charge interconversion experiments. Indeed, Ref. \cite{zhang_thermal_2019} demonstrated a charge current of $\sim$1 nA at 10 K in the EuO/KTaO$_3$ interface, generated through the Inverse Edelstein effect by thermally injecting a spin current from EuO (see Figure~\ref{fig2}b). We finally note that 111-oriented KTaO$_3$ 2DEGs were very recently shown to display superconductivity at 2 K, i.e. one order of magnitude higher than in STO 2DEGs \cite{liuDiscoveryTwodimensionalAnisotropic2020,chenElectricFieldControl2020}.

\begin{table}[ht]
\centering
\begin{tabular}{|l|l|l|}
\hline
Oxide system & Spin-charge interconversion efficiency & Ref. \\
\hline
\hline
LaAlO$_3$/SrTiO$_3$ & $\lambda_{IEE}$ = 6.4 nm at $T$ = 10 K & \cite{lesne_highly_2016} \\
\hline
LaAlO$_3$/SrTiO$_3$ & $\lambda_{IEE}$ = 1 nm at $T$ = 75 K & \cite{chauleau_efficient_2016} \\
\hline
LaAlO$_3$/SrTiO$_3$ & $\lambda_{IEE}$ = 6.7 nm at $T$ = 20 K & \cite{ohya_efficient_2020} \\
\hline
LaAlO$_3$/SrTiO$_3$ & $q_{DEE}$ = 0.63 nm$^{-1}$ at $T$ = 300 K &  \cite{wang_room-temperature_2017} \\
\hline
Al/SrTiO$_3$ & $\lambda_{IEE}$ = 28 nm at $T$ = 15 K & \cite{vaz_mapping_2019} \\
\hline
Ar$^+$-irradiated SrTiO$_3$ & $\lambda_{IEE}$ = 0.23 nm at $T$ = 300 K & \cite{zhang_spin_2016} \\
\hline
\hline
Indium Tin Oxide & $\theta_{SHE}$ = 0.65 $\%$ and $\lambda_{S}$ = 30 nm at $T$ = 300 K & \cite{qiu_experimental_2013} \\
\hline
Polycrystalline IrO$_{2}$ & $\theta_{SHE}$ = 4.0 $\%$ and $\lambda_{S}$ = 3.8 nm at $T$ = 300 K & \cite{fujiwara_5d_2013} \\
\hline
SrRuO$_{3}$ & $\theta_{SHE}$ = 2.7 $\%$ and $\lambda_{S}$ = 1.5 nm at $T$ = 190 K & \cite{wahler_inverse_2016} \\
\hline
SrIrO$_{3}$ & $\theta_{SHE}$ = 30-50$\%$ at $T$ = 300 K & \cite{everhardt_tunable_2019} \\
\hline
Sr$_{2}$IrO$_{4}$ & $\theta_{SHE}$ = 10$\%$ at $T$ = 300 K & \cite{everhardt_tunable_2019} \\
\hline
SrIrO$_{3}$ & $\theta_{SHE}$ = 86 $\%$ along [-110] at $T$ = 70 K & \cite{liu_current-induced_2019} \\
\hline
Cu/Bi$_{2}$O$_{3}$ & $\lambda_{IEE}$ = -0.17 nm at $T$ = 300 K & \cite{tsai_clear_2018} \\
\hline
Ag/Bi$_{2}$O$_{3}$ & $\lambda_{IEE}$ = 0.15 nm at $T$ = 300 K & \cite{tsai_clear_2018} \\
\hline
\end{tabular}
\caption{Spin-charge interconversion efficiencies of oxides. $\theta_{SHE}$ is the spin Hall angle, $\lambda_{S}$ the spin diffusion length and $q_{DEE}$ the charge-spin conversion efficiency by the direct Edelstein effect.}
\label{tab1}
\end{table}

\subsubsection*{BaSnO$_3$-based 2DEGs}
The promising transparent conductor BaSnO$_3$ represents yet another system which within the last years has attracted a great deal of attention. When covered with LaInO$_3$ \cite{kim_conducting_2016,kim_interface_2019}, the near-interface region of BaSnO$_3$ shows a 2DEG character. The most attractive feature of this 2DEG system is its excellent room temperature conductivity, owing to the combination of a very large electron mobility ($\sim$300 cm$^{2}$/Vs) \cite{kim_physical_2012}, exceeding that found in SrTiO$_3$ by two orders of magnitude \cite{trier_electron_2018}, and to a high carrier concentration ($\sim 10^{20}$ $\mathrm{cm}^{-3}$). Due to these unparalleled room-temperature transport properties, this 2DEG system is expected to hold a great potential for applications. Nonetheless, it is yet to be investigated as a possible spin/charge interconversion platform.

\subsubsection*{ZnO-based 2DEGs}
Moving beyond perovskite systems, several other oxide-based two-dimensional electron gases have been studied. The most prominent example is the high mobility interface between Mg$_x$Zn$_{1-x}$O and ZnO \cite{tsukazaki_quantum_2007}. For this 2DEG system, the carrier density and mobility vary greatly with the amount of magnesium substitution (i.e., $x$). For $x = 0.15$ and $x = 0.2$, sheet carrier densities of $n_s \sim 0.66\times10^{12}$ and $3.7\times10^{12}$ cm$^{-2}$, with corresponding electron mobilities of $\mu \sim$ 5500 and 2700 cm$^2$/Vs, were demonstrated at 1 K. These relative large mobility values, considering that the electron mass in ZnO of 0.3 m$_\mathrm{e}$, allowed the first demonstrations in an oxide of the quantum Hall effect~\cite{tsukazaki_quantum_2007} and of the fractional quantum Hall effect~\cite{tsukazaki_observation_2010} in samples grown by molecular beam epitaxy. These observations underline the cleanliness and extreme smoothness of the MgZnO/ZnO interface. Nonetheless, this 2DEG system only displays a moderately small Rashba spin orbit coupling strength of 0.7 meV.Å \cite{kozuka_rashba_2013}, and therefore has not yet served as the basis of any spin/charge interconversion experiment.

\subsection*{Ruthenates, iridates, and other oxide systems}

Apart from the interface 2DEGs covered up to this point, a large variety of bulk conductive oxides have recently emerged as promising systems for spintronic applications based on the spin Hall effect (SHE). The first experimental demonstration of a detectable spin-to-charge conversion in a conductive oxide was provided for indium tin oxide (ITO) \cite{qiu_all-oxide_2012}. These results paved the way for the possibility to use all-oxide systems in spintronics, with a ferromagnetic oxide as the spin current source and an oxide as the detector. Nonetheless, in ITO the spin-to-charge current conversion efficiency - the spin Hall angle ($\theta_{SHE}$) - is small, one order of magnitude smaller than that found in heavy metals \cite{qiu_experimental_2013}, cf Table \ref{tab1}.

\begin{figure}[h!]
    \centering
    \includegraphics[width=0.9\textwidth]{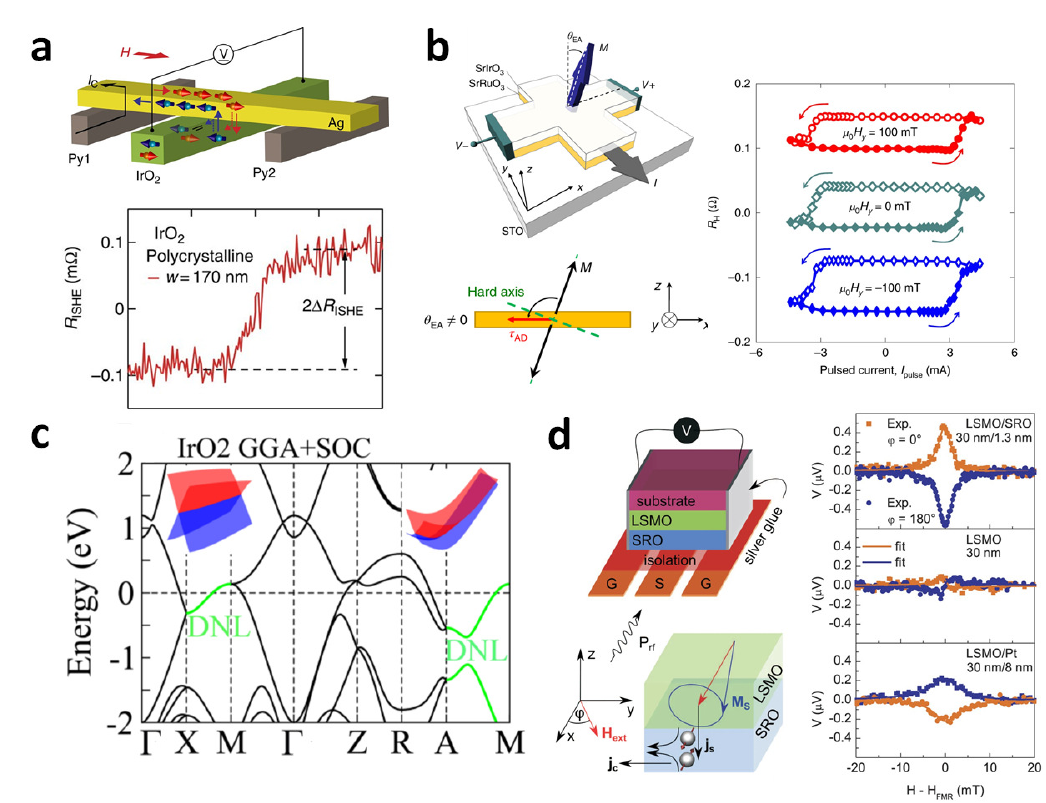}
    \caption{\textbf{Spin charge conversion in iridates and ruthenates.} (a) Inverse spin Hall effect measurement in polycrystalline IrO$_{2}$ using lateral spin valves \cite{fujiwara_5d_2013} (b) Current-induced magnetization switching in SrIrO$_3$/SrRuO$_3$ bilayers. Top left: schematic of the set-up for current-induced magnetization (M) switching in the SrIrO$_3$/SrRuO$_3$ bilayer. The magnetic easy-axis (EA) is tilted away from the $z$-axis with an angle of $\theta_{EA}$. Bottom left: schematic illustration of the spin torque switching behaviour. Here $\theta_{EA} \neq 0$ and a current was applied to switch the $x$-direction tilted magnetization without any external magnetic field. $\tau_{AD}$ refers to the spin transfer torque. The thick dotted green line refers to the magnetic hard axis. Right: current-induced magnetization switching in SrIrO$_3$/SrRuO$_3$ bilayers grown on STO(001). The curves in (b) are manually shifted for better visualization \cite{liu_current-induced_2019}. (c) Calculated band structures for IrO$_{2}$ with spin orbit coupling. The Dirac Nodal lines are highlighted by green lines. \cite{sun_dirac_2017} and (d) Inverse spin Hall effect measured using spin pumping by ferromagnetic resonance in La$_{0.7}$Sr$_{0.3}$MnO$_{3}$/SrRuO$_{3}$ at room temperature compared with the case of La$_{0.7}$Sr$_{0.3}$MnO$_{3}$/Pt. \cite{wahler_inverse_2016}}
    \label{Fig-iridate_ruthenate}
\end{figure}

This limitation can be circumvented by using other conducting oxides containing heavy elements. This is the case of iridium oxide IrO$_{2}$, which possesses a very high spin Hall conductivity, nearly ten times larger than Pt, making it an excellent spin detector \cite{fujiwara_5d_2013} (Fig. \ref{Fig-iridate_ruthenate}a). This large spin Hall conductivity is also accompanied by a low conductivity, and thus the spin Hall angle still remains smaller than that of Pt and is of the order of $\theta_{SHE}$ = 4\%. Recent spin-torque measurements have however indicated a $\theta_{SHE}$ = 9\%\cite{ueda_spin-orbit_2020}, with a spin diffustion length $\lambda_{S}$=1.7 nm. Spin Seebeck measurements yielded a product $\theta_{SHE} \times \lambda_{S}$=0.15 nm, consistent with these results\cite{qiu_all-oxide_2015}. The high spin Hall conductivity is connected to the presence of exotic features in the band-structure of IrO$_{2}$ known as Dirac nodal lines (Fig. \ref{Fig-iridate_ruthenate}c). In the presence of SOC, these Dirac nodal lines can induce a large intrinsic spin Hall effect \cite{sun_dirac_2017}. In strontium iridate SrIrO$_{3}$, similar topologically non-trivial states were also predicted~\cite{carter_semimetal_2012, zeb_interplay_2012}, and observed using angle resolved photo-emission spectroscopy~\cite{nie_interplay_2015, liu_direct_2016}. As shown from calculations of the spin Berry curvature\cite{patri_theory_2018}, the amplitude of the intrinsic spin Hall conductivity in strontium iridate is thus expected to be giant. One additional characteristic of the spin Hall effect in strontium iridate is that it is sensitive to oxygen octahedron tilting and to the lattice symmetry, which offers the possibility to tune the SHE with the thickness \cite{nan_anisotropic_2019}. This giant spin Hall effect was further evidenced using diverse experimental methods \cite{nan_anisotropic_2019, everhardt_tunable_2019, wang_large_2019}, and eventually could lead to current-induced magnetization switching, for instance for memory applications \cite{liu_current-induced_2019} (cf Fig. \ref{Fig-iridate_ruthenate}b). 

The presence of Dirac nodal lines in the band structure of conductive oxides is not limited to iridates and was also predicted in RuO$_{2}$ \cite{sun_dirac_2017}, whose associated large spin Hall angle was further confirmed through spin Seebeck effect measurements, in which the output voltage could be, after material optimization through annealing, larger than that of Pt  \cite{kirihara_annealing-temperature-dependent_2018}. Another ruthenate that attracted much attention recently is SrRuO$_{3}$, a conductive oxide with a ferromagnetic transition below 160 K. While several reports evidence the existence of the spin Hall effect in SrRuO$_{3}$ \cite{haidar_enhanced_2015, richter_spin_2017}, only few experiments quantitatively estimated the conversion efficiency. These results evidence that it possesses a short spin diffusion length \cite{emori_spin_2016} and a large spin Hall angle, comparable to that of Pt \cite{wahler_inverse_2016} (Fig. \ref{Fig-iridate_ruthenate}d). Similarly to the case of SrIrO$_{3}$, the degree of oxygen octahedron tilting plays an important role in the intensity of the spin Hall effect \cite{ ou_exceptionally_2019}. Following the recent rise of interest for the use of ferromagnets as efficient spin Hall effect materials \cite{davidson_perspectives_2020}, one can expect that SrRuO$_{3}$ will attract greater attention in the near future, as a platform to study the interplay between the magnetism and the spin Hall effect in oxide systems. The effect of the ferromagnetic transition is still not fully understood, and contradictory results have been reported \cite{wahler_inverse_2016, ou_exceptionally_2019}.
 
In such oxide systems, a high spin charge interconversion is not limited solely to the bulk or to the presence of a 2DEG but can also occur at interfaces, similarly to the case of the Ag/Bi Rashba interface \cite{rojas_sanchez_spin--charge_2013}. The use of a metallic interlayer can indeed lead to a substantial enhancement of the conversion efficiency in Ag/ITO compared to the bare ITO film \cite{kondou_efficient_2018}. Interestingly, the use of such a multilayer structure to harness the spin-charge conversion at an interface can be used for insulating oxides such as Cu/Bi$_{2}$O$_{3}$ \cite{karube_experimental_2016, tsai_clear_2018}. These results provide evidence that, in addition to the bulk properties or the presence of a 2DEG, the interface plays a critical role in the spin-charge conversion in oxides. Interestingly, iridates and ruthenates are also widely studied when associated, as skyrmionic structures can be observed in SrIrO$_{3}$/SrRuO$_{3}$ bilayers. This will be
discussed in more details in the second section of this review. 

\subsection*{Ferroelectric control of the spin-charge interconversion}

The polar nature of ferroelectrics makes them prime candidates to harbour a Rashba SOC, with the additional advantage that it could also be switchable by an electric field. The last few years have seen efforts towards the identification of single-phase ferroelectric Rashba semiconductors (FERESC)\cite{picozzi_ferroelectric_2014}, with the focus mainly directed towards GeTe, a low bandgap semiconductor and a ferroelectric with a Curie temperature of 700 K. GeTe has been predicted to be a bulk Rashba material\cite{di_sante_electric_2013}, with polarization switching causing a full reversal of the spin texture of the Rashba split Fermi contours. Experimentally, because of high leakage, indications of ferroelectricity have only been provided in thin films using piezo-response force microscopy (PFM)\cite{kolobov_ferroelectric_2014}. The surface band structure of GeTe has been mapped by ARPES, which provides evidence of a strong Rashba splitting depending on the ferroelectric polarization state (in two different samples\cite{rinaldi_ferroelectric_2018}, or on the same in-situ poled sample\cite{krempasky_effects_2008}). Early spin-charge conversion experiments in GeTe-based structures have however shown a low efficiency\cite{rinaldi_evidence_2016}.

\begin{table}[h!]
\centering
\begin{tabular}{|l|l|l|}
\hline
Material & Comment & Ref. \\
\hline
SrTiO$_3$ & Becomes ferroelectric-like at high electric field  & \cite{noel_non-volatile_2020}\\
\hline
SrBiO$_3$ & Predicted to become ferroelectric under epitaxial strain & \cite{varignon_electrically_2019} \\
\hline
BiAlO$_3$ & Switchable Rashba state near conduction band minimum & \cite{da_silveira_rashba-dresselhaus_2016} \\
\hline
PbTiO$_3$ & Switchable Rashba state near conduction band minimum  & \cite{arras_rashba-like_2019} \\
\hline
KTaO$_3$ & Predicted to become ferroelectric under epitaxial strain  & \cite{tao_strain-tunable_2016} \\
\hline
Bi$_2$WO$_6$ & Non perovskite  & \cite{djani_rationalizing_2019} \\
\hline
\end{tabular}
\caption{\label{tab2}Single-phase oxide materials expected to show a ferroelectricity-controlled Rashba SOC.}
\label{FE-Rashba}
\end{table}

In parallel, a switchable Rashba SOC has been predicted in compounds from the perovskite family including BiAlO$_3$ \cite{da_silveira_rashba-dresselhaus_2016}, strained KTaO$_3$\cite{tao_strain-tunable_2016}, strained SrBiO$_3$\cite{varignon_electrically_2019} and PbTiO$_3$\cite{arras_rashba-like_2019} -- see table \ref{FE-Rashba}. Importantly, it has been argued that the coexistence of a large spontaneous polarization and a sizeable spin–orbit coupling is not sufficient to have strong Rashba effects, and elucidated why simple ferroelectric oxide perovskites with transition metal at the B-site are typically not suitable FERSC candidates, i.e., compounds in which bands with a large and electrically switchable Rashba splitting would be present at the conduction band minimum (CBM) \cite{djani_rationalizing_2019}. Indeed, in several oxide FERSC, the Rashba split bands are typically higher in energy than the CBM, implying that when the Fermi level is positioned so as to intersect them, other, non Rashba-split bands will also contribute to the transport, and diminish the ensuing direct or inverse Edelstein effect (note, however, that this is also the situation in STO 2DEGs, that nevertheless show very high inverse Edelstein effect as discussed earlier in this review \cite{vaz_mapping_2019}). Instead, these authors proposed an Aurivillius compound, Bi$_2$WO$_6$, as a possible FERSC in which a different and larger crystal field causes the Rashba-split bands to lie at the CBM~\cite{djani_rationalizing_2019}.

Regarding interface systems, perhaps the first prediction of a ferroelectric control of Rashba SOC was done by Mirhosseini \emph{et al}. for Bi/BaTiO$_3$ interfaces~\cite{mirhosseini_toward_2010}, a system that was later probed experimentally~\cite{lutz_large_2017}. The dependence of the Rashba coefficient was modest, but other studies have also calculated a large (0.1-0.7 eV.Å) and fully switchable Rashba coefficient in various types of perovskite interfaces combining BaTiO$_3$ with BaRuO$_3$, BaIrO$_3$ or BaOsO$_3$~\cite{zhong_giant_2015}.

\begin{figure}[h!]
    \centering
    \includegraphics[width=0.8\textwidth]{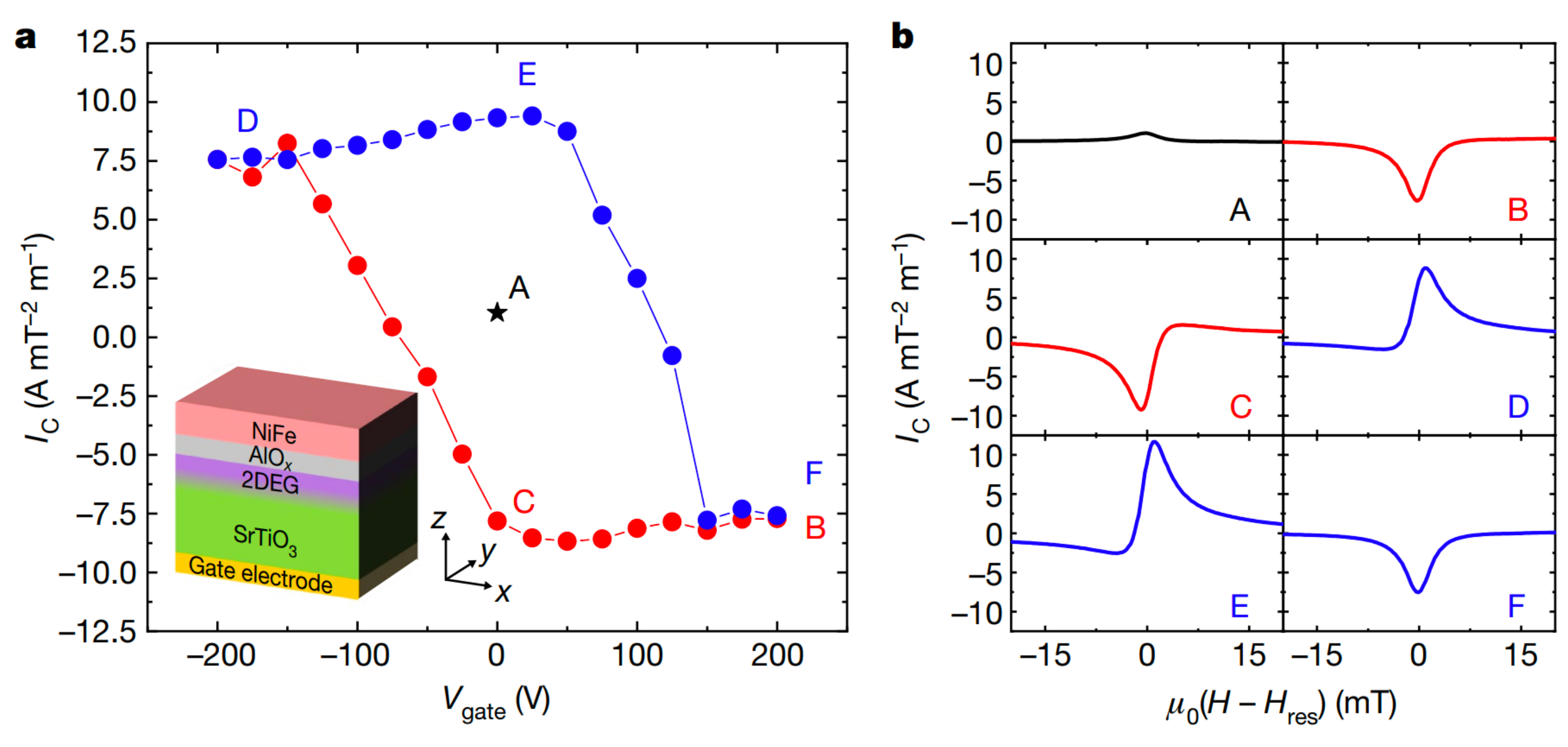}
    \caption{\textbf{Non-volatile electrical control of spin-charge conversion in STO 2DEGs} (a) Gate-voltage dependence of the normalized current produced by the inverse Edelstein effect in the sample sketched in the inset. (b) Magnetic field dependence of the normalized current produced in spin-pumping experiments, for different values of the gate voltage. \cite{noel_non-volatile_2020}
    \label{Fig-FESTO}}
\end{figure}

Experimental demonstrations of a ferroelectric control of the spin-charge interconversion were provided only very recently. Some of us \cite{noel_non-volatile_2020} showed that under high electric fields, the ferroelectric-like behaviour developing in STO -- a feature that had been previously reported, see e.g. \cite{hembergerElectricfielddependentDielectricConstant1995} -- could be harnessed to control the inverse Edelstein effect by ferroelectricity in STO 2DEGs. Working with NiFe/Al/STO samples, these authors observed that both the electrical conductivity spin-to-charge conversion displayed a hysteretic dependence with the gate voltage, cf Fig. \ref{Fig-FESTO}. The conversion was very large, with a record inverse Edelstein length to date, of around 60 nm, but also bipolar, the sign of the current changing with the ferroelectric polarization state. This bipolar property of the conversion shows electrical remanence up to $T \lesssim$ 50 K, which corresponds to the ferroelectric Curie temperature. By applying positive or negative voltage pulses across the STO substrate, the conversion current obtained in spin pumping experiments was fully reversed and stable for at least several hours.

The exact role of the induced ferroelectric state on the band structure and the position of the Fermi level within the band structure remains to be elucidated. A change of ferroelectric polarization, and of the associated electric field, at the interface with the 2DEGs, can lead to the reversal of the chirality of the Fermi contours. Nonetheless, the peculiar avoided crossing points in the band structure are known to play an important role in determining the intensity and the sign of the produced charge current~\cite{vaz_mapping_2019}. This is still an open question, but the data suggest that the ferroelectric state and additional electric field allow increasing the Rashba field, leading to higher conversion efficiencies.

Soon after the publication of these results, two groups showed that by minute Ca substitution for Sr in STO (Ca-STO) -- known to make STO ferroelectric \cite{bednorzSr1xCaxTiO3XYQuantum1984} -- and by depositing a reactive metal like Al \cite{brehin_switchable_2020} or an epitaxial LAO thin film\cite{tuviaFerroelectricExchangeBias2020}, it was possible to tune the electrical conductivity of the 2DEGs through the ferroelectric state of Ca-STO  \cite{brehin_switchable_2020,tuviaFerroelectricExchangeBias2020}. While in Noël \emph{et al.} \cite{noel_non-volatile_2020} the ferroelectric state disappears at around 50 K, Ca-STO exhibit the transition temperature at 28 K. An open question is whether the 2DEG is ferroelectric by itself, or if its vicinity with a ferroelectric material only tunes its carrier density. Indeed, with large doping values and quite extended 2DEGs, the electric field might be quickly screened in the 2DEG. 
More recently, Varotto \emph{et al} demonstrated a ferroelectric control of spin-charge conversion at room temperature with GeTe-based structures, with an efficiency comparable to that of Pt \cite{varotto_room-temperature_2021}. 

This remanent control of the spin-to-charge interconversion is highly interesting for spintronics applications, where the information would be stored in the ferroelectric state rather than on the ferromagnetic state. Beyond spin-orbit torques, the spin-to-charge conversion could for instance be used in devices derived from the MESO transistor proposed recently by Intel~\cite{manipatruni_scalable_2019}. Spin transfer and spin-orbit torque are fast ways to switch the magnetic state, but still require quite a lot of energy, typically in the 10--100 femto-Joule range, whereas atto-joule switching energy can be achieved for ferroelectric state~\cite{manipatruniCMOSComputingSpin2018}. This opens spintronics to the field of ultra low power devices, that add to the non-volatility ultra-low energy switching capabilities.

\section*{Chiral magnetism and skyrmionic structures in oxide systems}

Recent years have seen a renewal of interest for non-collinear spin structures, both in the oxide world -- many multiferroics are oxides and display spiral or cycloidal spin states -- and in metal-based systems. In most cases, spin-orbit coupling is a key ingredient leading to the stabilization of the non-collinear spin texture. While in both oxide and metallic systems most early work focused on bulk materials, thin film heterostructures are now under intense scrutiny and bring additional functionalities in terms of design and control.

\subsection*{Non-collinear spin order in oxides}

Several mechanisms can lead to non-collinear magnetic order~\cite{coeyNoncolllnearSpinStructures1987}. Let us first distinguish systems with inversion symmetry and those with broken inversion symmetry where antisymmetric exchange (Dyzaloshinskii-Moriya interaction, DMI, that depends linearly on the strength of the spin-orbit interaction) can exist and lead non chiral spin textures.

In centrosymmetric systems, non-collinear spin order can simply arise from the geometry of the spin lattice. A typical example is that of a triangular lattice of spins coupled by antiferromagnetic interactions. The impossibility to achieve an antiparallel alignment between all neighbouring spins leads to magnetic frustration and to a non-collinear spin arrangement corresponding for instance to spins aligned along the edges of the triangles, or pointing towards their centre (whilst maintaining a zero net magnetization). A similar situation occurs for spins arranged at the corners of tetrahedra, leading to a rich variety of spin arrangements within and between tetrahedra with (almost) degenerate energies (see e.g. this review on magnetic pyrochlores \cite{gardner_magnetic_2010}). Non-collinear magnetism can also arise when different magnetic interactions coexist and compete. In a system with two magnetic sublattices, at least three magnetic interactions are present, within each sublattice and between spins on different sublattices. If the strengths of these interactions are comparable, the ground state may not be a collinear ferrimagnetic or ferromagnetic state but a non-collinear state. This is the situation in some spinel ferrites~\cite{murthy_yafet-kittel_1969} in which the spins are canted according to the Yafet-Kittel model~\cite{yafet_antiferromagnetic_1952}. 

SrFeO$_3$ and CaFeO$_3$ are two rare examples of perovskite oxides with a spiral spin order. In these negative charge-transfer (CT) compounds, the magnetic order has been proposed to arise from the competition between ferromagnetic nearest-neighbour and antiferromagnetic next-nearest neighbour interactions~\cite{takeda_magnetic_1972} or due to the special situation of double-exchange in such negative CT systems~\cite{mostovoy_helicoidal_2005}. Upon increasing temperature and/or applying a large magnetic field, additional helical phases can be stabilized. Remarkably, some of these phases lead to the observation of a topological Hall effect, suggestive of a topological character for the spin texture, in analogy with the situation in the celebrated B20 alloys~\cite{muhlbauer_skyrmion_2009}. 

The competition between magnetic anisotropies can promote non-collinear magnetic states. A typical situation is that of thin films with a uniaxial anisotropy, for instance a perpendicular anisotropy (induced by epitaxial strain or interface anisotropy) that favors out-of-plane spins, competing with shape anisotropy, favoring spins lying in the film plane. Structural distortions can be another source of uniaxial anisotropy. This mechanism has been successfully used to engineer topological spin textures in oxides, as we will see in the following section.

\begin{figure}[h!]
    \centering
    \includegraphics[width=0.7\textwidth]{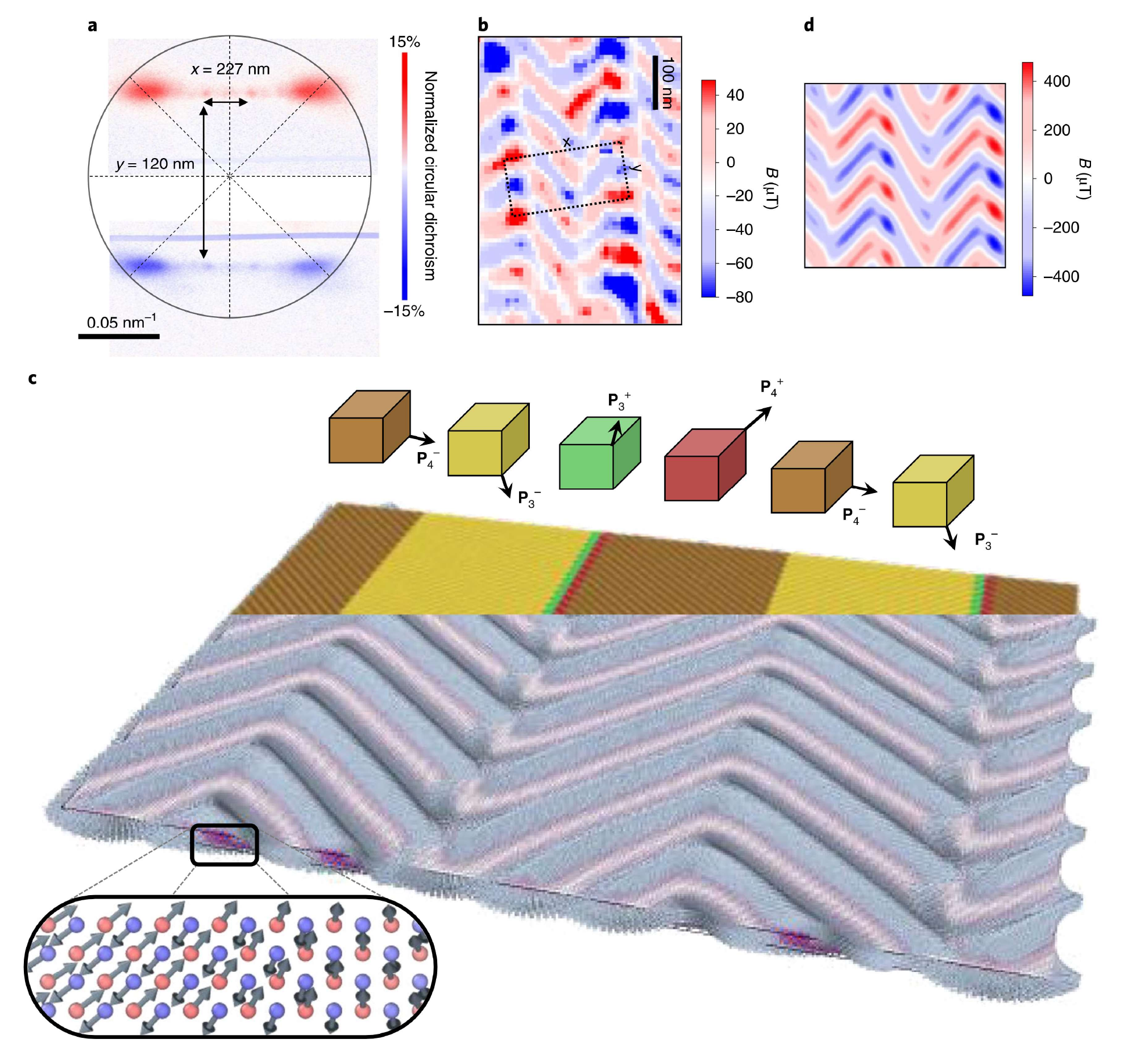}
    \caption{\textbf{Chiral magnetic textures at ferroelectric domain walls seen in reciprocal and real spaces in a BiFeO$_3$ thin film.} (a) Intensity difference between left and right X-ray circular polarizations for a two-cycloid system in X-ray magnetic scattering experiments. Each set of diffraction spots show opposite dichroism, indicating a chiral character. (b) Scanning NV magnetometry image of the same sample evidencing the presence of two cycloids, togteher with a rectangular array of bubbles at ferroelectric domain walls. (c) Magnetic simulations showing the cycloidal spin textures within the ferroelectric stripe domains and at the domain walls (c). Every second wall, the polarization (black arrows) rotates along a long winding consistent with the chirality measured for the ferroelectric order. Panel (d) shows a simulation of the stray field generating by the spin texture, in very good agreement with the NV microscopy image (b) \cite{chauleau_electric_2019}.}
    \label{Fig-BFO}
\end{figure}

The oxide family also hosts several non-centrosymmetric magnetic oxides that present non-collinear spin textures~\cite{izyumov_modulated_1984}. These compounds are usually insulating and their non-collinear spin order often leads to (or is associated with) ferroelectricity, making them multiferroic~\cite{kimuraSpiralMagnetsMagnetoelectrics2007}. A famous example is that of BiFeO$_3$ (BFO), one of the very few room-temperature multiferroics. In the bulk it displays a non-collinear antiferromagnetic state with a long-period ($\approx$ 62 nm) cycloidal modulation~\cite{sosnowska_spiral_1982}. The cycloid order arises from a Dzyaloshinskii-Moriya-like interaction caused by the magnetoelectric coupling between the large ferroelectric polarization and the spins carried by the Fe ions. An additional DMI due to the rotations of the FeO$_6$ octahedra induces a periodic canting of the spins leading to a weak ferromagnetic moment and to a spin density wave with the same periodicity as the cycloid.\cite{burnsExperimentalistGuideCycloid2020} In epitaxial thin films the cycloid state can see its period or propagation direction modified by strain \cite{sando_crafting_2013, agbelele_strain_2017, haykal_antiferromagnetic_2020}; very large strain even destroys it, inducing a transition to a weak-ferromagnetic state. The spin textures are also sensitive to the ferroelectric domain configurations (the cycloid propagation vector depends on the polarization direction) and at ferroelectric domain walls, interesting bubble-shaped chiral spin textures that are possible skyrmion embryos \cite{chauleau_electric_2019} have been observed, see Fig. \ref{Fig-BFO}. Engineering antiferromagnetic skyrmions in BFO heterostructure would open exciting possibilities given the superiority of antiferromagnetic skyrmions~\cite{legrand_room-temperature_2020} over ferromagnetic ones in terms of operation frequency and displacement speed~\cite{zhang_antiferromagnetic_2016}, for example, and of course due to the possibility to control them by electric field at low power.

In addition to BFO, several other multiferroic oxides display non-collinear magnetic order. This is the case of rare-earth manganite perovskites such as TbMnO$_3$ \cite{kimura_magnetic_2003} and more complicated compounds such as TbMn$_2$O$_5$ \cite{hur_electric_2004}. Unlike in BFO where ferroelectricity is caused by the lone-pair mechanism, leading to an off-centering of the Bi ions, in these materials the ferroelectric state is a consequence of the non-collinear (cycloidal) spin order (spin-driven ferroelectricity \cite{katsura_spin_2005}). The description of their rich physics goes beyond the scope of the present paper and we refer the reader to the review by Fiebig \textit{et al.} published in this journal in 2016 \cite{fiebig_evolution_2016}. Despite the obvious potential for skyrmion physics in multiferroics and efforts in this direction \cite{kurumaji_spiral_2020}, only Cu$_2$OSeO$_3$ and related compounds have been shown to exhibit skyrmions \cite{seki_observation_2012}. 

\begin{figure}[ht]
\centering
\includegraphics[width=\linewidth]{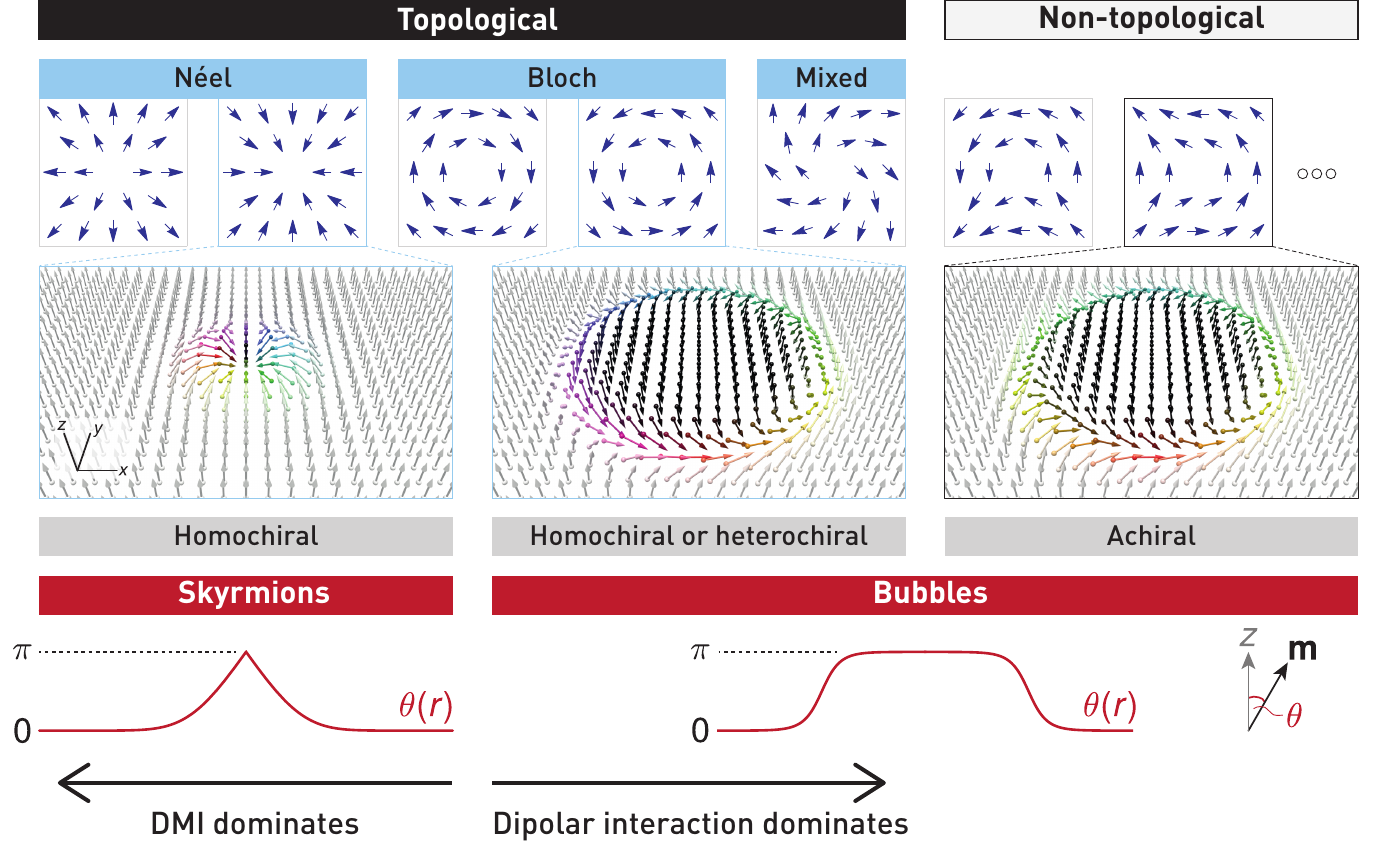}
\caption{\textbf{Box 2 | Skyrmions and magnetic bubbles.} These spin textures can be classified as topological if they possess a nonzero value of the topological charge, which is given by the quantity $\int dr^2 \; \mathbf{m} \cdot \left(\partial_x \mathbf{m} \times \partial_y\mathbf{m} \right)$, where $\mathbf{m}$ represents the magnetization vector. The in-plane component of the magnetization takes on a Néel-type configuration for interface-driven DMI, while Bloch-type configurations are favoured by bulk DMI, dipolar interactions, or both. Competing interactions can result in mixed Néel/Bloch states, while non-topological bubbles possess a mixture of opposite chiralities. Skyrmion states are favoured when the DMI dominates over dipolar interactions and are characterized by a compact core with a peak-like variation in the polar angle of the magnetization, $\theta(r)$, as a function of the radial distance $r$ from the core centre. Magnetic bubbles are favoured when the dipolar interaction is dominant and are characterized by a smooth plateau in $\theta(r)$ at their centre. Skyrmions are homochiral, while bubbles can either be homochiral if the DMI is sufficiently large, or heterochiral or achiral if the DMI is small or nonexistent.}
\label{Box2}
\end{figure}

\subsection*{Skyrmions and skyrmion bubbles in oxide heterostructures}

\subsubsection*{Manganites}

Mixed-valence manganites~\cite{tokura_colossal_1999} display rich phase diagrams where hole doping (by, e.g., Sr or Ca) into the parent compound LaMnO$_3$, an antiferromagnetic insulator, results first in charge and orbital-ordered ferromagnetic and insulating phases and then to ferromagnetic and metallic phases due to the onset of double-exchange interaction. At optimal doping (around 0.33) the Curie temperature is maximum and reaches 360 K in La$_{0.67}$Sr$_{0.33}$MnO$_3$. This compound is a half-metal~\cite{bowen_spin-polarized_2005} that has been widely used in spintronics devices such as magnetic tunnel junctions or spin filters~\cite{bibes_oxide_2007}. At high doping, manganites typically display insulating behavior and antiferromagnetism, but in some compounds, slight electron doping from the other end member (SrMnO$_3$ or CaMnO$_3$) produces a metallic phase with weak-ferromagnetism~\cite{sakai_electron_2010, caspi_structural_2004}.

Mixed-valence manganites are centrosymmetric systems in which a uniaxial anisotropy can arise at structural phase transitions or be engineered through epitaxial strain in thin films. Most compounds have Curie temperatures below 300 K, which makes the direct observation of nanoscale spin textures challenging, as compared with magnetic multilayers based on Co or Fe. Nevertheless, magnetic bubbles have been imaged by Lorenz microscopy in bulk manganite specimens \cite{nagai_formation_2012, yu_biskyrmion_2014, yu_variation_2017, kotani_observation_2016, nagao_direct_2013}. Despite being formed in the absence of DMI (but through long-range dipolar interactions), these bubbles, 100-200 nm in diameter, can possess a topological character and carry a finite topological charge albeit with different chiralities (see Box 2). 

\begin{figure}[h!]
    \centering
    \includegraphics[width=0.6\textwidth]{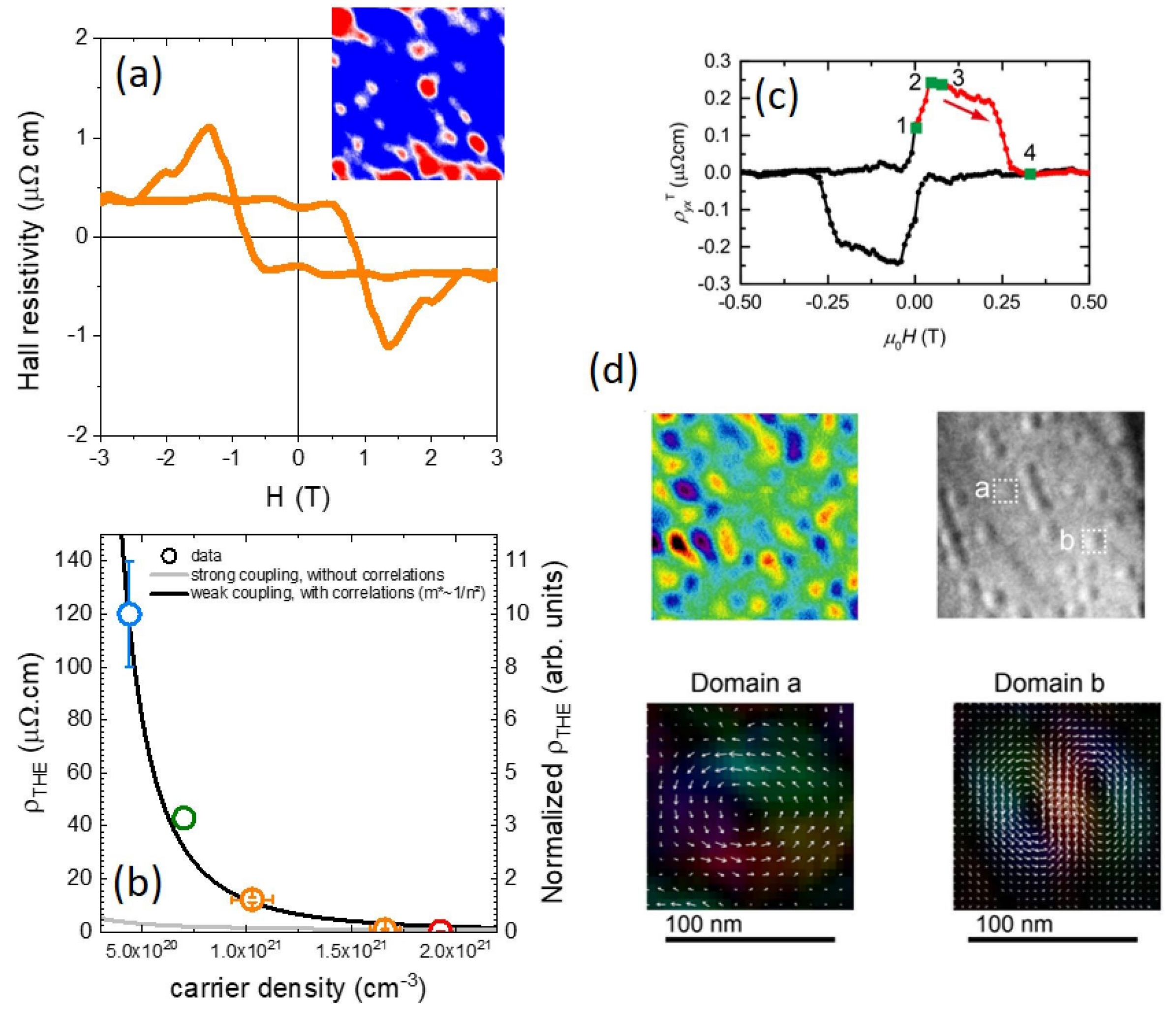}
    \caption{\textbf{Topological Hall effect in manganite heterostructures} (a) Hall effect at 15 K in a Ca$_{0.96}$Ce$_{0.04}$MnO$_3$ thin film. The inset shows a 3 $\mu$m $\times$ 3 $\mu$m MFM image. Relationship between the THE and the carrier density for various Ca$_{1-x}$Ce$_{x}$MnO$_3$ thin films. The grey and black lines correspond to a model in two different coupling regimes.~\cite{vistoli_giant_2019} (c) Topological Hall effect at 10 K for a La$_{0.7}$Sr$_{0.3}$Mn$_{0.95}$Ru$_{0.05}$O$_3$ thin film. (d) (top) 1 $\mu$m $\times$ 1 $\mu$m MFM (left) and Lorentz microscopy (right) images at the same temperature and 70 mT (point 3 in (c)) ; (bottom) Result of the transport-of-intensity equation analyses for the Lorentz TEM image in two specific areas labelled a and b. Domain a is a single skyrmion with skyrmion number $N=1$ and domain b is a biskyrmion with $N=2$.~\cite{nakamura_emergence_2018}}
    \label{Fig-THE-manganite}
\end{figure}

In manganite thin films, compressive strain promotes perpendicular anisotropy, which led to attempts to generate and observe topological bubbles and skyrmions. Ca$_{0.96}$Ce$_{0.04}$MnO$_3$ (CCMO) is a canted ferromagnet with a magnetization of $\sim0.8 \mu_B$/Mn that develops a strong PMA when grown on YAlO$_3$ substrates \cite{xiang_phase_2012}. Hall measurements below the Curie temperature ($\sim$110 K) revealed the presence of a hump associated with the topological Hall effect (THE) \cite{vistoli_giant_2019}, see Fig. \ref{Fig-THE-manganite}a. Magnetic force microscopy (MFM) showed the presence of bubble-like features, whose density is maximum at the peak of the THE. Interestingly, the amplitude of the THE was found to strongly increase with reducing carrier density in Ca$_{1-x}$Ce$_{x}$MnO$_3$ with various doping level, possibly due to enhanced electron correlations upon approaching the charge-transfer insulating state of the parent compound CaMnO$_3$ \cite{nakazawa_weak_2019,nakazawa_topological_2018}, \emph{cf} Fig.~\ref{Fig-THE-manganite}b. In La$_{0.7}$Sr$_{0.3}$Mn$_{0.95}$Ru$_{0.05}$O$_3$ thin films with tailored perpendicular anisotropy, another group also observed a sizeable topological Hall effect (Fig.~\ref{Fig-THE-manganite}c), ascribed to the presence of topological spin textures observed both by MFM and Lorenz microscopy. A variety of skyrmions were found, as illustrated in Fig. \ref{Fig-THE-manganite}d.

Even in the absence of DMI induced by interfacing with materials based on heavy elements, skyrmions and topological Hall effects have thus been observed in manganites. However, there have also been attempts to stabilize skyrmions in bilayers combining manganites with iridates -- where the interface breaks the inversion symmetry  \cite{skoropata_interfacial_2020,li_emergent_2019} -- in part motivated by theoretical predictions \cite{mohanta_topological_2019}. For instance, in ref \cite{skoropata_interfacial_2020} the authors combined LaMnO$_3$ with SrIrO$_3$ to craft chiral spin textures producing a THE signal. In spite of such efforts, unambiguous signatures of skyrmions in such heterostructures remain elusive.

\subsubsection*{SrRuO$_3$}
\begin{figure}[h!]
    \centering
    \includegraphics[width=\textwidth]{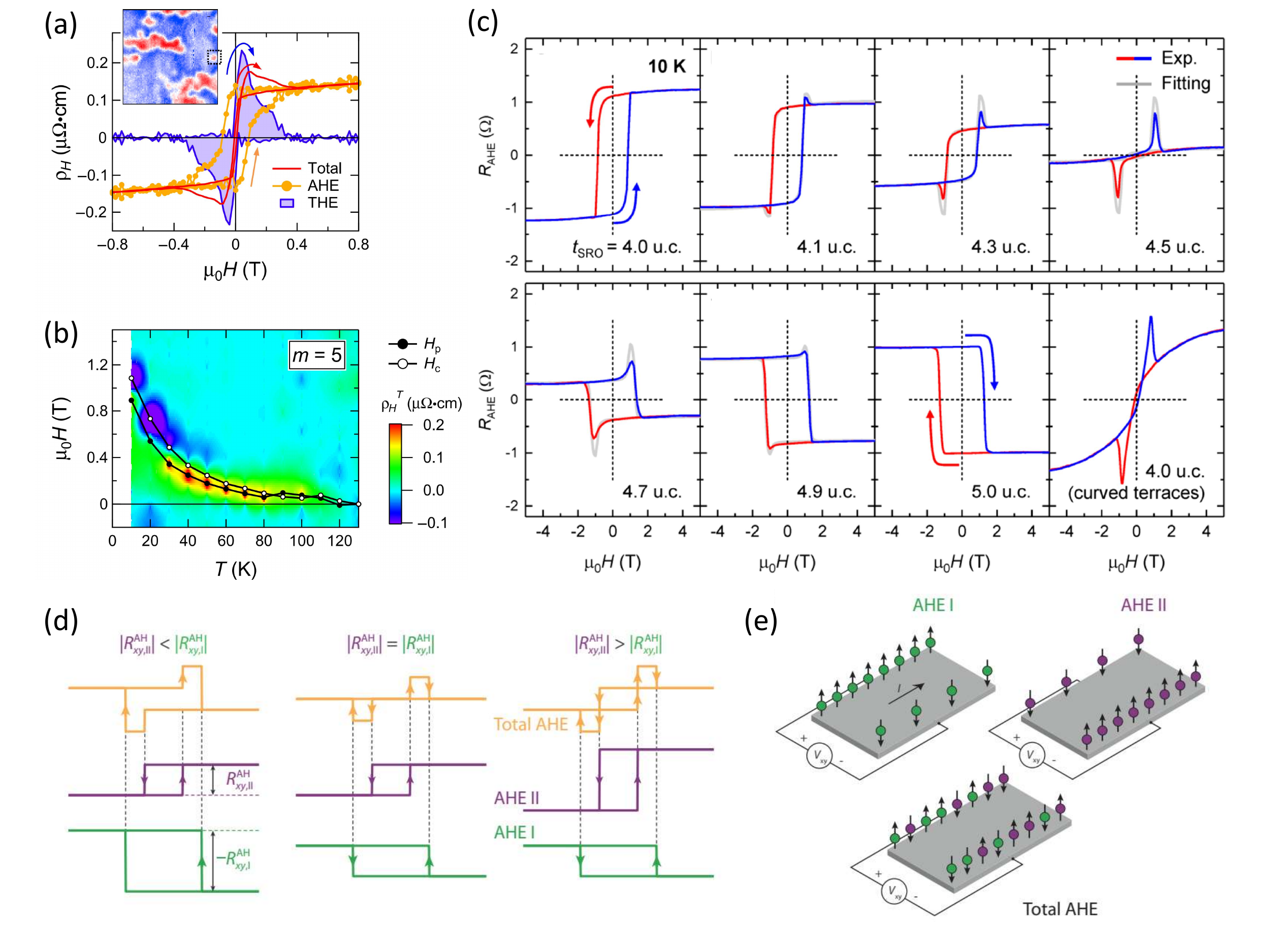}
    \caption{\textbf{Topological Hall effect in SrRuO$_3$ heterostructures} (a) Hall effect at 80 K in a SrRuO$_3$(5 u.c.)/SrIrO$_3$(2 u.c.) sample. The red curve shows the total Hall signal while the orange curve corresponds to the anomalous Hall effect and the blue curve to the topological Hall effect. The inset is a 600 nm $\times$ 600 nm MFM image at 0.2 T. The dotted square highlights a possible skyrmion. (b) Color map of topological Hall resistivity for the same sample. Black open and filled symbols represent the coercive field (H$_C$) and the field at which topological Hall resistivity reaches its maximum (H$_p$), respectively.~\cite{matsuno_interface-driven_2016} (c) Anomalous Hall resistance curves of SRO films with thickness increasing between 4 and 5 unit cells. The light gray curves are linear combinations of the AHE curves of individual 4 and 5 uc SRO films.\cite{wang_controllable_2020}. (d) Model illustrating how the addition of two AHE contributions with opposite sign can generate a signal resembling a THE. (e) Illustration showing the opposite spin accumulation of the two individual contributions. Reproduced from \cite{groenendijk_berry_2020}.}
    \label{Fig-THE-SRO}
\end{figure}

The case of SrRuO$_3$ (SRO) is perhaps the most complex. SRO is an itinerant ferromagnet with a Curie temperature of 160 K. It can be grown as epitaxial thin films of very high structural quality, notably on SrTiO$_3$ substrates \cite{koster_structure_2012} and generally displays strong perpendicular magnetic anisotropy (PMA). In 2016, the RIKEN group reported the observation of a topological effect in bilayers combining SRO and SrIrO$_3$ (SIO), interpreted as the result of 10-nm sized Néel skyrmions \cite{matsuno_interface-driven_2016}, cf Fig.~\ref{Fig-THE-SRO}a-b. Just as in metallic multilayers, a heavy metal-based layer -- here the SIO -- was used to generate DMI and induce skyrmions in the ferromagnet, SRO. These results were reproduced by other groups \cite{pang_spin-glass-like_2017, meng_observation_2019, ziese_unconventional_2019, gu_interfacial_2019} but a similar THE was also found in SRO single films~\cite{qin_emergence_2019, kan_electric_2020, wang_controllable_2020}, i.e., not interfaced with a layer with strong spin-orbit coupling such as SIO. This started to raise questions on the origin of the THE. Four years after the seminal paper on SRO/SIO \cite{matsuno_interface-driven_2016}, subsequent work has coalesced into two main lines of interpretation. 

A first set of works argue that the THE in SRO heterostructures is caused by topological spin textures. Several groups have performed magnetic force microscopy (MFM) as a function of temperature and/or magnetic field and observed contrast interpreted as bubbles or skyrmions~\cite{meng_observation_2019, wang_ferroelectrically_2018, matsuno_interface-driven_2016}. Huang \emph{et al}. also detected Néel-type chiral spin textures using resonant X-ray scattering~\cite{huang_detection_2020}. From their MFM data, Malsch \emph{et al}. however concluded that the detected features could not account for their observed THE \cite{malsch_correlating_2020}.

In a second body of results, it is argued that the THE is just the sum of two anomalous Hall effect (AHE) terms with opposite signs, in regions of the sample with different coercivity~\cite{wang_controllable_2020, miao_strain_2020, kimbell_two-channel_2020, van_thiel_extraordinary_2020, groenendijk_berry_2020, malsch_correlating_2020}. This is illustrated in Fig. \ref{Fig-THE-SRO}d and e. Indeed, in SRO the AHE is known to have an intrinsic, band-structure driven origin and to change sign with temperature or when alloyed with CaRuO$_3$, that both modify the magnetization~\cite{fang_anomalous_2003}. Ref. \cite{groenendijk_berry_2020} has provided a theoretical explanation for the existence of sign competing anomalous channels that derive from bands having a non-trivial topological character associated with non-zero Chern numbers, which was able to explain their and others' experimental data. One striking result is the very strong thickness dependence of the AHE in SRO single films at very low thickness. As visible in Fig.~\ref{Fig-THE-SRO}c the AHE is positive at 4 unit cells but negative at 5 unit cells. For intermediate, non integer thicknesses, the Hall signal displays a clear hump, usually interpreted as a THE, which the authors show can be rather well reproduced by a simple interpolation between the 4 and 5 unit cell AHE traces (grey lines in Fig.~\ref{Fig-THE-SRO}c). MFM on a 4.5 unit cell sample reveals an inhomogeneous magnetic response, with two types of regions having distinct coercive fields, leading to a bimodal distribution~\cite{fang_anomalous_2003}.

In summary, in SrRuO$_3$ clear-cut evidence of skyrmions with an unambiguous topological character -- that could be ascertained by Lorentz microscopy -- has not been provided yet. In fact, the latest results tend to suggest that the observed topological Hall effect may instead reflect reciprocal space-based mechanisms combined with magnetic disorder. More efforts are needed to disentangle this complicated problem.

\subsubsection*{Ferrites}
Heterostructures involving ferrimagnetic insulators such as yttrium iron garnet, Y$_3$Fe$_5$O$_{12}$ (YIG), thulium iron garnet, Tm$_3$Fe$_5$O$_{12}$ (TmIG), and terbium iron garnet, Tb$_3$F$_5$O$_{12}$ (TbIG), have recently been shown to possess the requisite properties for hosting topological spin textures such as skyrmions. The key advantage compared with other ferromagnetic oxides is the high Curie temperature of these compounds, in the 500 K range. TmIG, in particular, exhibits perpendicular magnetic anisotropy as a result of tensile strain when grown on (111)-oriented gadolinium gallium garnet (Gd$_3$Ga$_5$O$_{12}$, GGG) or their substituted variants (sGGG), which are commonly used substrates for such materials. When combined with a heavy-metal overlayer, such as platinum or tungsten, spin-orbit torques and interfacial Dzyaloshinskii-Moriya interactions (DMI) can be induced in the magnetic insulator in an analogous fashion to all-metallic counterparts such as Pt/Co-based systems~\cite{thiaville_dynamics_2012, ryu_chiral_2013, emori_current-driven_2013}. As a result, renewed focus on phenomena such as chiral domain wall motion under spin orbit torques and skyrmion bubble formation have been brought about by such garnet-based heterostructures.

Following observations of current-driven magnetization dynamics in YIG/Pt systems~\cite{hamadeh_full_2014, demidov_direct_2016, collet_generation_2016}, magnetization switching due to spin-orbit torques in TmIG was reported in 2017 with Pt overlayers~\cite{avci_current-induced_2017} and in 2018 with W overlayers~\cite{shao_role_2018}. Harmonic Hall effect measurements performed on these systems indicate that the strength of the spin-orbit torques is comparable to earlier work on YIG/Pt~\cite{avci_current-induced_2017}. Such torques have also been exploited to study current-driven domain wall motion in TmIG/Pt~\cite{velez_high-speed_2019, avci_interface-driven_2019} and TbIG/Pt~\cite{avci_interface-driven_2019}, where the observed high velocities reaching up to 800 m/s can be attributed to the interfacial DMI, which pushes the Walker transition to higher fields under field-driven motion and suppresses it entirely under pure SHE-driven motion~\cite{thiaville_dynamics_2012}. The homochiral Néel character of the domain walls has been deduced from how the wall velocities vary in the presence of in-plane applied magnetic fields along with nitrogen-vacancy centre magnetometry~\cite{velez_high-speed_2019}, where measurements with the latter suggest that the DMI induced at the TmIG/Pt interface is of the opposite sign as that of the sGGG/TmIG interface.

Probing the chiral magnetic order in TmIG with the topological Hall effect has also been reported. Ahmed~\emph{et al.}~\cite{ahmed_spin-hall_2019} and Shao~\emph{et al.}~\cite{shao_topological_2019} have revealed clear signatures of an additional topological contribution to the Hall effect in TmIG/Pt films, which appears over a temperature range around and above room temperature (depending on the TmIG film thickness) and for applied perpendicular fields up to about 0.5 T. The temperatures explored remain below but approach the Curie temperature of TmIG, which is about 560 K for bulk samples. The THE signal is attributed to the presence of magnetic skyrmions, where it is argued that the transition between a perpendicular magnetic anisotropy and an in-plane anisotropy, which occurs at a temperature coinciding with the appearance of the THE signal~\cite{shao_topological_2019}, favours the formation of a skyrmion lattice by reducing the domain wall energy. Because current only flows through the Pt overlayer, the topological spin structure induced in the TmIG film is probed indirectly. Ahmed~\emph{et al.} suggest that the THE signal arises from spin-torque interactions at the TmIG/Pt interface, leading to a phenomenon dubbed the spin-Hall topological Hall effect~\cite{ahmed_spin-hall_2019}. Shao~\emph{et al.} suggest instead that the THE arises from an imprinted spin structure in the Pt through the proximity effect~\cite{shao_exploring_2019}. Both of these viewpoints appear to be consistent with more recent experiments in which the THE signal disappears with the inclusion of an additional Cu buffer layer, i.e., TmIG/Cu/Pt~\cite{xia_interfacial_2020}, since both the proximity effect and spin-torque interactions would be suppressed when adding the buffer.

While direct correlations between the observed THE and spin structures is still lacking, separate studies have provided clear evidence of skyrmion states in TmIG films with different microscopy techniques. Besides the observation of chiral domain walls as mentioned above, Kerr microscopy has been used to reveal bubble-like states in TmIG/Pt films~\cite{ding_interfacial_2019-1}, which has been subsequently confirmed in greater detail in experiments involving scanning transmission x-ray microscopy~\cite{buttner_thermal_2020}. Complementary measurements with photo-emission electron microscopy reveal a variety of topological and non-topological configurations in the bubbles observed (Box 2), and the diameter of these micromagnetic states are found to be in the sub-micron range~\cite{buttner_thermal_2020}.

The origin of the Dzyaloshinskii-Moriya interaction in such garnet heterostructures remains a subject of debate. Evidence for the garnet/garnet interface being the primary source of the DMI has been put in Ref. ~\cite{velez_high-speed_2019} that showed that chiral domain walls in sGGG/TmIG transition from a pure Néel state to a mixed Bloch-Néel state with the addition of a Pt overlayer. This is corroborated by the observation of similar DMI values found for GGG/TbIG/Pt and GGG/TbIG/Cu/Pt~\cite{avci_interface-driven_2019, caretta_interfacial_2020-1}, where the insertion of a Cu spacer would be expected to suppress any DMI at the TbIG/Pt interface, and is also consistent with the finding that the DMI strength varies little with the Pt film thickness and with different capping layers~\cite{ding_identifying_2020}. Detailed experiments on current-driven domain wall motion suggest that the DMI arises primarily from the strong spin-orbit coupling in the rare-earth garnet itself, induced by broken inversion symmetry at an interface, rather than in the substrate or heavy metal overlayer~\cite{caretta_interfacial_2020-1} (although the latter does give a finite contribution). It can be noted that frequency nonreciprocity of spin wave propagation in GGG/YIG has also been reported and attributed to an induced DMI at the garnet/garnet interface~\cite{wang_chiral_2020}. The symmetry breaking at surfaces has also shown to enhance the DMI in BFO, for example~\cite{gross_real-space_2017}. Other studies suggest that the TmIG/Pt interface is more important for DMI. For example, Ref. ~\cite{lee_investigation_2020} shows that the inclusion of a Y$_3$Sc$_2$Al$_3$O$_{12}$ (YSAG) buffer layer between the GGG substrate and TmIG or YIG does not suppress the THE when Pt overlayers are present, from which the authors conclude that the DMI at the TmIG/Pt and YIG/Pt interface is sufficiently large to promote chiral spin textures. This is interpreted by the absence of $f$-band electrons in YSAG, which thereby minimizes the induced DMI at the YSAG/TmIG and YSAG/YIG interfaces. Ultimately, THE signals do not probe the existence of chiral interactions directly, but rather the presence of topological spin structures like skyrmions that may be stabilized by them.

Skyrmionic textures have also been reported in other ferrite compounds. A recent example involving a topological insulator is BaFe$_{12}$O$_{19}$/Bi$_2$Se$_3$, where low temperature signatures of the THE can be attributed to the presence of skyrmions~\cite{li_topological_2020}. Here, it is argued that the strong spin-momentum locking of the topological insulator Bi$_2$Se$_3$ induces a large DMI at the interface with BaFe$_{12}$O$_{19}$, where scattering of the surface states through an equilibrium or nonequilibrium proximity effect results in the observed THE signal. Simulations predict a window of applied magnetic field in which the skyrmion states should appear, which is found to correlate well with the appearance of the THE signal in experiment. Another example is $\alpha-$Fe$_2$O$_3$ (hematite), an insulating antiferromagnetic oxide of trigonal corundum structure that is composed of antiparallel aligned planes of ferromagnetic spins along the crystallographic $c$ axis. The magnetocrystalline anisotropy changes sign at the Morin temperature $T_m$, whereby the N{\'e}el vector characterising the antiferromagnetic (AFM) order reorients from the out of plane temperature for $T < T_m$ to in-plane for $T > T_m$. By examining the (AFM) order with linear magnetic dichroism in photoemission electron microscopy, it was shown that the temperature variations about $T_m$ can produce a variety of meron states, possessing half-skyrmion charges, which grow out of domain wall structures as the reorientation transition takes place~\cite{jani_antiferromagnetic_2021}. Despite the presence of a Pt capping layer, it is shown that the overall DMI is sufficiently weak to impose homochiral meron states, where both N{\'e}el- and Bloch-type merons are observed, along with antimerons, bimerons, and topologically trivial coupled meron states. However, the system may be promising for future studies of the electrical control of such states through spin-Hall effects.

\begin{figure}[h!]
    \centering\includegraphics[width=0.8\textwidth]{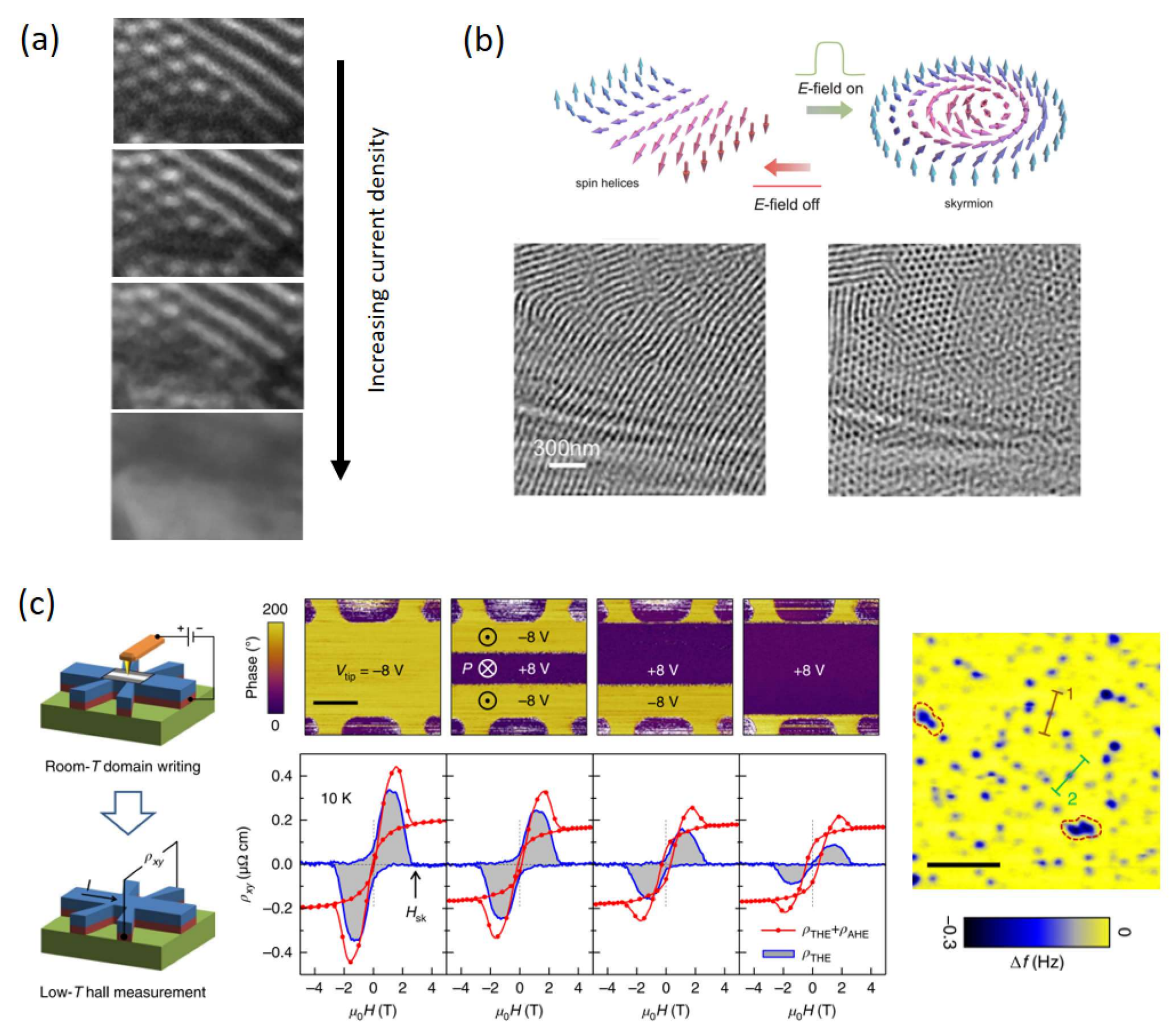}
    \caption{\textbf{Electrical control of skyrmions and topological Hall effect in oxide systems} (a) Changes in magnetic configurations with increasing current (from 4.8 to 9.0 10$^7$ A/m$^2$), obtained under a magnetic field of 0.3 T applied normal to the (001)-plane device plate at 20 K in a La$_{1.37}$Sr$_{1.63}$Mn$_2$O$_7$ sample\cite{yu_biskyrmion_2014}. (b) Top: Schematic illustration of the transition of the spin textures from the helical phase to the skyrmion phase upon the application of the E-field and vice versa in Cu$_2$OSeO$_3$. Bottom: Lorentz transmission electron microscopy image with no applied  electric field (left) and with a field of 3.6 V/ $\mu$m at 24.7 K (right)\cite{huang_situ_2018}. (c) (Left) Schematic diagram of the experimental set-up for ferroelectric domain switching and Hall measurements. (Centre) Piezoresponse force microscopy phase images (top panels), Hall and extracted topological Hall curves (bottom panels) of a SrRuO$_3$/BaTiO$_3$ sample for different ferroelectric poling states. The scale bar corresponds to 10 $\mu$m. (Right) Difference in MFM contrast between images taken at two different magnetic fields. See original paper~\cite{wang_ferroelectrically_2018} for details.}
    \label{Fig-Econtrol}
\end{figure}

\subsection*{Electrical control of skyrmions in oxide systems}

For potential devices based on skyrmions, the capacity to generate and displace these topological objects by electrical means is paramount. As discussed above, their electrical detection can be achieved through the THE, which can be much larger in oxides compared to typical values found in 3d metal-based systems, or through the AHE~\cite{maccariello_electrical_2018}. In metallic systems, skyrmions can be generated and displaced by charge currents through spin-transfer torques, and the literature abounds with examples in metallic multilayers, see, e.g., \cite{jiang_blowing_2015, legrand_room-temperature_2017}. An electric field can also be used, but then the material must be insulating, or the electric field must be applied across a gate oxide by which the interfacial anisotropy can be tuned~\cite{schott_skyrmion_2017}.

In oxides, both current- and voltage-based approaches have been used to manipulate skyrmions. In the layered manganite La$_{1.37}$Sr$_{1.63}$Mn$_2$O$_7$ Yu \emph{et al}. investigated the response of non-collinear spin textures comprising both stripes and biskyrmions to the application of a current. As shown in Fig.~\ref{Fig-Econtrol}a, upon increasing the current, the Lorentz image of the stripes and biskyrmions progressively become blurred, which the authors interpret as reflecting they dynamical motion, faster than the microscope frame rate~\cite{yu_biskyrmion_2014}. In garnets such as TmIG, controlled nucleation of the bubbles is possible with applied current pulses flowing in the Pt overlayer, but the absence of correlation between the bubble motion and current polarity suggests that Joule heating is the primary driving mechanism. This is consistent with the finding that bubble density increases with increase repetition of applied laser pulses~\cite{buttner_thermal_2020}.

In the magnetoelectric skyrmion system Cu$_2$OSeO$_3$~\cite{seki_observation_2012}, the application of an electric field was found to influence the skyrmion lattice. Using neutron diffraction, Ref. ~\cite{white_electric-field-induced_2014} evidenced a rotation of the skyrmion lattice induced by electric fields. In the same compound another group also reported a voltage-induced transition between a stripe phase and a skyrmion lattice~\cite{huang_situ_2018}, as shown in Fig.~\ref{Fig-Econtrol}b.

Thin film heterostructures enable the application of larger electric fields and also make it possible to combine skyrmion-hosting layers with ferroelectrics, which allow for a non-volatile electrical control of properties such as the skyrmion density. By working with SrRuO$_3$/SrIrO$_3$ bilayers grown on STO, it was shown that a backgate voltage can be applied to tune the topological and anomalous Hall effects. Finally, the Seoul group combined SrRuO$_3$ with ferroelectric BaTiO$_3$ to control the amplitude of the THE, associated with the presence of skyrmions in these heterostructures (see Fig. \ref{Fig-Econtrol}c and in particular the MFM image at the very right)~\cite{wang_ferroelectrically_2018}. This latter approach is particularly interesting as it takes advantage of a specific characteristic of oxides -- here the epitaxial combination with a ferroelectric material -- and suggests routes for future functional architectures harnessing spin-orbit properties in oxide materials.

\section*{Outlook}

The field of oxide spin-orbitronics is still in its infancy but has already pointed to exciting new research directions arising from the combination of spin-orbit interactions with the numerous coupled degrees of freedom present in oxide systems. Sizeable spin-charge interconversion effects have been observed, both through the spin Hall effect in thin films of e.g. IrO$_2$ or SrRuO$_3$ and in interface systems such as STO 2DEGs. The oxide family harbours many metals based on heavy elements, e.g. SrMoO$_3$ or doped BaSnO$_3$ and efforts should be made to characterize their spin-charge conversion properties~\cite{vazOxideSpinorbitronicsNew2018}.

STO 2DEGs show very efficient spin-charge conversion and a complex gate dependence linked to their peculiar band structure. KTO 2DEGs share a number of similarities with STO ones and could potentially display even higher responses; yet, their spin-conversion properties have barely been studied. In STO 2DEGs, most studies have focused on spin-charge conversion, with only a handful of papers exploring charge-spin conversion. The recently discovered bilinear magnetoresistance now provides a convenient method to explore this effect, even allowing for a quantitative estimation of the Rashba coefficient provided an appropriate formalism is used \cite{vaz_determining_2020}. Efforts should also be directed at spin-torque measurements and magnetization switching experiments, in the line of the work by Liu \emph{et al.} with SrIrO$_3$~\cite{liu_current-induced_2019}. As highlighted in a recent study \cite{noel_non-volatile_2020} bringing ferroelectricity as a new ingredient into spin-orbitronics opens many opportunities to design low-power devices operating on spin but whose non-volatility  would derive from ferroelectricity rather than from ferromagnetism. An important challenge is now to demonstrate the same type of control at room temperature. More generally, the interplay between ferroelectricity and the transport properties of 2DEGs emerges as an interesting topic, in particular in light of the enhanced superconducting $T_C$ in related systems \cite{ahadiEnhancingSuperconductivitySrTiO2019}.

As described in the second part of this review, spin-orbit coupling may also bestow oxides with non-collinear spin textures, some of which having a topological character. Some ambiguities in the interpretation of topological Hall signatures should be lifted by performing experiments able to directly assess the topological character of the spin structures. In addition, it would be interesting to explore materials with Curie temperatures above 300 K such as double perovskites or spinel ferrites. The recent results on garnets also suggest that combining oxides with other material families, such as heavy metals, is an efficient approach to achieve substantial DMI and non-collinear structures, in addition to controlling their dynamics with electrical currents.

Finally, in carefully designed heterostructures, it should be possible in the future to generate topological spin textures and control them by ferroelectricity and spin-orbit torques at room temperature, opening the way towards truly multifunctional devices based on oxides operating at low power. On the way to this goal, one may assume that novel, unexpected phenomena will also arise\cite{barthelemy_quasi-two-dimensional_2021}, further expanding the wealth of properties of these fascinating materials.

\hspace{1em}

\noindent\textbf{Acknowledgements}\\
We thank A. Caviglia and M. Cuoco for fruitful discussions and S. V{\'e}lez for a critical reading of this manuscript. This work received support from the ERC Advanced grant no. 833973 ``FRESCO'', the QUANTERA project ``QUANTOX'', the French Research Agency (ANR) as part of the ``Investissement d’Avenir'' programme (LABEX NanoSaclay, ref. ANR-10-LABX-0035) through project ``AXION'' and the Laboratoire d’Excellence LANEF (ANR-10-LABX-51-01), ANR project OISO (ANR-17-CE24-0026-03) and ANR project CONTRABASS (ANR-19-CE24-XXXX-XX). F. Trier acknowledges support by research grant 37338 (SANSIT) from VILLUM FONDEN. P.N. acknowledges the support of the ETH Zurich Postdoctoral fellowship program. \\

\noindent\textbf{Author contributions}\\
All authors contributed equally to the writing of this review paper.\\

\noindent\textbf{Competing interests}\\
The authors declare no competing interests. \\

\ifdefined\DeclarePrefChars\DeclarePrefChars{'’-}\else\fi


\begin{thebibliography}{100}
\urlstyle{rm}
\expandafter\ifx\csname url\endcsname\relax
  \def\url#1{\texttt{#1}}\fi
\expandafter\ifx\csname urlprefix\endcsname\relax\def\urlprefix{URL }\fi
\expandafter\ifx\csname doiprefix\endcsname\relax\def\doiprefix{DOI: }\fi
\providecommand{\bibinfo}[2]{#2}
\providecommand{\eprint}[2][]{\url{#2}}

\bibitem{soumyanarayanan_emergent_2016}
\bibinfo{author}{Soumyanarayanan, A.}, \bibinfo{author}{Reyren, N.},
  \bibinfo{author}{Fert, A.} \& \bibinfo{author}{Panagopoulos, C.}
\newblock \bibinfo{journal}{\bibinfo{title}{Emergent phenomena induced by
  spin--orbit coupling at surfaces and interfaces}}.
\newblock {\emph{\JournalTitle{Nature}}} \textbf{\bibinfo{volume}{539}},
  \bibinfo{pages}{509--517}, \doiprefix\url{10.1038/nature19820}
  (\bibinfo{year}{2016}).

\bibitem{sampaioNucleationStabilityCurrentinduced2013}
\bibinfo{author}{Sampaio, J.}, \bibinfo{author}{Cros, V.},
  \bibinfo{author}{Rohart, S.}, \bibinfo{author}{Thiaville, A.} \&
  \bibinfo{author}{Fert, A.}
\newblock \bibinfo{journal}{\bibinfo{title}{Nucleation, stability and
  current-induced motion of isolated magnetic skyrmions in nanostructures}}.
\newblock {\emph{\JournalTitle{Nature Nanotechnology}}}
  \textbf{\bibinfo{volume}{8}}, \bibinfo{pages}{839--844},
  \doiprefix\url{10.1038/nnano.2013.210} (\bibinfo{year}{2013}).

\bibitem{bibes_ultrathin_2011}
\bibinfo{author}{Bibes, M.}, \bibinfo{author}{Villegas, J.~E.} \&
  \bibinfo{author}{Barth{\'e}l{\'e}my, A.}
\newblock \bibinfo{journal}{\bibinfo{title}{Ultrathin oxide films and
  interfaces for electronics and spintronics}}.
\newblock {\emph{\JournalTitle{Advances in Physics}}}
  \textbf{\bibinfo{volume}{60}}, \bibinfo{pages}{5--84},
  \doiprefix\url{10.1080/00018732.2010.534865} (\bibinfo{year}{2011}).

\bibitem{hwang_emergent_2012}
\bibinfo{author}{Hwang, H.~Y.} \emph{et~al.}
\newblock \bibinfo{journal}{\bibinfo{title}{Emergent phenomena at oxide
  interfaces}}.
\newblock {\emph{\JournalTitle{Nature Materials}}}
  \textbf{\bibinfo{volume}{11}}, \bibinfo{pages}{103--113},
  \doiprefix\url{10.1038/nmat3223} (\bibinfo{year}{2012}).

\bibitem{gariglio_spinorbit_2019}
\bibinfo{author}{Gariglio, S.}, \bibinfo{author}{Caviglia, A.~D.},
  \bibinfo{author}{Triscone, J.-M.} \& \bibinfo{author}{Gabay, M.}
\newblock \bibinfo{journal}{\bibinfo{title}{A spin--orbit playground: Surfaces
  and interfaces of transition metal oxides}}.
\newblock {\emph{\JournalTitle{Reports on Progress in Physics}}}
  \textbf{\bibinfo{volume}{82}}, \bibinfo{pages}{012501},
  \doiprefix\url{10.1088/1361-6633/aad6ab} (\bibinfo{year}{2019}).

\bibitem{ramesh_creating_2019}
\bibinfo{author}{Ramesh, R.} \& \bibinfo{author}{Schlom, D.~G.}
\newblock \bibinfo{journal}{\bibinfo{title}{Creating emergent phenomena in
  oxide superlattices}}.
\newblock {\emph{\JournalTitle{Nature Reviews Materials}}}
  \textbf{\bibinfo{volume}{4}}, \bibinfo{pages}{257--268},
  \doiprefix\url{10.1038/s41578-019-0095-2} (\bibinfo{year}{2019}).

\bibitem{varignonNewSpinOxide2018}
\bibinfo{author}{Varignon, J.}, \bibinfo{author}{Vila, L.},
  \bibinfo{author}{Barth{\'e}l{\'e}my, A.} \& \bibinfo{author}{Bibes, M.}
\newblock \bibinfo{journal}{\bibinfo{title}{A new spin for oxide interfaces}}.
\newblock {\emph{\JournalTitle{Nature Physics}}} \textbf{\bibinfo{volume}{14}},
  \bibinfo{pages}{322--325}, \doiprefix\url{10.1038/s41567-018-0112-1}
  (\bibinfo{year}{2018}).

\bibitem{vaz_mapping_2019}
\bibinfo{author}{Vaz, D.~C.} \emph{et~al.}
\newblock \bibinfo{journal}{\bibinfo{title}{Mapping spin--charge conversion to
  the band structure in a topological oxide two-dimensional electron gas}}.
\newblock {\emph{\JournalTitle{Nature Materials}}}
  \textbf{\bibinfo{volume}{18}}, \bibinfo{pages}{1187--1193},
  \doiprefix\url{10.1038/s41563-019-0467-4} (\bibinfo{year}{2019}).

\bibitem{noel_non-volatile_2020}
\bibinfo{author}{No{\"e}l, P.} \emph{et~al.}
\newblock \bibinfo{journal}{\bibinfo{title}{{Non-Volatile Electric Control of
  Spin--Charge Conversion in a SrTiO$_3$ Rashba System}}}.
\newblock {\emph{\JournalTitle{Nature}}} \textbf{\bibinfo{volume}{580}},
  \bibinfo{pages}{483--486}, \doiprefix\url{10.1038/s41586-020-2197-9}
  (\bibinfo{year}{2020}).

\bibitem{wang_ferroelectrically_2018}
\bibinfo{author}{Wang, L.} \emph{et~al.}
\newblock \bibinfo{journal}{\bibinfo{title}{Ferroelectrically tunable magnetic
  skyrmions in ultrathin oxide heterostructures}}.
\newblock {\emph{\JournalTitle{Nature Materials}}}
  \textbf{\bibinfo{volume}{17}}, \bibinfo{pages}{1087--1094},
  \doiprefix\url{10.1038/s41563-018-0204-4} (\bibinfo{year}{2018}).

\bibitem{chauleau_electric_2019}
\bibinfo{author}{Chauleau, J.-Y.} \emph{et~al.}
\newblock \bibinfo{journal}{\bibinfo{title}{Electric and antiferromagnetic
  chiral textures at multiferroic domain walls}}.
\newblock {\emph{\JournalTitle{Nature Materials}}}
  \doiprefix\url{10.1038/s41563-019-0516-z} (\bibinfo{year}{2019}).

\bibitem{ohtomo_high-mobility_2004}
\bibinfo{author}{Ohtomo, A.} \& \bibinfo{author}{Hwang, H.~Y.}
\newblock \bibinfo{journal}{\bibinfo{title}{A high-mobility electron gas at the
  {{LaAlO$_3$}}/{{SrTiO$_3$}} heterointerface}}.
\newblock {\emph{\JournalTitle{Nature}}} \textbf{\bibinfo{volume}{427}},
  \bibinfo{pages}{423--426}, \doiprefix\url{10.1038/nature02308}
  (\bibinfo{year}{2004}).

\bibitem{stemmer_two-dimensional_2014}
\bibinfo{author}{Stemmer, S.} \& \bibinfo{author}{James~Allen, S.}
\newblock \bibinfo{journal}{\bibinfo{title}{Two-{{Dimensional Electron Gases}}
  at {{Complex Oxide Interfaces}}}}.
\newblock {\emph{\JournalTitle{Annual Review of Materials Research}}}
  \textbf{\bibinfo{volume}{44}}, \bibinfo{pages}{151--171},
  \doiprefix\url{10.1146/annurev-matsci-070813-113552} (\bibinfo{year}{2014}).

\bibitem{yu_polarity-induced_2014}
\bibinfo{author}{Yu, L.} \& \bibinfo{author}{Zunger, A.}
\newblock \bibinfo{journal}{\bibinfo{title}{A polarity-induced defect mechanism
  for conductivity and magnetism at polar--nonpolar oxide interfaces}}.
\newblock {\emph{\JournalTitle{Nature Communications}}}
  \textbf{\bibinfo{volume}{5}}, \bibinfo{pages}{5118},
  \doiprefix\url{10.1038/ncomms6118} (\bibinfo{year}{2014}).

\bibitem{bristowe_origin_2014}
\bibinfo{author}{Bristowe, N.~C.}, \bibinfo{author}{Ghosez, P.},
  \bibinfo{author}{Littlewood, P.~B.} \& \bibinfo{author}{Artacho, E.}
\newblock \bibinfo{journal}{\bibinfo{title}{The origin of two-dimensional
  electron gases at oxide interfaces: Insights from theory}}.
\newblock {\emph{\JournalTitle{Journal of Physics: Condensed Matter}}}
  \textbf{\bibinfo{volume}{26}}, \bibinfo{pages}{143201},
  \doiprefix\url{10.1088/0953-8984/26/14/143201} (\bibinfo{year}{2014}).

\bibitem{cantoni_electron_2012}
\bibinfo{author}{Cantoni, C.} \emph{et~al.}
\newblock \bibinfo{journal}{\bibinfo{title}{Electron {{Transfer}} and {{Ionic
  Displacements}} at the {{Origin}} of the {{2D Electron Gas}} at the
  {{LAO}}/{{STO Interface}}: {{Direct Measurements}} with {{Atomic}}-{{Column
  Spatial Resolution}}}}.
\newblock {\emph{\JournalTitle{Advanced Materials}}}
  \textbf{\bibinfo{volume}{24}}, \bibinfo{pages}{3952--3957},
  \doiprefix\url{10.1002/adma.201200667} (\bibinfo{year}{2012}).

\bibitem{li_very_2011}
\bibinfo{author}{Li, L.} \emph{et~al.}
\newblock \bibinfo{journal}{\bibinfo{title}{Very {{Large Capacitance
  Enhancement}} in a {{Two}}-{{Dimensional Electron System}}}}.
\newblock {\emph{\JournalTitle{Science}}} \textbf{\bibinfo{volume}{332}},
  \bibinfo{pages}{825--828}, \doiprefix\url{10.1126/science.1204168}
  (\bibinfo{year}{2011}).

\bibitem{chakhalian_whither_2012}
\bibinfo{author}{Chakhalian, J.}, \bibinfo{author}{Millis, A.~J.} \&
  \bibinfo{author}{Rondinelli, J.}
\newblock \bibinfo{journal}{\bibinfo{title}{Whither the oxide interface}}.
\newblock {\emph{\JournalTitle{Nature Materials}}}
  \textbf{\bibinfo{volume}{11}}, \bibinfo{pages}{92--94},
  \doiprefix\url{10.1038/nmat3225} (\bibinfo{year}{2012}).

\bibitem{cen_nanoscale_2008}
\bibinfo{author}{Cen, C.} \emph{et~al.}
\newblock \bibinfo{journal}{\bibinfo{title}{Nanoscale control of an interfacial
  metal--insulator transition at room temperature}}.
\newblock {\emph{\JournalTitle{Nature Materials}}}
  \textbf{\bibinfo{volume}{7}}, \bibinfo{pages}{298--302},
  \doiprefix\url{10.1038/nmat2136} (\bibinfo{year}{2008}).

\bibitem{thiel_tunable_2006}
\bibinfo{author}{Thiel, S.}
\newblock \bibinfo{journal}{\bibinfo{title}{Tunable
  {{Quasi}}-{{Two}}-{{Dimensional Electron Gases}} in {{Oxide
  Heterostructures}}}}.
\newblock {\emph{\JournalTitle{Science}}} \textbf{\bibinfo{volume}{313}},
  \bibinfo{pages}{1942--1945}, \doiprefix\url{10.1126/science.1131091}
  (\bibinfo{year}{2006}).

\bibitem{assmann_oxide_2013}
\bibinfo{author}{Assmann, E.} \emph{et~al.}
\newblock \bibinfo{journal}{\bibinfo{title}{Oxide {{Heterostructures}} for
  {{Efficient Solar Cells}}}}.
\newblock {\emph{\JournalTitle{Physical Review Letters}}}
  \textbf{\bibinfo{volume}{110}}, \bibinfo{pages}{078701},
  \doiprefix\url{10.1103/PhysRevLett.110.078701} (\bibinfo{year}{2013}).

\bibitem{li_coexistence_2011}
\bibinfo{author}{Li, L.}, \bibinfo{author}{Richter, C.},
  \bibinfo{author}{Mannhart, J.} \& \bibinfo{author}{Ashoori, R.~C.}
\newblock \bibinfo{journal}{\bibinfo{title}{Coexistence of magnetic order and
  two-dimensional superconductivity at {{LaAlO$_3$}}/{{SrTiO$_3$}}
  interfaces}}.
\newblock {\emph{\JournalTitle{Nature Physics}}} \textbf{\bibinfo{volume}{7}},
  \bibinfo{pages}{762--766}, \doiprefix\url{10.1038/nphys2080}
  (\bibinfo{year}{2011}).

\bibitem{reyren_superconducting_2007}
\bibinfo{author}{Reyren, N.} \emph{et~al.}
\newblock \bibinfo{journal}{\bibinfo{title}{Superconducting {{Interfaces
  Between Insulating Oxides}}}}.
\newblock {\emph{\JournalTitle{Science}}} \textbf{\bibinfo{volume}{317}},
  \bibinfo{pages}{1196--1199}, \doiprefix\url{10.1126/science.1146006}
  (\bibinfo{year}{2007}).

\bibitem{caviglia_electric_2008}
\bibinfo{author}{Caviglia, A.~D.} \emph{et~al.}
\newblock \bibinfo{journal}{\bibinfo{title}{Electric field control of the
  {{LaAlO$_3$}}/{{SrTiO$_3$}} interface ground state}}.
\newblock {\emph{\JournalTitle{Nature}}} \textbf{\bibinfo{volume}{456}},
  \bibinfo{pages}{624--627}, \doiprefix\url{10.1038/nature07576}
  (\bibinfo{year}{2008}).

\bibitem{brinkman_magnetic_2007}
\bibinfo{author}{Brinkman, A.} \emph{et~al.}
\newblock \bibinfo{journal}{\bibinfo{title}{Magnetic effects at the interface
  between non-magnetic oxides}}.
\newblock {\emph{\JournalTitle{Nature Materials}}}
  \textbf{\bibinfo{volume}{6}}, \bibinfo{pages}{493--496},
  \doiprefix\url{10.1038/nmat1931} (\bibinfo{year}{2007}).

\bibitem{trier_quantization_2016}
\bibinfo{author}{Trier, F.} \emph{et~al.}
\newblock \bibinfo{journal}{\bibinfo{title}{Quantization of {Hall} {Resistance}
  at the {Metallic} {Interface} between an {Oxide} {Insulator} and
  {SrTiO}$_{\textrm{3}}$}}.
\newblock {\emph{\JournalTitle{Physical Review Letters}}}
  \textbf{\bibinfo{volume}{117}}, \bibinfo{pages}{096804},
  \doiprefix\url{10.1103/PhysRevLett.117.096804} (\bibinfo{year}{2016}).

\bibitem{nakagawa_why_2006}
\bibinfo{author}{Nakagawa, N.}, \bibinfo{author}{Hwang, H.~Y.} \&
  \bibinfo{author}{Muller, D.~A.}
\newblock \bibinfo{journal}{\bibinfo{title}{Why some interfaces cannot be
  sharp}}.
\newblock {\emph{\JournalTitle{Nature Materials}}}
  \textbf{\bibinfo{volume}{5}}, \bibinfo{pages}{204--209},
  \doiprefix\url{10.1038/nmat1569} (\bibinfo{year}{2006}).

\bibitem{zhong_theory_2013}
\bibinfo{author}{Zhong, Z.}, \bibinfo{author}{T{\'o}th, A.} \&
  \bibinfo{author}{Held, K.}
\newblock \bibinfo{journal}{\bibinfo{title}{Theory of spin-orbit coupling at
  {{LaAlO$_3$}}/{{SrTiO$_3$}} interfaces and {{SrTiO$_3$}} surfaces}}.
\newblock {\emph{\JournalTitle{Physical Review B}}}
  \textbf{\bibinfo{volume}{87}}, \bibinfo{pages}{161102},
  \doiprefix\url{10.1103/PhysRevB.87.161102} (\bibinfo{year}{2013}).

\bibitem{king_quasiparticle_2014}
\bibinfo{author}{King, P. D.~C.} \emph{et~al.}
\newblock \bibinfo{journal}{\bibinfo{title}{Quasiparticle dynamics and
  spin--orbital texture of the {{SrTiO$_3$}} two-dimensional electron gas}}.
\newblock {\emph{\JournalTitle{Nature Communications}}}
  \textbf{\bibinfo{volume}{5}}, \doiprefix\url{10.1038/ncomms4414}
  (\bibinfo{year}{2014}).

\bibitem{chen_metallic_2011}
\bibinfo{author}{Chen, Y.}, \bibinfo{author}{Pryds, N.}, \bibinfo{author}{Shen,
  B.}, \bibinfo{author}{Rijnders, G.} \& \bibinfo{author}{Linderoth, S.}
\newblock \bibinfo{journal}{\bibinfo{title}{Metallic and {{Insulating
  Interfaces}} of {{Amorphous SrTiO$_3$}}-{{Based Oxide Heterostructures}}}}.
\newblock {\emph{\JournalTitle{Nano Letters}}} \bibinfo{pages}{5}
  (\bibinfo{year}{2011}).

\bibitem{lee_creation_2012}
\bibinfo{author}{Lee, S.~W.}, \bibinfo{author}{Liu, Y.}, \bibinfo{author}{Heo,
  J.} \& \bibinfo{author}{Gordon, R.~G.}
\newblock \bibinfo{journal}{\bibinfo{title}{Creation and {{Control}} of
  {{Two}}-{{Dimensional Electron Gas Using Al}}-{{Based Amorphous
  Oxides}}/{{SrTiO$_3$}} {{Heterostructures Grown}} by {{Atomic Layer
  Deposition}}}}.
\newblock {\emph{\JournalTitle{Nano Letters}}} \textbf{\bibinfo{volume}{12}},
  \bibinfo{pages}{4775--4783}, \doiprefix\url{10.1021/nl302214x}
  (\bibinfo{year}{2012}).

\bibitem{liu_origin_2013}
\bibinfo{author}{Liu, Z.~Q.} \emph{et~al.}
\newblock \bibinfo{journal}{\bibinfo{title}{Origin of the {{Two}}-{{Dimensional
  Electron Gas}} at {{LaAlO$_3$}}/{{SrTiO$_3$}} {{Interfaces}}: {{The Role}} of
  {{Oxygen Vacancies}} and {{Electronic Reconstruction}}}}.
\newblock {\emph{\JournalTitle{Physical Review X}}}
  \textbf{\bibinfo{volume}{3}}, \bibinfo{pages}{021010},
  \doiprefix\url{10.1103/PhysRevX.3.021010} (\bibinfo{year}{2013}).

\bibitem{sing_profiling_2009}
\bibinfo{author}{Sing, M.} \emph{et~al.}
\newblock \bibinfo{journal}{\bibinfo{title}{Profiling the {{Interface Electron
  Gas}} of {{LaAlO}}$_3$/{{SrTiO}}$_3$ {{Heterostructures}} with {{Hard
  X}}-{{Ray Photoelectron Spectroscopy}}}}.
\newblock {\emph{\JournalTitle{Physical Review Letters}}}
  \textbf{\bibinfo{volume}{102}}, \bibinfo{pages}{176805},
  \doiprefix\url{10.1103/PhysRevLett.102.176805} (\bibinfo{year}{2009}).

\bibitem{takizawa_electronic_2011}
\bibinfo{author}{Takizawa, M.}, \bibinfo{author}{Tsuda, S.},
  \bibinfo{author}{Susaki, T.}, \bibinfo{author}{Hwang, H.~Y.} \&
  \bibinfo{author}{Fujimori, A.}
\newblock \bibinfo{journal}{\bibinfo{title}{Electronic charges and electric
  potential at {{LaAlO}}$_3$/{{SrTiO}}$_3$ interfaces studied by core-level
  photoemission spectroscopy}}.
\newblock {\emph{\JournalTitle{Physical Review B}}}
  \textbf{\bibinfo{volume}{84}}, \bibinfo{pages}{245124},
  \doiprefix\url{10.1103/PhysRevB.84.245124} (\bibinfo{year}{2011}).

\bibitem{rubano_spectral_2011}
\bibinfo{author}{Rubano, A.} \emph{et~al.}
\newblock \bibinfo{journal}{\bibinfo{title}{Spectral and spatial distribution
  of polarization at the {{LaAlO}}$_3$/{{SrTiO}}$_3$ interface}}.
\newblock {\emph{\JournalTitle{Physical Review B}}}
  \textbf{\bibinfo{volume}{83}}, \bibinfo{pages}{155405},
  \doiprefix\url{10.1103/PhysRevB.83.155405} (\bibinfo{year}{2011}).

\bibitem{slooten_hard_2013}
\bibinfo{author}{Slooten, E.} \emph{et~al.}
\newblock \bibinfo{journal}{\bibinfo{title}{Hard x-ray photoemission and
  density functional theory study of the internal electric field in
  {{SrTiO}}$_3$/{{LaAlO}}$_3$ oxide heterostructures}}.
\newblock {\emph{\JournalTitle{Physical Review B}}}
  \textbf{\bibinfo{volume}{87}}, \bibinfo{pages}{085128},
  \doiprefix\url{10.1103/PhysRevB.87.085128} (\bibinfo{year}{2013}).

\bibitem{xu_reversible_2017}
\bibinfo{author}{Xu, P.} \emph{et~al.}
\newblock \bibinfo{journal}{\bibinfo{title}{Reversible {{Formation}} of {{2D
  Electron Gas}} at the {{LaFeO}}$_3$/{{SrTiO}}$_3$ {{Interface}} via
  {{Control}} of {{Oxygen Vacancies}}}}.
\newblock {\emph{\JournalTitle{Advanced Materials}}}
  \textbf{\bibinfo{volume}{29}}, \bibinfo{pages}{1604447},
  \doiprefix\url{10.1002/adma.201604447} (\bibinfo{year}{2017}).

\bibitem{maznichenko_formation_2020}
\bibinfo{author}{Maznichenko, I.~V.}, \bibinfo{author}{Ostanin, S.},
  \bibinfo{author}{Ernst, A.}, \bibinfo{author}{Henk, J.} \&
  \bibinfo{author}{Mertig, I.}
\newblock \bibinfo{journal}{\bibinfo{title}{Formation and {{Tuning}} of {{2D
  Electron Gas}} in {{Perovskite Heterostructures}}}}.
\newblock {\emph{\JournalTitle{Physica Status Solidi B}}}
  \textbf{\bibinfo{volume}{257}}, \bibinfo{pages}{1900540},
  \doiprefix\url{10.1002/pssb.201900540} (\bibinfo{year}{2020}).

\bibitem{oja_d0_2012}
\bibinfo{author}{Oja, R.} \emph{et~al.}
\newblock \bibinfo{journal}{\bibinfo{title}{$d^0$ {{Ferromagnetic Interface}}
  between {{Nonmagnetic Perovskites}}}}.
\newblock {\emph{\JournalTitle{Physical Review Letters}}}
  \textbf{\bibinfo{volume}{19}}, \bibinfo{pages}{127207},
  \doiprefix\url{10.1103/PhysRevLett.109.127207} (\bibinfo{year}{2012}).

\bibitem{ohtomo_artificial_2002}
\bibinfo{author}{Ohtomo, A.}, \bibinfo{author}{Muller, D.~A.},
  \bibinfo{author}{Grazul, J.~L.} \& \bibinfo{author}{Hwang, H.~Y.}
\newblock \bibinfo{journal}{\bibinfo{title}{Artificial charge-modulationin
  atomic-scale perovskite titanate superlattices}}.
\newblock {\emph{\JournalTitle{Nature}}} \textbf{\bibinfo{volume}{419}},
  \bibinfo{pages}{378--380}, \doiprefix\url{10.1038/nature00977}
  (\bibinfo{year}{2002}).

\bibitem{ohtsuka_transport_2010}
\bibinfo{author}{Ohtsuka, R.}, \bibinfo{author}{Matvejeff, M.},
  \bibinfo{author}{Nishio, K.}, \bibinfo{author}{Takahashi, R.} \&
  \bibinfo{author}{Lippmaa, M.}
\newblock \bibinfo{journal}{\bibinfo{title}{Transport properties of
  {{LaTiO$_3$}}/{{SrTiO$_3$}} heterostructures}}.
\newblock {\emph{\JournalTitle{Appl. Phys. Lett.}}}
  \textbf{\bibinfo{volume}{96}}, \bibinfo{pages}{192111}
  (\bibinfo{year}{2010}).

\bibitem{perna_conducting_2010}
\bibinfo{author}{Perna, P.} \emph{et~al.}
\newblock \bibinfo{journal}{\bibinfo{title}{Conducting interfaces between band
  insulating oxides: The {{LaGaO$_3$}}/{{SrTiO$_3$}}}}.
\newblock {\emph{\JournalTitle{Applied Physics Letters}}}
  \textbf{\bibinfo{volume}{97}}, \bibinfo{pages}{152111},
  \doiprefix\url{10.1063/1.3496440} (\bibinfo{year}{2010}).

\bibitem{hotta_polar_2007}
\bibinfo{author}{Hotta, Y.}, \bibinfo{author}{Susaki, T.} \&
  \bibinfo{author}{Hwang, H.~Y.}
\newblock \bibinfo{journal}{\bibinfo{title}{Polar {{Discontinuity Doping}} of
  the {{LaVO}}$_3$/{{SrTiO}}$_3$ {{Interface}}}}.
\newblock {\emph{\JournalTitle{Physical Review Letters}}}
  \textbf{\bibinfo{volume}{99}}, \bibinfo{pages}{236805},
  \doiprefix\url{10.1103/PhysRevLett.99.236805} (\bibinfo{year}{2007}).

\bibitem{moetakef_transport_2011}
\bibinfo{author}{Moetakef, P.} \emph{et~al.}
\newblock \bibinfo{journal}{\bibinfo{title}{Transport in ferromagnetic
  {{GdTiO$_3$}}/{{SrTiO$_3$}} heterostructures}}.
\newblock {\emph{\JournalTitle{Applied Physics Letters}}}
  \textbf{\bibinfo{volume}{98}}, \bibinfo{pages}{112110},
  \doiprefix\url{10.1063/1.3568894} (\bibinfo{year}{2011}).

\bibitem{he_metal-insulator_2012}
\bibinfo{author}{He, C.} \emph{et~al.}
\newblock \bibinfo{journal}{\bibinfo{title}{Metal-insulator transitions in
  epitaxial {{LaVO}}$_3$ and {{LaTiO}}$_3$ films}}.
\newblock {\emph{\JournalTitle{Physical Review B}}}
  \textbf{\bibinfo{volume}{86}}, \bibinfo{pages}{081401},
  \doiprefix\url{10.1103/PhysRevB.86.081401} (\bibinfo{year}{2012}).

\bibitem{annadi_electronic_2012}
\bibinfo{author}{Annadi, A.} \emph{et~al.}
\newblock \bibinfo{journal}{\bibinfo{title}{Electronic correlation and strain
  effects at the interfaces between polar and nonpolar complex oxides}}.
\newblock {\emph{\JournalTitle{Physical Review B}}}
  \textbf{\bibinfo{volume}{86}}, \bibinfo{pages}{085450},
  \doiprefix\url{10.1103/PhysRevB.86.085450} (\bibinfo{year}{2012}).

\bibitem{nazir_charge_2011}
\bibinfo{author}{Nazir, S.}, \bibinfo{author}{Singh, N.} \&
  \bibinfo{author}{Schwingenschl{\"o}gl, U.}
\newblock \bibinfo{journal}{\bibinfo{title}{Charge transfer mechanism for the
  formation of metallic states at the {{KTaO}}$_3$/{{SrTiO}}$_3$ interface}}.
\newblock {\emph{\JournalTitle{Physical Review B}}}
  \textbf{\bibinfo{volume}{83}}, \bibinfo{pages}{113107},
  \doiprefix\url{10.1103/PhysRevB.83.113107} (\bibinfo{year}{2011}).

\bibitem{chenHighmobilityTwodimensionalElectron2013}
\bibinfo{author}{Chen, Y.~Z.} \emph{et~al.}
\newblock \bibinfo{journal}{\bibinfo{title}{{A High-Mobility Two-Dimensional
  Electron Gas at the Spinel/Perovskite Interface of
  $\gamma$-{{Al$_2$O$_3$}}/{{SrTiO$_3$}}}}}.
\newblock {\emph{\JournalTitle{Nature Communications}}}
  \textbf{\bibinfo{volume}{4}}, \bibinfo{pages}{1371},
  \doiprefix\url{10.1038/ncomms2394} (\bibinfo{year}{2013}).

\bibitem{santander-syro_two-dimensional_2011}
\bibinfo{author}{Santander-Syro, A.~F.} \emph{et~al.}
\newblock \bibinfo{journal}{\bibinfo{title}{Two-dimensional electron gas with
  universal subbands at the surface of {{SrTiO$_3$}}}}.
\newblock {\emph{\JournalTitle{Nature}}} \textbf{\bibinfo{volume}{469}},
  \bibinfo{pages}{189--193}, \doiprefix\url{10.1038/nature09720}
  (\bibinfo{year}{2011}).

\bibitem{meevasana_creation_2011}
\bibinfo{author}{Meevasana, W.} \emph{et~al.}
\newblock \bibinfo{journal}{\bibinfo{title}{Creation and control of a
  two-dimensional electron liquid at the bare {{SrTiO$_3$}} surface}}.
\newblock {\emph{\JournalTitle{Nature Materials}}}
  \textbf{\bibinfo{volume}{10}}, \bibinfo{pages}{114--118},
  \doiprefix\url{10.1038/nmat2943} (\bibinfo{year}{2011}).

\bibitem{rodel_universal_2016}
\bibinfo{author}{R{\"o}del, T.~C.} \emph{et~al.}
\newblock \bibinfo{journal}{\bibinfo{title}{Universal {{Fabrication}} of {{2D
  Electron Systems}} in {{Functional Oxides}}}}.
\newblock {\emph{\JournalTitle{Advanced Materials}}}
  \textbf{\bibinfo{volume}{28}}, \bibinfo{pages}{1976--1980},
  \doiprefix\url{10.1002/adma.201505021} (\bibinfo{year}{2016}).

\bibitem{vaz_tuning_2017}
\bibinfo{author}{Vaz, D.~C.} \emph{et~al.}
\newblock \bibinfo{journal}{\bibinfo{title}{Tuning {{Up}} or {{Down}} the
  {{Critical Thickness}} in {{LaAlO$_3$}}/{{SrTiO$_3$}} through {{In Situ
  Deposition}} of {{Metal Overlayers}}}}.
\newblock {\emph{\JournalTitle{Advanced Materials}}}
  \textbf{\bibinfo{volume}{29}}, \bibinfo{pages}{1700486},
  \doiprefix\url{10.1002/adma.201700486} (\bibinfo{year}{2017}).

\bibitem{trier_electron_2018}
\bibinfo{author}{Trier, F.}, \bibinfo{author}{Christensen, D.~V.} \&
  \bibinfo{author}{Pryds, N.}
\newblock \bibinfo{journal}{\bibinfo{title}{Electron mobility in oxide
  heterostructures}}.
\newblock {\emph{\JournalTitle{Journal of Physics D: Applied Physics}}}
  \textbf{\bibinfo{volume}{51}}, \bibinfo{pages}{293002},
  \doiprefix\url{10.1088/1361-6463/aac9aa} (\bibinfo{year}{2018}).

\bibitem{hurand_field-effect_2015}
\bibinfo{author}{Hurand, S.} \emph{et~al.}
\newblock \bibinfo{journal}{\bibinfo{title}{Field-effect control of
  superconductivity and {{Rashba}} spin-orbit coupling in top-gated
  {{LaAlO$_3$}}/{{SrTiO$_3$}} devices}}.
\newblock {\emph{\JournalTitle{Scientific Reports}}}
  \textbf{\bibinfo{volume}{5}}, \bibinfo{pages}{12751},
  \doiprefix\url{10.1038/srep12751} (\bibinfo{year}{2015}).

\bibitem{lesne_highly_2016}
\bibinfo{author}{Lesne, E.} \emph{et~al.}
\newblock \bibinfo{journal}{\bibinfo{title}{Highly efficient and tunable
  spin-to-charge conversion through {{Rashba}} coupling at oxide interfaces}}.
\newblock {\emph{\JournalTitle{Nature Materials}}}
  \textbf{\bibinfo{volume}{15}}, \bibinfo{pages}{1261--1266},
  \doiprefix\url{10.1038/nmat4726} (\bibinfo{year}{2016}).

\bibitem{khalsa_theory_2012}
\bibinfo{author}{Khalsa, G.} \& \bibinfo{author}{MacDonald, A.~H.}
\newblock \bibinfo{journal}{\bibinfo{title}{Theory of the {{SrTiO}}$_3$ surface
  state two-dimensional electron gas}}.
\newblock {\emph{\JournalTitle{Physical Review B}}}
  \textbf{\bibinfo{volume}{86}}, \doiprefix\url{10.1103/PhysRevB.86.125121}
  (\bibinfo{year}{2012}).

\bibitem{caviglia_tunable_2010}
\bibinfo{author}{Caviglia, A.~D.} \emph{et~al.}
\newblock \bibinfo{journal}{\bibinfo{title}{Tunable {{Rashba Spin}}-{{Orbit
  Interaction}} at {{Oxide Interfaces}}}}.
\newblock {\emph{\JournalTitle{Physical Review Letters}}}
  \textbf{\bibinfo{volume}{104}}, \bibinfo{pages}{126803},
  \doiprefix\url{10.1103/PhysRevLett.104.126803} (\bibinfo{year}{2010}).

\bibitem{ben_shalom_tuning_2010}
\bibinfo{author}{Ben~Shalom, M.}, \bibinfo{author}{Sachs, M.},
  \bibinfo{author}{Rakhmilevitch, D.}, \bibinfo{author}{Palevski, A.} \&
  \bibinfo{author}{Dagan, Y.}
\newblock \bibinfo{journal}{\bibinfo{title}{Tuning {{Spin}}-{{Orbit Coupling}}
  and {{Superconductivity}} at the {{SrTiO}}$_3$ / {{LaAlO}}$_3$ {{Interface}}:
  {{A Magnetotransport Study}}}}.
\newblock {\emph{\JournalTitle{Physical Review Letters}}}
  \textbf{\bibinfo{volume}{104}}, \bibinfo{pages}{126802},
  \doiprefix\url{10.1103/PhysRevLett.104.126802} (\bibinfo{year}{2010}).

\bibitem{trier_electric-field_2019}
\bibinfo{author}{Trier, F.} \emph{et~al.}
\newblock \bibinfo{journal}{\bibinfo{title}{Electric-{{Field Control}} of
  {{Spin Current Generation}} and {{Detection}} in {{Ferromagnet}}-{{Free
  SrTiO$_3$}}‑{{Based Nanodevices}}}}.
\newblock {\emph{\JournalTitle{Nano Letters}}} \textbf{\bibinfo{volume}{20}},
  \bibinfo{pages}{395--401}, \doiprefix\url{10.1021/acs.nanolett.9b04079}
  (\bibinfo{year}{2020}).

\bibitem{shen_microscopic_2014}
\bibinfo{author}{Shen, K.}, \bibinfo{author}{Vignale, G.} \&
  \bibinfo{author}{Raimondi, R.}
\newblock \bibinfo{journal}{\bibinfo{title}{Microscopic {{Theory}} of the
  {{Inverse Edelstein Effect}}}}.
\newblock {\emph{\JournalTitle{Physical Review Letters}}}
  \textbf{\bibinfo{volume}{112}}, \bibinfo{pages}{096601},
  \doiprefix\url{10.1103/PhysRevLett.112.096601} (\bibinfo{year}{2014}).

\bibitem{zhang_conversion_2016}
\bibinfo{author}{Zhang, S.} \& \bibinfo{author}{Fert, A.}
\newblock \bibinfo{journal}{\bibinfo{title}{Conversion between spin and charge
  currents with topological insulators}}.
\newblock {\emph{\JournalTitle{Physical Review B}}}
  \textbf{\bibinfo{volume}{94}}, \doiprefix\url{10.1103/PhysRevB.94.184423}
  (\bibinfo{year}{2016}).

\bibitem{chauleau_efficient_2016}
\bibinfo{author}{Chauleau, J.-Y.} \emph{et~al.}
\newblock \bibinfo{journal}{\bibinfo{title}{Efficient spin-to-charge conversion
  in the {{2D}} electron liquid at the {{LAO}}/{{STO}} interface}}.
\newblock {\emph{\JournalTitle{EPL (Europhysics Letters)}}}
  \textbf{\bibinfo{volume}{116}}, \bibinfo{pages}{17006},
  \doiprefix\url{10.1209/0295-5075/116/17006} (\bibinfo{year}{2016}).

\bibitem{sahin_strain_2019}
\bibinfo{author}{{\c S}ahin, C.}, \bibinfo{author}{Vignale, G.} \&
  \bibinfo{author}{Flatt{\'e}, M.~E.}
\newblock \bibinfo{journal}{\bibinfo{title}{Strain {{Engineering}} of the
  {{Intrinsic Spin Hall Conductivity}} in a {{SrTiO}}$_3$ {{Quantum Well}}}}.
\newblock {\emph{\JournalTitle{Physical Review Materials}}}
  \textbf{\bibinfo{volume}{3}}, \bibinfo{pages}{014401},
  \doiprefix\url{10.1103/PhysRevMaterials.3.014401} (\bibinfo{year}{2019}).
\newblock \eprint{1804.00061}.

\bibitem{telesio_study_2018}
\bibinfo{author}{Telesio, F.} \emph{et~al.}
\newblock \bibinfo{journal}{\bibinfo{title}{Study of equilibrium carrier
  transfer in {LaAlO}$_3$/{SrTiO}$_3$ from an epitaxial
  {La}$_{1-x}${Sr}$_x${MnO}$_3$ ferromagnetic layer}}.
\newblock {\emph{\JournalTitle{Journal of Physics Communications}}}
  \textbf{\bibinfo{volume}{2}}, \bibinfo{pages}{025010},
  \doiprefix\url{10.1088/2399-6528/aaa943} (\bibinfo{year}{2018}).

\bibitem{manipatruni_scalable_2019}
\bibinfo{author}{Manipatruni, S.} \emph{et~al.}
\newblock \bibinfo{journal}{\bibinfo{title}{Scalable energy-efficient
  magnetoelectric spin--orbit logic}}.
\newblock {\emph{\JournalTitle{Nature}}} \textbf{\bibinfo{volume}{565}},
  \bibinfo{pages}{35--42}, \doiprefix\url{10.1038/s41586-018-0770-2}
  (\bibinfo{year}{2019}).

\bibitem{miron_perpendicular_2011}
\bibinfo{author}{Miron, I.~M.} \emph{et~al.}
\newblock \bibinfo{journal}{\bibinfo{title}{Perpendicular switching of a single
  ferromagnetic layer induced by in-plane current injection}}.
\newblock {\emph{\JournalTitle{Nature}}} \textbf{\bibinfo{volume}{476}},
  \bibinfo{pages}{189--193}, \doiprefix\url{10.1038/nature10309}
  (\bibinfo{year}{2011}).

\bibitem{liu_spin-torque_2012}
\bibinfo{author}{Liu, L.} \emph{et~al.}
\newblock \bibinfo{journal}{\bibinfo{title}{Spin-{{Torque Switching}} with the
  {{Giant Spin Hall Effect}} of {{Tantalum}}}}.
\newblock {\emph{\JournalTitle{Science}}} \textbf{\bibinfo{volume}{336}},
  \bibinfo{pages}{555--558}, \doiprefix\url{10.1126/science.1218197}
  (\bibinfo{year}{2012}).

\bibitem{dieny_opportunities_2020}
\bibinfo{author}{Dieny, B.} \emph{et~al.}
\newblock \bibinfo{journal}{\bibinfo{title}{Opportunities and challenges for
  spintronics in the microelectronics industry}}.
\newblock {\emph{\JournalTitle{Nature Electronics}}}
  \textbf{\bibinfo{volume}{3}}, \bibinfo{pages}{446--459},
  \doiprefix\url{10.1038/s41928-020-0461-5} (\bibinfo{year}{2020}).

\bibitem{grollier_neuromorphic_2020}
\bibinfo{author}{Grollier, J.} \emph{et~al.}
\newblock \bibinfo{journal}{\bibinfo{title}{Neuromorphic spintronics}}.
\newblock {\emph{\JournalTitle{Nature Electronics}}}
  \textbf{\bibinfo{volume}{3}}, \bibinfo{pages}{360--370},
  \doiprefix\url{10.1038/s41928-019-0360-9} (\bibinfo{year}{2020}).

\bibitem{wang_room-temperature_2017}
\bibinfo{author}{Wang, Y.} \emph{et~al.}
\newblock \bibinfo{journal}{\bibinfo{title}{Room-{{Temperature Giant
  Charge}}-to-{{Spin Conversion}} at the {{SrTiO}}$_3$--{{LaAlO}}$_3$ {{Oxide
  Interface}}}}.
\newblock {\emph{\JournalTitle{Nano Letters}}} \textbf{\bibinfo{volume}{17}},
  \bibinfo{pages}{7659--7664}, \doiprefix\url{10.1021/acs.nanolett.7b03714}
  (\bibinfo{year}{2017}).

\bibitem{yamanouchiCurrentinducedEffectiveMagnetic2020}
\bibinfo{author}{Yamanouchi, M.}, \bibinfo{author}{Oyamada, T.} \&
  \bibinfo{author}{Ohta, H.}
\newblock \bibinfo{journal}{\bibinfo{title}{Current-induced effective magnetic
  field in
  {{La}}$_{0.67}${{Sr}}$_{0.33}${{MnO}}$_3$/{{LaAlO}}$_3$/{{SrTiO}}$_3$
  structures}}.
\newblock {\emph{\JournalTitle{AIP Advances}}} \textbf{\bibinfo{volume}{10}},
  \bibinfo{pages}{015129}, \doiprefix\url{10.1063/1.5129283}
  (\bibinfo{year}{2020}).

\bibitem{sinova_universal_2004}
\bibinfo{author}{Sinova, J.} \emph{et~al.}
\newblock \bibinfo{journal}{\bibinfo{title}{Universal {{Intrinsic Spin Hall
  Effect}}}}.
\newblock {\emph{\JournalTitle{Physical Review Letters}}}
  \textbf{\bibinfo{volume}{92}}, \bibinfo{pages}{126603},
  \doiprefix\url{10.1103/PhysRevLett.92.126603} (\bibinfo{year}{2004}).

\bibitem{jinNonlocalSpinDiffusion2017}
\bibinfo{author}{Jin, M.-J.} \emph{et~al.}
\newblock \bibinfo{journal}{\bibinfo{title}{Nonlocal {{Spin Diffusion Driven}}
  by {{Giant Spin Hall Effect}} at {{Oxide Heterointerfaces}}}}.
\newblock {\emph{\JournalTitle{Nano Letters}}} \textbf{\bibinfo{volume}{17}},
  \bibinfo{pages}{36--43}, \doiprefix\url{10.1021/acs.nanolett.6b03050}
  (\bibinfo{year}{2017}).

\bibitem{tokuraNonreciprocalResponsesNoncentrosymmetric2018}
\bibinfo{author}{Tokura, Y.} \& \bibinfo{author}{Nagaosa, N.}
\newblock \bibinfo{journal}{\bibinfo{title}{Nonreciprocal responses from
  non-centrosymmetric quantum materials}}.
\newblock {\emph{\JournalTitle{Nature Communications}}}
  \textbf{\bibinfo{volume}{9}}, \bibinfo{pages}{3740},
  \doiprefix\url{10.1038/s41467-018-05759-4} (\bibinfo{year}{2018}).

\bibitem{dyrdal_spin-momentum-locking_2020}
\bibinfo{author}{Dyrda{\l}, A.}, \bibinfo{author}{Barna{\'s}, J.} \&
  \bibinfo{author}{Fert, A.}
\newblock \bibinfo{journal}{\bibinfo{title}{Spin-{{Momentum}}-{{Locking
  Inhomogeneities}} as a {{Source}} of {{Bilinear Magnetoresistance}} in
  {{Topological Insulators}}}}.
\newblock {\emph{\JournalTitle{Physical Review Letters}}}
  \textbf{\bibinfo{volume}{124}}, \bibinfo{pages}{046802},
  \doiprefix\url{10.1103/PhysRevLett.124.046802} (\bibinfo{year}{2020}).

\bibitem{vaz_determining_2020}
\bibinfo{author}{Vaz, D.~C.} \emph{et~al.}
\newblock \bibinfo{journal}{\bibinfo{title}{Determining the {{Rashba}}
  parameter from the bilinear magnetoresistance response in a two-dimensional
  electron gas}}.
\newblock {\emph{\JournalTitle{Physical Review Materials}}}
  \textbf{\bibinfo{volume}{4}}, \bibinfo{pages}{071001},
  \doiprefix\url{10.1103/PhysRevMaterials.4.071001} (\bibinfo{year}{2020}).

\bibitem{choe_gate-tunable_2019}
\bibinfo{author}{Choe, D.} \emph{et~al.}
\newblock \bibinfo{journal}{\bibinfo{title}{Gate-tunable giant nonreciprocal
  charge transport in noncentrosymmetric oxide interfaces}}.
\newblock {\emph{\JournalTitle{Nature Communications}}}
  \textbf{\bibinfo{volume}{10}}, \bibinfo{pages}{4510},
  \doiprefix\url{10.1038/s41467-019-12466-1} (\bibinfo{year}{2019}).

\bibitem{he_observation_2018}
\bibinfo{author}{He, P.} \emph{et~al.}
\newblock \bibinfo{journal}{\bibinfo{title}{Observation of {{Out}}-of-{{Plane
  Spin Texture}} in a {{SrTiO}}$_3$ ( 111 ) {{Two}}-{{Dimensional Electron
  Gas}}}}.
\newblock {\emph{\JournalTitle{Physical Review Letters}}}
  \textbf{\bibinfo{volume}{120}}, \bibinfo{pages}{266802},
  \doiprefix\url{10.1103/PhysRevLett.120.266802} (\bibinfo{year}{2018}).

\bibitem{zou_latio_2015}
\bibinfo{author}{Zou, K.} \emph{et~al.}
\newblock \bibinfo{journal}{\bibinfo{title}{{{LaTiO}}$_3$/{{KTaO}}$_3$
  interfaces: {{A}} new two-dimensional electron gas system}}.
\newblock {\emph{\JournalTitle{APL Materials}}} \textbf{\bibinfo{volume}{3}},
  \bibinfo{pages}{036104}, \doiprefix\url{10.1063/1.4914310}
  (\bibinfo{year}{2015}).

\bibitem{zhang_highly_2017}
\bibinfo{author}{Zhang, H.} \emph{et~al.}
\newblock \bibinfo{journal}{\bibinfo{title}{Highly {{Mobile Two}}-{{Dimensional
  Electron Gases}} with a {{Strong Gating Effect}} at the {{Amorphous
  LaAlO}}$_3$/{{KTaO}}$_3$ {{Interface}}}}.
\newblock {\emph{\JournalTitle{ACS Applied Materials \& Interfaces}}}
  \textbf{\bibinfo{volume}{9}}, \bibinfo{pages}{36456--36461},
  \doiprefix\url{10.1021/acsami.7b12814} (\bibinfo{year}{2017}).

\bibitem{zhang_high-mobility_2018}
\bibinfo{author}{Zhang, H.} \emph{et~al.}
\newblock \bibinfo{journal}{\bibinfo{title}{High-{{Mobility Spin}}-{{Polarized
  Two}}-{{Dimensional Electron Gases}} at {{EuO}}/{{KTaO}}$_3$
  {{Interfaces}}}}.
\newblock {\emph{\JournalTitle{Physical Review Letters}}}
  \textbf{\bibinfo{volume}{121}},
  \doiprefix\url{10.1103/PhysRevLett.121.116803} (\bibinfo{year}{2018}).

\bibitem{wadehra_planar_2020}
\bibinfo{author}{Wadehra, N.} \emph{et~al.}
\newblock \bibinfo{journal}{\bibinfo{title}{Planar {{Hall}} effect and
  anisotropic magnetoresistance in polar-polar interface of
  {{LaVO$_3$}}-{{KTaO$_3$}} with strong spin-orbit coupling}}.
\newblock {\emph{\JournalTitle{Nature Communications}}}
  \textbf{\bibinfo{volume}{11}}, \bibinfo{pages}{874},
  \doiprefix\url{10.1038/s41467-020-14689-z} (\bibinfo{year}{2020}).

\bibitem{nakamura_electric_2009}
\bibinfo{author}{Nakamura, H.} \& \bibinfo{author}{Kimura, T.}
\newblock \bibinfo{journal}{\bibinfo{title}{Electric field tuning of spin-orbit
  coupling in {{KTaO}}$_3$ field-effect transistors}}.
\newblock {\emph{\JournalTitle{Physical Review B}}}
  \textbf{\bibinfo{volume}{80}}, \bibinfo{pages}{121308},
  \doiprefix\url{10.1103/PhysRevB.80.121308} (\bibinfo{year}{2009}).

\bibitem{king_subband_2012}
\bibinfo{author}{King, P. D.~C.} \emph{et~al.}
\newblock \bibinfo{journal}{\bibinfo{title}{Subband {{Structure}} of a
  {{Two}}-{{Dimensional Electron Gas Formed}} at the {{Polar Surface}} of the
  {{Strong Spin}}-{{Orbit Perovskite KTaO}}$_3$}}.
\newblock {\emph{\JournalTitle{Physical Review Letters}}}
  \textbf{\bibinfo{volume}{108}},
  \doiprefix\url{10.1103/PhysRevLett.108.117602} (\bibinfo{year}{2012}).

\bibitem{santander-syro_orbital_2012}
\bibinfo{author}{Santander-Syro, A.~F.} \emph{et~al.}
\newblock \bibinfo{journal}{\bibinfo{title}{Orbital symmetry reconstruction and
  strong mass renormalization in the two-dimensional electron gas at the
  surface of {{KTaO}}$_3$}}.
\newblock {\emph{\JournalTitle{Physical Review B}}}
  \textbf{\bibinfo{volume}{86}}, \bibinfo{pages}{121107},
  \doiprefix\url{10.1103/PhysRevB.86.121107} (\bibinfo{year}{2012}).

\bibitem{zhang_thermal_2019}
\bibinfo{author}{Zhang, H.} \emph{et~al.}
\newblock \bibinfo{journal}{\bibinfo{title}{Thermal {{Spin Injection}} and
  {{Inverse Edelstein Effect}} of the {{Two}}-{{Dimensional Electron Gas}} at
  {{EuO}}--{{KTaO}}$_3$ {{Interfaces}}}}.
\newblock {\emph{\JournalTitle{Nano Letters}}} \textbf{\bibinfo{volume}{19}},
  \bibinfo{pages}{1605--1612}, \doiprefix\url{10.1021/acs.nanolett.8b04509}
  (\bibinfo{year}{2019}).

\bibitem{liuDiscoveryTwodimensionalAnisotropic2020}
\bibinfo{author}{Liu, C.} \emph{et~al.}
\newblock \bibinfo{title}{Discovery of two-dimensional anisotropic
  superconductivity at {{KTaO}}$_3$ (111) interfaces} (\bibinfo{year}{2020}).
\newblock \eprint{2004.07416}.

\bibitem{chenElectricFieldControl2020}
\bibinfo{author}{Chen, Z.} \emph{et~al.}
\newblock \bibinfo{title}{Electric field control of disorder-tunable
  superconductivity and the emergence of quantum metal at an oxide interface}
  (\bibinfo{year}{2020}).
\newblock \eprint{2009.05896}.

\bibitem{ohya_efficient_2020}
\bibinfo{author}{Ohya, S.} \emph{et~al.}
\newblock \bibinfo{journal}{\bibinfo{title}{Efficient intrinsic spin-to-charge
  current conversion in an all-epitaxial single-crystal perovskite-oxide
  heterostructure of
  {{La}}$_{0.67}${{Sr}}$_{0.33}${{MnO}}$_3$/{{LaAlO}}$_3$/{{SrTiO}}$_3$}}.
\newblock {\emph{\JournalTitle{Physical Review Research}}}
  \textbf{\bibinfo{volume}{2}}, \bibinfo{pages}{012014},
  \doiprefix\url{10.1103/PhysRevResearch.2.012014} (\bibinfo{year}{2020}).

\bibitem{zhang_spin_2016}
\bibinfo{author}{Zhang, W.} \emph{et~al.}
\newblock \bibinfo{journal}{\bibinfo{title}{Spin galvanic effect at the
  conducting {{SrTiO}}$_3$ surfaces}}.
\newblock {\emph{\JournalTitle{Applied Physics Letters}}}
  \textbf{\bibinfo{volume}{109}}, \bibinfo{pages}{262402},
  \doiprefix\url{10.1063/1.4973479} (\bibinfo{year}{2016}).

\bibitem{qiu_experimental_2013}
\bibinfo{author}{Qiu, Z.} \emph{et~al.}
\newblock \bibinfo{journal}{\bibinfo{title}{Experimental investigation of spin
  {{Hall}} effect in indium tin oxide thin film}}.
\newblock {\emph{\JournalTitle{Applied Physics Letters}}}
  \textbf{\bibinfo{volume}{103}}, \bibinfo{pages}{182404},
  \doiprefix\url{10.1063/1.4827808} (\bibinfo{year}{2013}).

\bibitem{fujiwara_5d_2013}
\bibinfo{author}{Fujiwara, K.} \emph{et~al.}
\newblock \bibinfo{journal}{\bibinfo{title}{5d iridium oxide as a material for
  spin-current detection}}.
\newblock {\emph{\JournalTitle{Nature Communications}}}
  \textbf{\bibinfo{volume}{4}}, \doiprefix\url{10.1038/ncomms3893}
  (\bibinfo{year}{2013}).

\bibitem{wahler_inverse_2016}
\bibinfo{author}{Wahler, M.} \emph{et~al.}
\newblock \bibinfo{journal}{\bibinfo{title}{Inverse spin {{Hall}} effect in a
  complex ferromagnetic oxide heterostructure}}.
\newblock {\emph{\JournalTitle{Scientific Reports}}}
  \textbf{\bibinfo{volume}{6}}, \bibinfo{pages}{28727},
  \doiprefix\url{10.1038/srep28727} (\bibinfo{year}{2016}).

\bibitem{everhardt_tunable_2019}
\bibinfo{author}{Everhardt, A.~S.} \emph{et~al.}
\newblock \bibinfo{journal}{\bibinfo{title}{Tunable charge to spin conversion
  in strontium iridate thin films}}.
\newblock {\emph{\JournalTitle{Physical Review Materials}}}
  \textbf{\bibinfo{volume}{3}}, \bibinfo{pages}{051201},
  \doiprefix\url{10.1103/PhysRevMaterials.3.051201} (\bibinfo{year}{2019}).

\bibitem{liu_current-induced_2019}
\bibinfo{author}{Liu, L.} \emph{et~al.}
\newblock \bibinfo{journal}{\bibinfo{title}{Current-induced magnetization
  switching in all-oxide heterostructures}}.
\newblock {\emph{\JournalTitle{Nature Nanotechnology}}}
  \textbf{\bibinfo{volume}{14}}, \bibinfo{pages}{939--944},
  \doiprefix\url{10.1038/s41565-019-0534-7} (\bibinfo{year}{2019}).

\bibitem{tsai_clear_2018}
\bibinfo{author}{Tsai, H.} \emph{et~al.}
\newblock \bibinfo{journal}{\bibinfo{title}{Clear variation of spin splitting
  by changing electron distribution at non-magnetic metal/{{Bi$_2$O$_3$}}
  interfaces}}.
\newblock {\emph{\JournalTitle{Scientific Reports}}}
  \textbf{\bibinfo{volume}{8}}, \bibinfo{pages}{5564},
  \doiprefix\url{10.1038/s41598-018-23787-4} (\bibinfo{year}{2018}).

\bibitem{kim_conducting_2016}
\bibinfo{author}{Kim, U.}, \bibinfo{author}{Park, C.}, \bibinfo{author}{Kim,
  Y.~M.}, \bibinfo{author}{Shin, J.} \& \bibinfo{author}{Char, K.}
\newblock \bibinfo{journal}{\bibinfo{title}{Conducting interface states at
  {{LaInO}}$_3$/{{BaSnO}}$_3$ polar interface controlled by {{Fermi}} level}}.
\newblock {\emph{\JournalTitle{APL Materials}}} \textbf{\bibinfo{volume}{4}},
  \bibinfo{pages}{071102}, \doiprefix\url{10.1063/1.4959960}
  (\bibinfo{year}{2016}).

\bibitem{kim_interface_2019}
\bibinfo{author}{Kim, Y.~M.} \emph{et~al.}
\newblock \bibinfo{journal}{\bibinfo{title}{Interface polarization model for a
  2-dimensional electron gas at the {{BaSnO3}}/{{LaInO3}} interface}}.
\newblock {\emph{\JournalTitle{Scientific Reports}}}
  \textbf{\bibinfo{volume}{9}}, \bibinfo{pages}{16202},
  \doiprefix\url{10.1038/s41598-019-52772-8} (\bibinfo{year}{2019}).

\bibitem{kim_physical_2012}
\bibinfo{author}{Kim, H.~J.} \emph{et~al.}
\newblock \bibinfo{journal}{\bibinfo{title}{Physical properties of transparent
  perovskite oxides ({{Ba}},{{La}}){{SnO}}$_3$ with high electrical mobility at
  room temperature}}.
\newblock {\emph{\JournalTitle{Physical Review B}}}
  \textbf{\bibinfo{volume}{86}}, \bibinfo{pages}{165205},
  \doiprefix\url{10.1103/PhysRevB.86.165205} (\bibinfo{year}{2012}).

\bibitem{tsukazaki_quantum_2007}
\bibinfo{author}{Tsukazaki, A.} \emph{et~al.}
\newblock \bibinfo{journal}{\bibinfo{title}{Quantum {{Hall Effect}} in {{Polar
  Oxide Heterostructures}}}}.
\newblock {\emph{\JournalTitle{Science}}} \textbf{\bibinfo{volume}{315}},
  \bibinfo{pages}{1388--1391}, \doiprefix\url{10.1126/science.1137430}
  (\bibinfo{year}{2007}).

\bibitem{tsukazaki_observation_2010}
\bibinfo{author}{Tsukazaki, A.} \emph{et~al.}
\newblock \bibinfo{journal}{\bibinfo{title}{Observation of the fractional
  quantum {{Hall}} effect in an oxide}}.
\newblock {\emph{\JournalTitle{Nature Materials}}}
  \textbf{\bibinfo{volume}{9}}, \bibinfo{pages}{889--893},
  \doiprefix\url{10.1038/nmat2874} (\bibinfo{year}{2010}).

\bibitem{kozuka_rashba_2013}
\bibinfo{author}{Kozuka, Y.} \emph{et~al.}
\newblock \bibinfo{journal}{\bibinfo{title}{Rashba spin-orbit interaction in a
  {Mg}$_{x}${Zn}$_{1\ensuremath{-}x}${O}/{ZnO} two-dimensional electron gas
  studied by electrically detected electron spin resonance}}.
\newblock {\emph{\JournalTitle{Physical Review B}}}
  \textbf{\bibinfo{volume}{87}}, \bibinfo{pages}{205411},
  \doiprefix\url{10.1103/PhysRevB.87.205411} (\bibinfo{year}{2013}).

\bibitem{qiu_all-oxide_2012}
\bibinfo{author}{Qiu, Z.} \emph{et~al.}
\newblock \bibinfo{journal}{\bibinfo{title}{All-oxide system for spin
  pumping}}.
\newblock {\emph{\JournalTitle{Applied Physics Letters}}}
  \textbf{\bibinfo{volume}{100}}, \bibinfo{pages}{022402},
  \doiprefix\url{10.1063/1.3675463} (\bibinfo{year}{2012}).

\bibitem{ueda_spin-orbit_2020}
\bibinfo{author}{Ueda, K.} \emph{et~al.}
\newblock \bibinfo{journal}{\bibinfo{title}{Spin-orbit torque generation in ni
  fe / ir o 2 bilayers}}.
\newblock {\emph{\JournalTitle{Physical Review B}}}
  \textbf{\bibinfo{volume}{102}}, \bibinfo{pages}{134432},
  \doiprefix\url{10.1103/PhysRevB.102.134432} (\bibinfo{year}{2020}).

\bibitem{qiu_all-oxide_2015}
\bibinfo{author}{Qiu, Z.}, \bibinfo{author}{Hou, D.}, \bibinfo{author}{Kikkawa,
  T.}, \bibinfo{author}{Uchida, K.-i.} \& \bibinfo{author}{Saitoh, E.}
\newblock \bibinfo{journal}{\bibinfo{title}{All-oxide spin {{Seebeck}}
  effects}}.
\newblock {\emph{\JournalTitle{Applied Physics Express}}}
  \textbf{\bibinfo{volume}{8}}, \bibinfo{pages}{083001},
  \doiprefix\url{10.7567/APEX.8.083001} (\bibinfo{year}{2015}).

\bibitem{sun_dirac_2017}
\bibinfo{author}{Sun, Y.}, \bibinfo{author}{Zhang, Y.}, \bibinfo{author}{Liu,
  C.-X.}, \bibinfo{author}{Felser, C.} \& \bibinfo{author}{Yan, B.}
\newblock \bibinfo{journal}{\bibinfo{title}{Dirac nodal lines and induced spin
  {{Hall}} effect in metallic rutile oxides}}.
\newblock {\emph{\JournalTitle{Physical Review B}}}
  \textbf{\bibinfo{volume}{95}}, \bibinfo{pages}{235104},
  \doiprefix\url{10.1103/PhysRevB.95.235104} (\bibinfo{year}{2017}).

\bibitem{carter_semimetal_2012}
\bibinfo{author}{Carter, J.-M.}, \bibinfo{author}{Shankar, V.~V.},
  \bibinfo{author}{Zeb, M.~A.} \& \bibinfo{author}{Kee, H.-Y.}
\newblock \bibinfo{journal}{\bibinfo{title}{Semimetal and {{Topological
  Insulator}} in {{Perovskite Iridates}}}}.
\newblock {\emph{\JournalTitle{Physical Review B}}}
  \textbf{\bibinfo{volume}{85}}, \doiprefix\url{10.1103/PhysRevB.85.115105}
  (\bibinfo{year}{2012}).

\bibitem{zeb_interplay_2012}
\bibinfo{author}{Zeb, M.~A.} \& \bibinfo{author}{Kee, H.-Y.}
\newblock \bibinfo{journal}{\bibinfo{title}{Interplay between spin-orbit
  coupling and {{Hubbard}} interaction in {{SrIrO$_3$}} and related {{Pbnm}}
  perovskites}}.
\newblock {\emph{\JournalTitle{Physical Review B}}}
  \textbf{\bibinfo{volume}{86}}, \doiprefix\url{10.1103/PhysRevB.86.085149}
  (\bibinfo{year}{2012}).
\newblock \eprint{1206.5836}.

\bibitem{nie_interplay_2015}
\bibinfo{author}{Nie, Y.~F.} \emph{et~al.}
\newblock \bibinfo{journal}{\bibinfo{title}{Interplay of {{Spin}}-{{Orbit
  Interactions}}, {{Dimensionality}}, and {{Octahedral Rotations}} in
  {{Semimetallic SrIrO}}$_3$}}.
\newblock {\emph{\JournalTitle{Physical Review Letters}}}
  \textbf{\bibinfo{volume}{114}}, \bibinfo{pages}{016401},
  \doiprefix\url{10.1103/PhysRevLett.114.016401} (\bibinfo{year}{2015}).

\bibitem{liu_direct_2016}
\bibinfo{author}{Liu, Z.~T.} \emph{et~al.}
\newblock \bibinfo{journal}{\bibinfo{title}{Direct observation of the {{Dirac}}
  nodes lifting in semimetallic perovskite {{SrIrO}}$_3$ thin films}}.
\newblock {\emph{\JournalTitle{Scientific Reports}}}
  \textbf{\bibinfo{volume}{6}}, \bibinfo{pages}{30309},
  \doiprefix\url{10.1038/srep30309} (\bibinfo{year}{2016}).

\bibitem{patri_theory_2018}
\bibinfo{author}{Patri, A.~S.}, \bibinfo{author}{Hwang, K.},
  \bibinfo{author}{Lee, H.-W.} \& \bibinfo{author}{Kim, Y.~B.}
\newblock \bibinfo{journal}{\bibinfo{title}{Theory of {{Large Intrinsic Spin
  Hall Effect}} in {{Iridate Semimetals}}}}.
\newblock {\emph{\JournalTitle{Scientific Reports}}}
  \textbf{\bibinfo{volume}{8}}, \doiprefix\url{10.1038/s41598-018-26355-y}
  (\bibinfo{year}{2018}).

\bibitem{nan_anisotropic_2019}
\bibinfo{author}{Nan, T.} \emph{et~al.}
\newblock \bibinfo{journal}{\bibinfo{title}{Anisotropic spin-orbit torque
  generation in epitaxial {SrIrO}$_3$ by symmetry design}}.
\newblock {\emph{\JournalTitle{Proceedings of the National Academy of
  Sciences}}} \textbf{\bibinfo{volume}{116}}, \bibinfo{pages}{16186--16191},
  \doiprefix\url{10.1073/pnas.1812822116} (\bibinfo{year}{2019}).

\bibitem{wang_large_2019}
\bibinfo{author}{Wang, H.} \emph{et~al.}
\newblock \bibinfo{journal}{\bibinfo{title}{Large spin-orbit torque observed in
  epitaxial {{SrIrO}}$_3$ thin films}}.
\newblock {\emph{\JournalTitle{Applied Physics Letters}}}
  \textbf{\bibinfo{volume}{114}}, \bibinfo{pages}{232406},
  \doiprefix\url{10.1063/1.5097699} (\bibinfo{year}{2019}).

\bibitem{kirihara_annealing-temperature-dependent_2018}
\bibinfo{author}{Kirihara, A.} \emph{et~al.}
\newblock \bibinfo{journal}{\bibinfo{title}{Annealing-temperature-dependent
  voltage-sign reversal in all-oxide spin {{Seebeck}} devices using
  {{RuO$_2$}}}}.
\newblock {\emph{\JournalTitle{Journal of Physics D: Applied Physics}}}
  \textbf{\bibinfo{volume}{51}}, \bibinfo{pages}{154002},
  \doiprefix\url{10.1088/1361-6463/aab2cd} (\bibinfo{year}{2018}).

\bibitem{haidar_enhanced_2015}
\bibinfo{author}{Haidar, S.~M.}, \bibinfo{author}{Shiomi, Y.},
  \bibinfo{author}{Lustikova, J.} \& \bibinfo{author}{Saitoh, E.}
\newblock \bibinfo{journal}{\bibinfo{title}{Enhanced inverse spin {{Hall}}
  contribution at high microwave power levels in
  {{La}}$_{0.67}${{Sr}}$_{0.33}${{MnO$_3$}}/{{SrRuO$_3$}} epitaxial bilayers}}.
\newblock {\emph{\JournalTitle{Applied Physics Letters}}}
  \textbf{\bibinfo{volume}{107}}, \bibinfo{pages}{152408},
  \doiprefix\url{10.1063/1.4933379} (\bibinfo{year}{2015}).

\bibitem{richter_spin_2017}
\bibinfo{author}{Richter, T.} \emph{et~al.}
\newblock \bibinfo{journal}{\bibinfo{title}{Spin pumping and inverse spin
  {{Hall}} effect in ultrathin {{SrRuO}}$_3$ films around the percolation
  limit}}.
\newblock {\emph{\JournalTitle{Physical Review B}}}
  \textbf{\bibinfo{volume}{96}}, \bibinfo{pages}{184407},
  \doiprefix\url{10.1103/PhysRevB.96.184407} (\bibinfo{year}{2017}).

\bibitem{emori_spin_2016}
\bibinfo{author}{Emori, S.} \emph{et~al.}
\newblock \bibinfo{journal}{\bibinfo{title}{Spin transport and dynamics in
  all-oxide perovskite {{La}}$_{2/3}${{Sr}}$_{1/3}${{MnO}}$_3$/{{SrRuO}}$_3$
  bilayers probed by ferromagnetic resonance}}.
\newblock {\emph{\JournalTitle{Physical Review B}}}
  \textbf{\bibinfo{volume}{94}}, \bibinfo{pages}{224423},
  \doiprefix\url{10.1103/PhysRevB.94.224423} (\bibinfo{year}{2016}).

\bibitem{ou_exceptionally_2019}
\bibinfo{author}{Ou, Y.} \emph{et~al.}
\newblock \bibinfo{journal}{\bibinfo{title}{Exceptionally {{High}}, {{Strongly
  Temperature Dependent}}, {{Spin Hall Conductivity}} of {{SrRuO}}$_3$}}.
\newblock {\emph{\JournalTitle{Nano Letters}}} \textbf{\bibinfo{volume}{19}},
  \bibinfo{pages}{3663--3670}, \doiprefix\url{10.1021/acs.nanolett.9b00729}
  (\bibinfo{year}{2019}).

\bibitem{davidson_perspectives_2020}
\bibinfo{author}{Davidson, A.}, \bibinfo{author}{Amin, V.~P.},
  \bibinfo{author}{Aljuaid, W.~S.}, \bibinfo{author}{Haney, P.~M.} \&
  \bibinfo{author}{Fan, X.}
\newblock \bibinfo{journal}{\bibinfo{title}{Perspectives of electrically
  generated spin currents in ferromagnetic materials}}.
\newblock {\emph{\JournalTitle{Physics Letters A}}}
  \textbf{\bibinfo{volume}{384}}, \bibinfo{pages}{126228},
  \doiprefix\url{10.1016/j.physleta.2019.126228} (\bibinfo{year}{2020}).

\bibitem{rojas_sanchez_spin--charge_2013}
\bibinfo{author}{Rojas~S{\'a}nchez, J.~C.} \emph{et~al.}
\newblock \bibinfo{journal}{\bibinfo{title}{Spin-to-charge conversion using
  {{Rashba}} coupling at the interface between non-magnetic materials}}.
\newblock {\emph{\JournalTitle{Nature Communications}}}
  \textbf{\bibinfo{volume}{4}}, \doiprefix\url{10.1038/ncomms3944}
  (\bibinfo{year}{2013}).

\bibitem{kondou_efficient_2018}
\bibinfo{author}{Kondou, K.}, \bibinfo{author}{Tsai, H.},
  \bibinfo{author}{Isshiki, H.} \& \bibinfo{author}{Otani, Y.}
\newblock \bibinfo{journal}{\bibinfo{title}{Efficient spin current generation
  and suppression of magnetic damping due to fast spin ejection from
  nonmagnetic metal/indium-tin-oxide interfaces}}.
\newblock {\emph{\JournalTitle{APL Materials}}} \textbf{\bibinfo{volume}{6}},
  \bibinfo{pages}{101105}, \doiprefix\url{10.1063/1.5050848}
  (\bibinfo{year}{2018}).

\bibitem{karube_experimental_2016}
\bibinfo{author}{Karube, S.}, \bibinfo{author}{Kondou, K.} \&
  \bibinfo{author}{Otani, Y.}
\newblock \bibinfo{journal}{\bibinfo{title}{Experimental observation of
  spin-to-charge current conversion at non-magnetic metal/{{Bi}}$_2${{O}}$_3$
  interfaces}}.
\newblock {\emph{\JournalTitle{Applied Physics Express}}}
  \textbf{\bibinfo{volume}{9}}, \bibinfo{pages}{033001},
  \doiprefix\url{10.7567/APEX.9.033001} (\bibinfo{year}{2016}).

\bibitem{picozzi_ferroelectric_2014}
\bibinfo{author}{Picozzi, S.}
\newblock \bibinfo{journal}{\bibinfo{title}{Ferroelectric {{Rashba}}
  semiconductors as a novel class of multifunctional materials}}.
\newblock {\emph{\JournalTitle{Frontiers in Physics}}}
  \textbf{\bibinfo{volume}{2}}, \doiprefix\url{10.3389/fphy.2014.00010}
  (\bibinfo{year}{2014}).

\bibitem{di_sante_electric_2013}
\bibinfo{author}{Di~Sante, D.}, \bibinfo{author}{Barone, P.},
  \bibinfo{author}{Bertacco, R.} \& \bibinfo{author}{Picozzi, S.}
\newblock \bibinfo{journal}{\bibinfo{title}{Electric {{Control}} of the {{Giant
  Rashba Effect}} in {{Bulk GeTe}}}}.
\newblock {\emph{\JournalTitle{Advanced Materials}}}
  \textbf{\bibinfo{volume}{25}}, \bibinfo{pages}{509--513},
  \doiprefix\url{10.1002/adma.201203199} (\bibinfo{year}{2013}).

\bibitem{kolobov_ferroelectric_2014}
\bibinfo{author}{Kolobov, A.~V.} \emph{et~al.}
\newblock \bibinfo{journal}{\bibinfo{title}{Ferroelectric switching in
  epitaxial {{GeTe}} films}}.
\newblock {\emph{\JournalTitle{APL Materials}}} \textbf{\bibinfo{volume}{2}},
  \bibinfo{pages}{066101}, \doiprefix\url{10.1063/1.4881735}
  (\bibinfo{year}{2014}).

\bibitem{rinaldi_ferroelectric_2018}
\bibinfo{author}{Rinaldi, C.} \emph{et~al.}
\newblock \bibinfo{journal}{\bibinfo{title}{Ferroelectric {{Control}} of the
  {{Spin Texture}} in {{GeTe}}}}.
\newblock {\emph{\JournalTitle{Nano Letters}}} \textbf{\bibinfo{volume}{18}},
  \bibinfo{pages}{2751--2758}, \doiprefix\url{10.1021/acs.nanolett.7b04829}
  (\bibinfo{year}{2018}).

\bibitem{krempasky_effects_2008}
\bibinfo{author}{Krempask{\'y}, J.} \emph{et~al.}
\newblock \bibinfo{journal}{\bibinfo{title}{Effects of three-dimensional band
  structure in angle- and spin-resolved photoemission from half-metallic
  {La}$_{2/3}${Sr}$_{1/3}${MnO}$_3$}}.
\newblock {\emph{\JournalTitle{Physical Review B}}}
  \textbf{\bibinfo{volume}{77}}, \bibinfo{pages}{165120},
  \doiprefix\url{10.1103/PhysRevB.77.165120} (\bibinfo{year}{2008}).

\bibitem{rinaldi_evidence_2016}
\bibinfo{author}{Rinaldi, C.} \emph{et~al.}
\newblock \bibinfo{journal}{\bibinfo{title}{Evidence for spin to charge
  conversion in {{GeTe}}(111)}}.
\newblock {\emph{\JournalTitle{APL Materials}}} \textbf{\bibinfo{volume}{4}},
  \bibinfo{pages}{032501}, \doiprefix\url{10.1063/1.4941276}
  (\bibinfo{year}{2016}).

\bibitem{varignon_electrically_2019}
\bibinfo{author}{Varignon, J.}, \bibinfo{author}{Santamaria, J.} \&
  \bibinfo{author}{Bibes, M.}
\newblock \bibinfo{journal}{\bibinfo{title}{Electrically {{Switchable}} and
  {{Tunable Rashba}}-{{Type Spin Splitting}} in {{Covalent Perovskite
  Oxides}}}}.
\newblock {\emph{\JournalTitle{Physical Review Letters}}}
  \textbf{\bibinfo{volume}{122}}, \bibinfo{pages}{116401},
  \doiprefix\url{10.1103/PhysRevLett.122.116401} (\bibinfo{year}{2019}).

\bibitem{da_silveira_rashba-dresselhaus_2016}
\bibinfo{author}{da~Silveira, L. G.~D.}, \bibinfo{author}{Barone, P.} \&
  \bibinfo{author}{Picozzi, S.}
\newblock \bibinfo{journal}{\bibinfo{title}{Rashba-{{Dresselhaus}}
  spin-splitting in the bulk ferroelectric oxide {{BiAlO}}$_3$}}.
\newblock {\emph{\JournalTitle{Physical Review B}}}
  \textbf{\bibinfo{volume}{93}}, \bibinfo{pages}{245159},
  \doiprefix\url{10.1103/PhysRevB.93.245159} (\bibinfo{year}{2016}).

\bibitem{arras_rashba-like_2019}
\bibinfo{author}{Arras, R.} \emph{et~al.}
\newblock \bibinfo{journal}{\bibinfo{title}{Rashba-like spin-orbit and strain
  effects in tetragonal {{PbTiO}}$_3$}}.
\newblock {\emph{\JournalTitle{Physical Review B}}}
  \textbf{\bibinfo{volume}{100}}, \bibinfo{pages}{174415},
  \doiprefix\url{10.1103/PhysRevB.100.174415} (\bibinfo{year}{2019}).

\bibitem{tao_strain-tunable_2016}
\bibinfo{author}{Tao, L.~L.} \& \bibinfo{author}{Wang, J.}
\newblock \bibinfo{journal}{\bibinfo{title}{Strain-tunable ferroelectricity and
  its control of {{Rashba}} effect in {{KTaO}}$_3$}}.
\newblock {\emph{\JournalTitle{Journal of Applied Physics}}}
  \textbf{\bibinfo{volume}{120}}, \bibinfo{pages}{234101},
  \doiprefix\url{10.1063/1.4972198} (\bibinfo{year}{2016}).

\bibitem{djani_rationalizing_2019}
\bibinfo{author}{Djani, H.} \emph{et~al.}
\newblock \bibinfo{journal}{\bibinfo{title}{Rationalizing and engineering
  {{Rashba}} spin-splitting in ferroelectric oxides}}.
\newblock {\emph{\JournalTitle{npj Quantum Materials}}}
  \textbf{\bibinfo{volume}{4}}, \bibinfo{pages}{51},
  \doiprefix\url{10.1038/s41535-019-0190-z} (\bibinfo{year}{2019}).

\bibitem{mirhosseini_toward_2010}
\bibinfo{author}{Mirhosseini, H.} \emph{et~al.}
\newblock \bibinfo{journal}{\bibinfo{title}{Toward a ferroelectric control of
  {{Rashba}} spin-orbit coupling: {{Bi}} on {{BaTiO}}$_3$ (001) from first
  principles}}.
\newblock {\emph{\JournalTitle{Physical Review B}}}
  \textbf{\bibinfo{volume}{81}}, \bibinfo{pages}{073406},
  \doiprefix\url{10.1103/PhysRevB.81.073406} (\bibinfo{year}{2010}).

\bibitem{lutz_large_2017}
\bibinfo{author}{Lutz, P.}, \bibinfo{author}{Figgemeier, T.},
  \bibinfo{author}{El-Fattah, Z. M.~A.}, \bibinfo{author}{Bentmann, H.} \&
  \bibinfo{author}{Reinert, F.}
\newblock \bibinfo{journal}{\bibinfo{title}{Large {{Spin Splitting}} and
  {{Interfacial States}} in a {{Bi}}/{{BaTiO}}$_3$ (001) {{Rashba Ferroelectric
  Heterostructure}}}}.
\newblock {\emph{\JournalTitle{Physical Review Applied}}}
  \textbf{\bibinfo{volume}{7}}, \bibinfo{pages}{044011},
  \doiprefix\url{10.1103/PhysRevApplied.7.044011} (\bibinfo{year}{2017}).

\bibitem{zhong_giant_2015}
\bibinfo{author}{Zhong, Z.} \emph{et~al.}
\newblock \bibinfo{journal}{\bibinfo{title}{Giant {{Switchable Rashba Effect}}
  in {{Oxide Heterostructures}}}}.
\newblock {\emph{\JournalTitle{Advanced Materials Interfaces}}}
  \textbf{\bibinfo{volume}{2}}, \bibinfo{pages}{1400445},
  \doiprefix\url{10.1002/admi.201400445} (\bibinfo{year}{2015}).

\bibitem{hembergerElectricfielddependentDielectricConstant1995}
\bibinfo{author}{Hemberger, J.}, \bibinfo{author}{Lunkenheimer, P.},
  \bibinfo{author}{Viana, R.}, \bibinfo{author}{B{\"o}hmer, R.} \&
  \bibinfo{author}{Loidl, A.}
\newblock \bibinfo{journal}{\bibinfo{title}{Electric-field-dependent dielectric
  constant and nonlinear susceptibility in {{SrTiO}}$_3$}}.
\newblock {\emph{\JournalTitle{Physical Review B}}}
  \textbf{\bibinfo{volume}{52}}, \bibinfo{pages}{13159--13162},
  \doiprefix\url{10.1103/PhysRevB.52.13159} (\bibinfo{year}{1995}).

\bibitem{bednorzSr1xCaxTiO3XYQuantum1984}
\bibinfo{author}{Bednorz, J.~G.} \& \bibinfo{author}{M{\"u}ller, K.~A.}
\newblock
  \bibinfo{journal}{\bibinfo{title}{{{Sr}}$_{1-x}${{Ca}}$_x${{TiO}}$_3$. {{An
  XY Quantum Ferroelectric}} with {{Transition}} to {{Randomness}}}}.
\newblock {\emph{\JournalTitle{Physical Review Letters}}}
  \textbf{\bibinfo{volume}{52}}, \bibinfo{pages}{2289--2292}
  (\bibinfo{year}{1984}).

\bibitem{brehin_switchable_2020}
\bibinfo{author}{Br\'ehin, J.} \emph{et~al.}
\newblock \bibinfo{journal}{\bibinfo{title}{Switchable two-dimensional electron
  gas based on ferroelectric {{Ca}}:{{SrTi}}{{O}}$_{3}$}}.
\newblock {\emph{\JournalTitle{Phys. Rev. Mater.}}}
  \textbf{\bibinfo{volume}{4}}, \bibinfo{pages}{041002},
  \doiprefix\url{10.1103/PhysRevMaterials.4.041002} (\bibinfo{year}{2020}).

\bibitem{tuviaFerroelectricExchangeBias2020}
\bibinfo{author}{Tuvia, G.} \emph{et~al.}
\newblock \bibinfo{journal}{\bibinfo{title}{Ferroelectric {{Exchange Bias
  Affects Interfacial Electronic States}}}}.
\newblock {\emph{\JournalTitle{Advanced Materials}}} \bibinfo{pages}{2000216},
  \doiprefix\url{10.1002/adma.202000216} (\bibinfo{year}{2020}).

\bibitem{varotto_room-temperature_2021}
\bibinfo{author}{Varotto, S.} \emph{et~al.}
\newblock \bibinfo{journal}{\bibinfo{title}{Troom-temperature ferroelectric
  switching of spin-to-charge conversion in gete}}.
\newblock {\emph{\JournalTitle{ArXiv}}} \bibinfo{pages}{2103.07646}.

\bibitem{manipatruniCMOSComputingSpin2018}
\bibinfo{author}{Manipatruni, S.}, \bibinfo{author}{Nikonov, D.~E.} \&
  \bibinfo{author}{Young, I.~A.}
\newblock \bibinfo{journal}{\bibinfo{title}{Beyond {{CMOS}} computing with spin
  and polarization}}.
\newblock {\emph{\JournalTitle{Nature Physics}}} \textbf{\bibinfo{volume}{14}},
  \bibinfo{pages}{338--343}, \doiprefix\url{10.1038/s41567-018-0101-4}
  (\bibinfo{year}{2018}).

\bibitem{coeyNoncolllnearSpinStructures1987}
\bibinfo{author}{Coey, M.~D.}
\newblock \bibinfo{journal}{\bibinfo{title}{Noncolllnear spin structures'}}.
\newblock {\emph{\JournalTitle{Can. J. Phys.}}} \textbf{\bibinfo{volume}{65}},
  \bibinfo{pages}{23}, \doiprefix\url{10.1139/p87-197} (\bibinfo{year}{1987}).

\bibitem{gardner_magnetic_2010}
\bibinfo{author}{Gardner, J.~S.}, \bibinfo{author}{Gingras, M. J.~P.} \&
  \bibinfo{author}{Greedan, J.~E.}
\newblock \bibinfo{journal}{\bibinfo{title}{Magnetic pyrochlore oxides}}.
\newblock {\emph{\JournalTitle{Reviews of Modern Physics}}}
  \textbf{\bibinfo{volume}{82}}, \bibinfo{pages}{55} (\bibinfo{year}{2010}).

\bibitem{murthy_yafet-kittel_1969}
\bibinfo{author}{Murthy, N. S.~S.}, \bibinfo{author}{Natera, M.~G.},
  \bibinfo{author}{Youssef, S.~I.}, \bibinfo{author}{Begum, R.~J.} \&
  \bibinfo{author}{Srivastava, C.~M.}
\newblock \bibinfo{journal}{\bibinfo{title}{Yafet-{{Kittel Angles}} in
  {{Zinc}}-{{Nickel Ferrites}}}}.
\newblock {\emph{\JournalTitle{Physical Review}}}
  \textbf{\bibinfo{volume}{181}}, \bibinfo{pages}{969--977},
  \doiprefix\url{10.1103/PhysRev.181.969} (\bibinfo{year}{1969}).

\bibitem{yafet_antiferromagnetic_1952}
\bibinfo{author}{Yafet, Y.} \& \bibinfo{author}{Kittel, C.}
\newblock \bibinfo{journal}{\bibinfo{title}{Antiferromagnetic {{Arrangements}}
  in {{Ferrites}}}}.
\newblock {\emph{\JournalTitle{Physical Review}}}
  \textbf{\bibinfo{volume}{87}}, \bibinfo{pages}{290--294},
  \doiprefix\url{10.1103/PhysRev.87.290} (\bibinfo{year}{1952}).

\bibitem{takeda_magnetic_1972}
\bibinfo{author}{Takeda, T.}, \bibinfo{author}{Yamaguchi, Y.} \&
  \bibinfo{author}{Watanabe, H.}
\newblock \bibinfo{journal}{\bibinfo{title}{Magnetic {{Structure}} of
  {{SrFeO}}$_3$}}.
\newblock {\emph{\JournalTitle{Journal of the Physical Society of Japan}}}
  \textbf{\bibinfo{volume}{33}}, \bibinfo{pages}{967--969},
  \doiprefix\url{10.1143/JPSJ.33.967} (\bibinfo{year}{1972}).

\bibitem{mostovoy_helicoidal_2005}
\bibinfo{author}{Mostovoy, M.}
\newblock \bibinfo{journal}{\bibinfo{title}{Helicoidal {{Ordering}} in {{Iron
  Perovskites}}}}.
\newblock {\emph{\JournalTitle{Physical Review Letters}}}
  \textbf{\bibinfo{volume}{94}}, \bibinfo{pages}{137205},
  \doiprefix\url{10.1103/PhysRevLett.94.137205} (\bibinfo{year}{2005}).

\bibitem{muhlbauer_skyrmion_2009}
\bibinfo{author}{Muhlbauer, S.} \emph{et~al.}
\newblock \bibinfo{journal}{\bibinfo{title}{Skyrmion {{Lattice}} in a {{Chiral
  Magnet}}}}.
\newblock {\emph{\JournalTitle{Science}}} \textbf{\bibinfo{volume}{323}},
  \bibinfo{pages}{915--919}, \doiprefix\url{10.1126/science.1166767}
  (\bibinfo{year}{2009}).

\bibitem{izyumov_modulated_1984}
\bibinfo{author}{Izyumov, Y.~A.}
\newblock \bibinfo{journal}{\bibinfo{title}{Modulated, or long-periodic,
  magnetic structures of crystals}}.
\newblock {\emph{\JournalTitle{Soviet Physics Uspekhi}}}
  \textbf{\bibinfo{volume}{27}}, \bibinfo{pages}{845} (\bibinfo{year}{1984}).

\bibitem{kimuraSpiralMagnetsMagnetoelectrics2007}
\bibinfo{author}{Kimura, T.}
\newblock \bibinfo{journal}{\bibinfo{title}{{Spiral Magnets as
  Magnetoelectrics}}}.
\newblock {\emph{\JournalTitle{Annual Review of Materials Research}}}
  \textbf{\bibinfo{volume}{37}}, \bibinfo{pages}{387--413},
  \doiprefix\url{10.1146/annurev.matsci.37.052506.084259}
  (\bibinfo{year}{2007}).

\bibitem{sosnowska_spiral_1982}
\bibinfo{author}{Sosnowska, I.}, \bibinfo{author}{Neumaier, T.~P.} \&
  \bibinfo{author}{Steichele, E.}
\newblock \bibinfo{journal}{\bibinfo{title}{Spiral magnetic ordering in bismuth
  ferrite}}.
\newblock {\emph{\JournalTitle{Journal of Physics C: Solid State Physics}}}
  \textbf{\bibinfo{volume}{15}}, \bibinfo{pages}{4835--4846},
  \doiprefix\url{10.1088/0022-3719/15/23/020} (\bibinfo{year}{1982}).

\bibitem{burnsExperimentalistGuideCycloid2020}
\bibinfo{author}{Burns, S.~R.}, \bibinfo{author}{Paull, O.},
  \bibinfo{author}{Juraszek, J.}, \bibinfo{author}{Nagarajan, V.} \&
  \bibinfo{author}{Sando, D.}
\newblock \bibinfo{journal}{\bibinfo{title}{The {{Experimentalist}}'s {{Guide}}
  to the {{Cycloid}}, or {{Noncollinear Antiferromagnetism}} in {{Epitaxial
  BiFeO}}$_3$}}.
\newblock {\emph{\JournalTitle{Advanced Materials}}}
  \textbf{\bibinfo{volume}{32}}, \bibinfo{pages}{2003711},
  \doiprefix\url{10.1002/adma.202003711} (\bibinfo{year}{2020}).

\bibitem{sando_crafting_2013}
\bibinfo{author}{Sando, D.} \emph{et~al.}
\newblock \bibinfo{journal}{\bibinfo{title}{Crafting the magnonic and
  spintronic response of {{BiFeO}}$_3$ films by epitaxial strain}}.
\newblock {\emph{\JournalTitle{Nature Materials}}}
  \textbf{\bibinfo{volume}{12}}, \bibinfo{pages}{641--646},
  \doiprefix\url{10.1038/nmat3629} (\bibinfo{year}{2013}).

\bibitem{agbelele_strain_2017}
\bibinfo{author}{Agbelele, A.} \emph{et~al.}
\newblock \bibinfo{journal}{\bibinfo{title}{Strain and {{Magnetic Field Induced
  Spin}}-{{Structure Transitions}} in {{Multiferroic BiFeO}}$_3$}}.
\newblock {\emph{\JournalTitle{Advanced Materials}}}
  \textbf{\bibinfo{volume}{29}}, \bibinfo{pages}{1602327},
  \doiprefix\url{10.1002/adma.201602327} (\bibinfo{year}{2017}).

\bibitem{haykal_antiferromagnetic_2020}
\bibinfo{author}{Haykal, A.} \emph{et~al.}
\newblock \bibinfo{journal}{\bibinfo{title}{Antiferromagnetic textures in
  {{BiFeO}}$_3$ controlled by strain and electric field}}.
\newblock {\emph{\JournalTitle{Nature Communications}}}
  \textbf{\bibinfo{volume}{11}}, \bibinfo{pages}{1704},
  \doiprefix\url{10.1038/s41467-020-15501-8} (\bibinfo{year}{2020}).

\bibitem{legrand_room-temperature_2020}
\bibinfo{author}{Legrand, W.} \emph{et~al.}
\newblock \bibinfo{journal}{\bibinfo{title}{Room-temperature stabilization of
  antiferromagnetic skyrmions in synthetic antiferromagnets}}.
\newblock {\emph{\JournalTitle{Nature Materials}}}
  \textbf{\bibinfo{volume}{19}}, \bibinfo{pages}{34--42},
  \doiprefix\url{10.1038/s41563-019-0468-3} (\bibinfo{year}{2020}).

\bibitem{zhang_antiferromagnetic_2016}
\bibinfo{author}{Zhang, X.}, \bibinfo{author}{Zhou, Y.} \&
  \bibinfo{author}{Ezawa, M.}
\newblock \bibinfo{journal}{\bibinfo{title}{Antiferromagnetic {{Skyrmion}}:
  {{Stability}}, {{Creation}} and {{Manipulation}}}}.
\newblock {\emph{\JournalTitle{Scientific Reports}}}
  \textbf{\bibinfo{volume}{6}}, \doiprefix\url{10.1038/srep24795}
  (\bibinfo{year}{2016}).

\bibitem{kimura_magnetic_2003}
\bibinfo{author}{Kimura, T.} \emph{et~al.}
\newblock \bibinfo{journal}{\bibinfo{title}{Magnetic control of ferroelectric
  polarization}}.
\newblock {\emph{\JournalTitle{Nature}}} \textbf{\bibinfo{volume}{426}},
  \bibinfo{pages}{55--58}, \doiprefix\url{10.1038/nature02018}
  (\bibinfo{year}{2003}).

\bibitem{hur_electric_2004}
\bibinfo{author}{Hur, N.} \emph{et~al.}
\newblock \bibinfo{journal}{\bibinfo{title}{Electric polarization reversal and
  memory in a multiferroic material induced by magnetic fields}}.
\newblock {\emph{\JournalTitle{Nature}}} \textbf{\bibinfo{volume}{429}},
  \bibinfo{pages}{392--395}, \doiprefix\url{10.1038/nature02572}
  (\bibinfo{year}{2004}).

\bibitem{katsura_spin_2005}
\bibinfo{author}{Katsura, H.}, \bibinfo{author}{Nagaosa, N.} \&
  \bibinfo{author}{Balatsky, A.~V.}
\newblock \bibinfo{journal}{\bibinfo{title}{Spin {{Current}} and
  {{Magnetoelectric Effect}} in {{Noncollinear Magnets}}}}.
\newblock {\emph{\JournalTitle{Physical Review Letters}}}
  \textbf{\bibinfo{volume}{95}}, \bibinfo{pages}{057205},
  \doiprefix\url{10.1103/PhysRevLett.95.057205} (\bibinfo{year}{2005}).

\bibitem{fiebig_evolution_2016}
\bibinfo{author}{Fiebig, M.}, \bibinfo{author}{Lottermoser, T.},
  \bibinfo{author}{Meier, D.} \& \bibinfo{author}{Trassin, M.}
\newblock \bibinfo{journal}{\bibinfo{title}{The evolution of multiferroics}}.
\newblock {\emph{\JournalTitle{Nature Reviews Materials}}}
  \textbf{\bibinfo{volume}{1}}, \bibinfo{pages}{16046},
  \doiprefix\url{10.1038/natrevmats.2016.46} (\bibinfo{year}{2016}).

\bibitem{kurumaji_spiral_2020}
\bibinfo{author}{Kurumaji, T.}
\newblock \bibinfo{journal}{\bibinfo{title}{Spiral spin structures and
  skyrmions in multiferroics}}.
\newblock {\emph{\JournalTitle{Physical Sciences Reviews}}}
  \textbf{\bibinfo{volume}{5}}, \doiprefix\url{10.1515/psr-2019-0016}
  (\bibinfo{year}{2020}).

\bibitem{seki_observation_2012}
\bibinfo{author}{Seki, S.}, \bibinfo{author}{Yu, X.~Z.},
  \bibinfo{author}{Ishiwata, S.} \& \bibinfo{author}{Tokura, Y.}
\newblock \bibinfo{journal}{\bibinfo{title}{Observation of {{Skyrmions}} in a
  {{Multiferroic Material}}}}.
\newblock {\emph{\JournalTitle{Science}}} \textbf{\bibinfo{volume}{336}},
  \bibinfo{pages}{198--201}, \doiprefix\url{10.1126/science.1214143}
  (\bibinfo{year}{2012}).

\bibitem{tokura_colossal_1999}
\bibinfo{author}{Tokura, Y.} \& \bibinfo{author}{Tomioka, Y.}
\newblock \bibinfo{journal}{\bibinfo{title}{Colossal magnetoresistive
  manganites}}.
\newblock {\emph{\JournalTitle{Journal of Magnetism and Magnetic Materials}}}
  \bibinfo{pages}{23} (\bibinfo{year}{1999}).

\bibitem{bowen_spin-polarized_2005}
\bibinfo{author}{Bowen, M.} \emph{et~al.}
\newblock \bibinfo{journal}{\bibinfo{title}{Spin-{{Polarized Tunneling
  Spectroscopy}} in {{Tunnel Junctions}} with {{Half}}-{{Metallic
  Electrodes}}}}.
\newblock {\emph{\JournalTitle{Physical Review Letters}}}
  \textbf{\bibinfo{volume}{95}}, \bibinfo{pages}{137203},
  \doiprefix\url{10.1103/PhysRevLett.95.137203} (\bibinfo{year}{2005}).

\bibitem{bibes_oxide_2007}
\bibinfo{author}{Bibes, M.} \& \bibinfo{author}{Barthelemy, A.}
\newblock \bibinfo{journal}{\bibinfo{title}{Oxide {{Spintronics}}}}.
\newblock {\emph{\JournalTitle{IEEE Transactions on Electron Devices}}}
  \textbf{\bibinfo{volume}{54}}, \bibinfo{pages}{1003--1023},
  \doiprefix\url{10.1109/TED.2007.894366} (\bibinfo{year}{2007}).

\bibitem{sakai_electron_2010}
\bibinfo{author}{Sakai, H.} \emph{et~al.}
\newblock \bibinfo{journal}{\bibinfo{title}{Electron doping in the cubic
  perovskite {{SrMnO}}$_3$: {{Isotropic}} metal versus chainlike ordering of
  {{Jahn}}-{{Teller}} polarons}}.
\newblock {\emph{\JournalTitle{Physical Review B}}}
  \textbf{\bibinfo{volume}{82}}, \bibinfo{pages}{180409},
  \doiprefix\url{10.1103/PhysRevB.82.180409} (\bibinfo{year}{2010}).

\bibitem{caspi_structural_2004}
\bibinfo{author}{Caspi, E.~N.} \emph{et~al.}
\newblock \bibinfo{journal}{\bibinfo{title}{Structural and magnetic phase
  diagram of the two-electron-doped ({Ca}$_{1-x}${Ce}$_x$){MnO}$_3$ system:
  Effects of competition among charge, orbital, and spin ordering}}.
\newblock {\emph{\JournalTitle{Physical Review B}}}
  \textbf{\bibinfo{volume}{69}}, \bibinfo{pages}{104402},
  \doiprefix\url{10.1103/PhysRevB.69.104402} (\bibinfo{year}{2004}).

\bibitem{nagai_formation_2012}
\bibinfo{author}{Nagai, T.} \emph{et~al.}
\newblock \bibinfo{journal}{\bibinfo{title}{Formation of nanoscale magnetic
  bubbles in ferromagnetic insulating manganite
  {{La}}$_{7/8}${{Sr}}$_{1/8}${{MnO}}$_{3}$}}.
\newblock {\emph{\JournalTitle{Applied Physics Letters}}}
  \textbf{\bibinfo{volume}{101}}, \bibinfo{pages}{162401},
  \doiprefix\url{10.1063/1.4760266} (\bibinfo{year}{2012}).

\bibitem{yu_biskyrmion_2014}
\bibinfo{author}{Yu, X.~Z.} \emph{et~al.}
\newblock \bibinfo{journal}{\bibinfo{title}{Biskyrmion states and their
  current-driven motion in a layered manganite}}.
\newblock {\emph{\JournalTitle{Nature Communications}}}
  \textbf{\bibinfo{volume}{5}}, \doiprefix\url{10.1038/ncomms4198}
  (\bibinfo{year}{2014}).

\bibitem{yu_variation_2017}
\bibinfo{author}{Yu, X.}, \bibinfo{author}{Tokunaga, Y.},
  \bibinfo{author}{Taguchi, Y.} \& \bibinfo{author}{Tokura, Y.}
\newblock \bibinfo{journal}{\bibinfo{title}{Variation of {{Topology}} in
  {{Magnetic Bubbles}} in a {{Colossal Magnetoresistive Manganite}}}}.
\newblock {\emph{\JournalTitle{Advanced Materials}}}
  \textbf{\bibinfo{volume}{29}}, \bibinfo{pages}{1603958},
  \doiprefix\url{10.1002/adma.201603958} (\bibinfo{year}{2017}).

\bibitem{kotani_observation_2016}
\bibinfo{author}{Kotani, A.}, \bibinfo{author}{Nakajima, H.},
  \bibinfo{author}{Ishii, Y.}, \bibinfo{author}{Harada, K.} \&
  \bibinfo{author}{Mori, S.}
\newblock \bibinfo{journal}{\bibinfo{title}{Observation of spin textures in
  {La}${}_{1-x}${Sr}${}_x${MnO}${}_3$ ($x = 0.175$)}}.
\newblock {\emph{\JournalTitle{AIP Advances}}} \textbf{\bibinfo{volume}{6}},
  \bibinfo{pages}{056403}, \doiprefix\url{10.1063/1.4943611}
  (\bibinfo{year}{2016}).

\bibitem{nagao_direct_2013}
\bibinfo{author}{Nagao, M.} \emph{et~al.}
\newblock \bibinfo{journal}{\bibinfo{title}{Direct observation and dynamics of
  spontaneous skyrmion-like magnetic domains in a ferromagnet}}.
\newblock {\emph{\JournalTitle{Nature Nanotechnology}}}
  \textbf{\bibinfo{volume}{8}}, \bibinfo{pages}{325--328},
  \doiprefix\url{10.1038/nnano.2013.69} (\bibinfo{year}{2013}).

\bibitem{vistoli_giant_2019}
\bibinfo{author}{Vistoli, L.} \emph{et~al.}
\newblock \bibinfo{journal}{\bibinfo{title}{Giant topological {{Hall}} effect
  in correlated oxide thin films}}.
\newblock {\emph{\JournalTitle{Nature Physics}}} \textbf{\bibinfo{volume}{15}},
  \bibinfo{pages}{67--72}, \doiprefix\url{10.1038/s41567-018-0307-5}
  (\bibinfo{year}{2019}).

\bibitem{nakamura_emergence_2018}
\bibinfo{author}{Nakamura, M.} \emph{et~al.}
\newblock \bibinfo{journal}{\bibinfo{title}{Emergence of topological {{Hall}}
  effect in half-metallic manganite thin films by tuning perpendicular magnetic
  anisotropy}}.
\newblock {\emph{\JournalTitle{Journal of the Physical Society of Japan}}}
  \textbf{\bibinfo{volume}{87}}, \bibinfo{pages}{074704},
  \doiprefix\url{10.7566/JPSJ.87.074704} (\bibinfo{year}{2018}).
\newblock \eprint{1804.01425}.

\bibitem{xiang_phase_2012}
\bibinfo{author}{Xiang, P.-H.}, \bibinfo{author}{Yamada, H.},
  \bibinfo{author}{Akoh, H.} \& \bibinfo{author}{Sawa, A.}
\newblock \bibinfo{journal}{\bibinfo{title}{Phase diagrams of strained
  {Ca}$_{1-x}${Ce}$_x${MnO}$_3$ films}}.
\newblock {\emph{\JournalTitle{Journal of Applied Physics}}}
  \textbf{\bibinfo{volume}{112}}, \bibinfo{pages}{113703},
  \doiprefix\url{10.1063/1.4768198} (\bibinfo{year}{2012}).

\bibitem{nakazawa_weak_2019}
\bibinfo{author}{Nakazawa, K.} \& \bibinfo{author}{Kohno, H.}
\newblock \bibinfo{journal}{\bibinfo{title}{Weak coupling theory of topological
  {{Hall}} effect}}.
\newblock {\emph{\JournalTitle{Physical Review B}}}
  \textbf{\bibinfo{volume}{99}}, \bibinfo{pages}{174425},
  \doiprefix\url{10.1103/PhysRevB.99.174425} (\bibinfo{year}{2019}).

\bibitem{nakazawa_topological_2018}
\bibinfo{author}{Nakazawa, K.}, \bibinfo{author}{Bibes, M.} \&
  \bibinfo{author}{Kohno, H.}
\newblock \bibinfo{journal}{\bibinfo{title}{Topological {{Hall Effect}} from
  {{Strong}} to {{Weak Coupling}}}}.
\newblock {\emph{\JournalTitle{Journal of the Physical Society of Japan}}}
  \textbf{\bibinfo{volume}{87}}, \bibinfo{pages}{033705},
  \doiprefix\url{10.7566/JPSJ.87.033705} (\bibinfo{year}{2018}).

\bibitem{skoropata_interfacial_2020}
\bibinfo{author}{Skoropata, E.} \emph{et~al.}
\newblock \bibinfo{journal}{\bibinfo{title}{Interfacial tuning of chiral
  magnetic interactions for large topological {{Hall}} effects in
  {{LaMnO}}$_3$/{{SrIrO}}$_3$ heterostructures}}.
\newblock {\emph{\JournalTitle{Science Advances}}}
  \textbf{\bibinfo{volume}{6}}, \bibinfo{pages}{eaaz3902},
  \doiprefix\url{10.1126/sciadv.aaz3902} (\bibinfo{year}{2020}).

\bibitem{li_emergent_2019}
\bibinfo{author}{Li, Y.} \emph{et~al.}
\newblock \bibinfo{journal}{\bibinfo{title}{Emergent {{Topological Hall
  Effect}} in {{La}}$_{0.7}${{Sr}}$_{0.3}${{MnO}}$_3$/{{SrIrO}}$_3$
  {{Heterostructures}}}}.
\newblock {\emph{\JournalTitle{ACS Applied Materials \& Interfaces}}}
  \textbf{\bibinfo{volume}{11}}, \bibinfo{pages}{21268--21274},
  \doiprefix\url{10.1021/acsami.9b05562} (\bibinfo{year}{2019}).

\bibitem{mohanta_topological_2019}
\bibinfo{author}{Mohanta, N.}, \bibinfo{author}{Dagotto, E.} \&
  \bibinfo{author}{Okamoto, S.}
\newblock \bibinfo{journal}{\bibinfo{title}{Topological {{Hall}} effect and
  emergent skyrmion crystal at manganite-iridate oxide interfaces}}.
\newblock {\emph{\JournalTitle{Physical Review B}}}
  \textbf{\bibinfo{volume}{100}}, \bibinfo{pages}{064429},
  \doiprefix\url{10.1103/PhysRevB.100.064429} (\bibinfo{year}{2019}).

\bibitem{koster_structure_2012}
\bibinfo{author}{Koster, G.} \emph{et~al.}
\newblock \bibinfo{journal}{\bibinfo{title}{Structure, physical properties, and
  applications of {{SrRuO}}$_3$ thin films}}.
\newblock {\emph{\JournalTitle{Reviews of Modern Physics}}}
  \textbf{\bibinfo{volume}{84}}, \bibinfo{pages}{253--298},
  \doiprefix\url{10.1103/RevModPhys.84.253} (\bibinfo{year}{2012}).

\bibitem{matsuno_interface-driven_2016}
\bibinfo{author}{Matsuno, J.} \emph{et~al.}
\newblock \bibinfo{journal}{\bibinfo{title}{Interface-driven topological
  {{Hall}} effect in {{SrRuO}}$_3$-{{SrIrO}}$_3$ bilayer}}.
\newblock {\emph{\JournalTitle{Science Advances}}}
  \textbf{\bibinfo{volume}{2}}, \bibinfo{pages}{e1600304},
  \doiprefix\url{10.1126/sciadv.1600304} (\bibinfo{year}{2016}).

\bibitem{pang_spin-glass-like_2017}
\bibinfo{author}{Pang, B.} \emph{et~al.}
\newblock \bibinfo{journal}{\bibinfo{title}{Spin-{{Glass}}-{{Like Behavior}}
  and {{Topological Hall Effect}} in {{SrRuO}}$_3$/{{SrIrO}}$_3$
  {{Superlattices}} for {{Oxide Spintronics Applications}}}}.
\newblock {\emph{\JournalTitle{ACS Applied Materials \& Interfaces}}}
  \textbf{\bibinfo{volume}{9}}, \bibinfo{pages}{3201--3207},
  \doiprefix\url{10.1021/acsami.7b00150} (\bibinfo{year}{2017}).

\bibitem{meng_observation_2019}
\bibinfo{author}{Meng, K.-Y.} \emph{et~al.}
\newblock \bibinfo{journal}{\bibinfo{title}{Observation of {{Nanoscale
  Skyrmions}} in {{SrIrO}}$_3$/{{SrRuO}}$_3$ {{Bilayers}}}}.
\newblock {\emph{\JournalTitle{Nano Letters}}} \textbf{\bibinfo{volume}{19}},
  \bibinfo{pages}{3169--3175}, \doiprefix\url{10.1021/acs.nanolett.9b00596}
  (\bibinfo{year}{2019}).

\bibitem{ziese_unconventional_2019}
\bibinfo{author}{Ziese, M.}, \bibinfo{author}{Jin, L.} \&
  \bibinfo{author}{Lindfors-Vrejoiu, I.}
\newblock \bibinfo{journal}{\bibinfo{title}{Unconventional anomalous {{Hall}}
  effect driven by oxygen-octahedra-tailoring of the {{SrRuO}}$_3$ structure}}.
\newblock {\emph{\JournalTitle{Journal of Physics: Materials}}}
  \textbf{\bibinfo{volume}{2}}, \bibinfo{pages}{034008},
  \doiprefix\url{10.1088/2515-7639/ab1aef} (\bibinfo{year}{2019}).

\bibitem{gu_interfacial_2019}
\bibinfo{author}{Gu, Y.} \emph{et~al.}
\newblock \bibinfo{journal}{\bibinfo{title}{Interfacial
  oxygen-octahedral-tilting-driven electrically tunable topological {{Hall}}
  effect in ultrathin {{SrRuO}}$_3$ films}}.
\newblock {\emph{\JournalTitle{Journal of Physics D: Applied Physics}}}
  \textbf{\bibinfo{volume}{52}}, \bibinfo{pages}{404001},
  \doiprefix\url{10.1088/1361-6463/ab2fe8} (\bibinfo{year}{2019}).

\bibitem{qin_emergence_2019}
\bibinfo{author}{Qin, Q.} \emph{et~al.}
\newblock \bibinfo{journal}{\bibinfo{title}{Emergence of {{Topological Hall
  Effect}} in a {{SrRuO$_3$}} {{Single Layer}}}}.
\newblock {\emph{\JournalTitle{Advanced Materials}}}
  \textbf{\bibinfo{volume}{31}}, \bibinfo{pages}{1807008},
  \doiprefix\url{10.1002/adma.201807008} (\bibinfo{year}{2019}).

\bibitem{kan_electric_2020}
\bibinfo{author}{Kan, D.}, \bibinfo{author}{Kobayashi, K.} \&
  \bibinfo{author}{Shimakawa, Y.}
\newblock \bibinfo{journal}{\bibinfo{title}{Electric field induced modulation
  of transverse resistivity anomalies in ultrathin {{SrRu O}}$_3$ epitaxial
  films}}.
\newblock {\emph{\JournalTitle{Physical Review B}}}
  \textbf{\bibinfo{volume}{101}}, \bibinfo{pages}{144405},
  \doiprefix\url{10.1103/PhysRevB.101.144405} (\bibinfo{year}{2020}).

\bibitem{wang_controllable_2020}
\bibinfo{author}{Wang, L.} \emph{et~al.}
\newblock \bibinfo{journal}{\bibinfo{title}{Controllable {{Thickness
  Inhomogeneity}} and {{Berry Curvature Engineering}} of {{Anomalous Hall
  Effect}} in {{SrRuO}} {\textsubscript{3}} {{Ultrathin Films}}}}.
\newblock {\emph{\JournalTitle{Nano Letters}}} \textbf{\bibinfo{volume}{20}},
  \bibinfo{pages}{2468--2477}, \doiprefix\url{10.1021/acs.nanolett.9b05206}
  (\bibinfo{year}{2020}).

\bibitem{groenendijk_berry_2020}
\bibinfo{author}{Groenendijk, D.~J.} \emph{et~al.}
\newblock \bibinfo{journal}{\bibinfo{title}{Berry phase engineering at oxide
  interfaces}}.
\newblock {\emph{\JournalTitle{Physical Review Research}}}
  \textbf{\bibinfo{volume}{2}}, \bibinfo{pages}{023404},
  \doiprefix\url{10.1103/PhysRevResearch.2.023404} (\bibinfo{year}{2020}).

\bibitem{huang_detection_2020}
\bibinfo{author}{Huang, H.} \emph{et~al.}
\newblock \bibinfo{journal}{\bibinfo{title}{Detection of the {{Chiral Spin
  Structure}} in {{Ferromagnetic SrRuO}}$_3$ {{Thin Film}}}}.
\newblock {\emph{\JournalTitle{ACS Applied Materials \& Interfaces}}}
  \textbf{\bibinfo{volume}{12}}, \bibinfo{pages}{37757--37763},
  \doiprefix\url{10.1021/acsami.0c10545} (\bibinfo{year}{2020}).

\bibitem{malsch_correlating_2020}
\bibinfo{author}{Malsch, G.} \emph{et~al.}
\newblock \bibinfo{journal}{\bibinfo{title}{Correlating the {{Nanoscale
  Structural}}, {{Magnetic}}, and {{Magneto}}-{{Transport Properties}} in
  {{SrRuO}}$_3$-{{Based Perovskite Thin Films}}: {{Implications}} for {{Oxide
  Skyrmion Devices}}}}.
\newblock {\emph{\JournalTitle{ACS Applied Nano Materials}}}
  \textbf{\bibinfo{volume}{3}}, \bibinfo{pages}{1182--1190},
  \doiprefix\url{10.1021/acsanm.9b01918} (\bibinfo{year}{2020}).

\bibitem{miao_strain_2020}
\bibinfo{author}{Miao, L.} \emph{et~al.}
\newblock \bibinfo{journal}{\bibinfo{title}{Strain relaxation induced
  transverse resistivity anomalies in {{SrRuO}}$_3$ thin films}}.
\newblock {\emph{\JournalTitle{Physical Review B}}}
  \textbf{\bibinfo{volume}{102}}, \bibinfo{pages}{064406},
  \doiprefix\url{10.1103/PhysRevB.102.064406} (\bibinfo{year}{2020}).

\bibitem{kimbell_two-channel_2020}
\bibinfo{author}{Kimbell, G.} \emph{et~al.}
\newblock \bibinfo{journal}{\bibinfo{title}{Two-channel anomalous {{Hall}}
  effect in {{SrRuO}}$_3$}}.
\newblock {\emph{\JournalTitle{Physical Review Materials}}}
  \textbf{\bibinfo{volume}{4}}, \bibinfo{pages}{054414},
  \doiprefix\url{10.1103/PhysRevMaterials.4.054414} (\bibinfo{year}{2020}).

\bibitem{van_thiel_extraordinary_2020}
\bibinfo{author}{van Thiel, T.~C.}, \bibinfo{author}{Groenendijk, D.~J.} \&
  \bibinfo{author}{Caviglia, A.~D.}
\newblock \bibinfo{journal}{\bibinfo{title}{Extraordinary {{Hall}} balance in
  ultrathin {{SrRuO}}$_3$ bilayers}}.
\newblock {\emph{\JournalTitle{Journal of Physics: Materials}}}
  \textbf{\bibinfo{volume}{3}}, \bibinfo{pages}{025005},
  \doiprefix\url{10.1088/2515-7639/ab7a03} (\bibinfo{year}{2020}).

\bibitem{fang_anomalous_2003}
\bibinfo{author}{Fang, Z.}
\newblock \bibinfo{journal}{\bibinfo{title}{The {{Anomalous Hall Effect}} and
  {{Magnetic Monopoles}} in {{Momentum Space}}}}.
\newblock {\emph{\JournalTitle{Science}}} \textbf{\bibinfo{volume}{302}},
  \bibinfo{pages}{92--95}, \doiprefix\url{10.1126/science.1089408}
  (\bibinfo{year}{2003}).

\bibitem{thiaville_dynamics_2012}
\bibinfo{author}{Thiaville, A.}, \bibinfo{author}{Rohart, S.},
  \bibinfo{author}{Ju{\'e}, {\'E}.}, \bibinfo{author}{Cros, V.} \&
  \bibinfo{author}{Fert, A.}
\newblock \bibinfo{journal}{\bibinfo{title}{Dynamics of {{Dzyaloshinskii}}
  domain walls in ultrathin magnetic films}}.
\newblock {\emph{\JournalTitle{EPL (Europhysics Letters)}}}
  \textbf{\bibinfo{volume}{100}}, \bibinfo{pages}{57002},
  \doiprefix\url{10.1209/0295-5075/100/57002} (\bibinfo{year}{2012}).

\bibitem{ryu_chiral_2013}
\bibinfo{author}{Ryu, K.-S.}, \bibinfo{author}{Thomas, L.},
  \bibinfo{author}{Yang, S.-H.} \& \bibinfo{author}{Parkin, S.}
\newblock \bibinfo{journal}{\bibinfo{title}{Chiral spin torque at magnetic
  domain walls}}.
\newblock {\emph{\JournalTitle{Nature Nanotechnology}}}
  \textbf{\bibinfo{volume}{8}}, \bibinfo{pages}{527--533},
  \doiprefix\url{10.1038/nnano.2013.102} (\bibinfo{year}{2013}).

\bibitem{emori_current-driven_2013}
\bibinfo{author}{Emori, S.}, \bibinfo{author}{Bauer, U.}, \bibinfo{author}{Ahn,
  S.-M.}, \bibinfo{author}{Martinez, E.} \& \bibinfo{author}{Beach, G. S.~D.}
\newblock \bibinfo{journal}{\bibinfo{title}{Current-driven dynamics of chiral
  ferromagnetic domain walls}}.
\newblock {\emph{\JournalTitle{Nature Materials}}}
  \textbf{\bibinfo{volume}{12}}, \bibinfo{pages}{611--616},
  \doiprefix\url{10.1038/nmat3675} (\bibinfo{year}{2013}).

\bibitem{hamadeh_full_2014}
\bibinfo{author}{Hamadeh, A.} \emph{et~al.}
\newblock \bibinfo{journal}{\bibinfo{title}{Full {{Control}} of the
  {{Spin}}-{{Wave Damping}} in a {{Magnetic Insulator Using Spin}}-{{Orbit
  Torque}}}}.
\newblock {\emph{\JournalTitle{Physical Review Letters}}}
  \textbf{\bibinfo{volume}{113}}, \bibinfo{pages}{197203},
  \doiprefix\url{10.1103/PhysRevLett.113.197203} (\bibinfo{year}{2014}).

\bibitem{demidov_direct_2016}
\bibinfo{author}{Demidov, V.~E.} \emph{et~al.}
\newblock \bibinfo{journal}{\bibinfo{title}{Direct observation of dynamic modes
  excited in a magnetic insulator by pure spin current}}.
\newblock {\emph{\JournalTitle{Scientific Reports}}}
  \textbf{\bibinfo{volume}{6}}, \bibinfo{pages}{32781},
  \doiprefix\url{10.1038/srep32781} (\bibinfo{year}{2016}).

\bibitem{collet_generation_2016}
\bibinfo{author}{Collet, M.} \emph{et~al.}
\newblock \bibinfo{journal}{\bibinfo{title}{Generation of coherent spin-wave
  modes in yttrium iron garnet microdiscs by spin--orbit torque}}.
\newblock {\emph{\JournalTitle{Nature Communications}}}
  \textbf{\bibinfo{volume}{7}}, \bibinfo{pages}{10377},
  \doiprefix\url{10.1038/ncomms10377} (\bibinfo{year}{2016}).

\bibitem{avci_current-induced_2017}
\bibinfo{author}{Avci, C.~O.} \emph{et~al.}
\newblock \bibinfo{journal}{\bibinfo{title}{Current-induced switching in a
  magnetic insulator}}.
\newblock {\emph{\JournalTitle{Nature Materials}}}
  \textbf{\bibinfo{volume}{16}}, \bibinfo{pages}{309--314},
  \doiprefix\url{10.1038/nmat4812} (\bibinfo{year}{2017}).

\bibitem{shao_role_2018}
\bibinfo{author}{Shao, Q.} \emph{et~al.}
\newblock \bibinfo{journal}{\bibinfo{title}{Role of dimensional crossover on
  spin-orbit torque efficiency in magnetic insulator thin films}}.
\newblock {\emph{\JournalTitle{Nature Communications}}}
  \textbf{\bibinfo{volume}{9}}, \bibinfo{pages}{3612} (\bibinfo{year}{2018}).

\bibitem{velez_high-speed_2019}
\bibinfo{author}{V{\'e}lez, S.} \emph{et~al.}
\newblock \bibinfo{journal}{\bibinfo{title}{High-speed domain wall racetracks
  in a magnetic insulator}}.
\newblock {\emph{\JournalTitle{Nature Communications}}}
  \textbf{\bibinfo{volume}{10}}, \bibinfo{pages}{4750},
  \doiprefix\url{10.1038/s41467-019-12676-7} (\bibinfo{year}{2019}).

\bibitem{avci_interface-driven_2019}
\bibinfo{author}{Avci, C.~O.}
\newblock \bibinfo{journal}{\bibinfo{title}{Interface-driven chiral magnetism
  and current-driven domain walls in insulating magnetic garnets}}.
\newblock {\emph{\JournalTitle{Nature Nanotechnology}}}
  \textbf{\bibinfo{volume}{14}}, \bibinfo{pages}{7} (\bibinfo{year}{2019}).

\bibitem{ahmed_spin-hall_2019}
\bibinfo{author}{Ahmed, A.~S.} \emph{et~al.}
\newblock \bibinfo{journal}{\bibinfo{title}{Spin-{{Hall Topological Hall
  Effect}} in {{Highly Tunable Pt}}/{{Ferrimagnetic}}-{{Insulator Bilayers}}}}.
\newblock {\emph{\JournalTitle{Nano Letters}}} \textbf{\bibinfo{volume}{19}},
  \bibinfo{pages}{5683--5688}, \doiprefix\url{10.1021/acs.nanolett.9b02265}
  (\bibinfo{year}{2019}).

\bibitem{shao_topological_2019}
\bibinfo{author}{Shao, Q.} \emph{et~al.}
\newblock \bibinfo{journal}{\bibinfo{title}{Topological {{Hall}} effect at
  above room temperature in heterostructures composed of a magnetic insulator
  and a heavy metal}}.
\newblock {\emph{\JournalTitle{Nature Electronics}}}
  \textbf{\bibinfo{volume}{2}}, \bibinfo{pages}{182--186},
  \doiprefix\url{10.1038/s41928-019-0246-x} (\bibinfo{year}{2019}).

\bibitem{shao_exploring_2019}
\bibinfo{author}{Shao, Q.} \emph{et~al.}
\newblock \bibinfo{journal}{\bibinfo{title}{Exploring interfacial exchange
  coupling and sublattice effect in heavy metal/ferrimagnetic insulator
  heterostructures using {{Hall}} measurements, x-ray magnetic circular
  dichroism, and neutron reflectometry}}.
\newblock {\emph{\JournalTitle{Physical Review B}}}
  \textbf{\bibinfo{volume}{99}}, \bibinfo{pages}{104401},
  \doiprefix\url{10.1103/PhysRevB.99.104401} (\bibinfo{year}{2019}).

\bibitem{xia_interfacial_2020}
\bibinfo{author}{Xia, S.} \emph{et~al.}
\newblock \bibinfo{journal}{\bibinfo{title}{Interfacial
  {{Dzyaloshinskii}}-{{Moriya}} interaction between ferromagnetic insulator and
  heavy metal}}.
\newblock {\emph{\JournalTitle{Applied Physics Letters}}}
  \textbf{\bibinfo{volume}{116}}, \bibinfo{pages}{052404},
  \doiprefix\url{10.1063/1.5134762} (\bibinfo{year}{2020}).

\bibitem{ding_interfacial_2019-1}
\bibinfo{author}{Ding, S.} \emph{et~al.}
\newblock \bibinfo{journal}{\bibinfo{title}{Interfacial
  {{Dzyaloshinskii}}-{{Moriya}} interaction and chiral magnetic textures in a
  ferrimagnetic insulator}}.
\newblock {\emph{\JournalTitle{Physical Review B}}}
  \textbf{\bibinfo{volume}{100}}, \bibinfo{pages}{100406},
  \doiprefix\url{10.1103/PhysRevB.100.100406} (\bibinfo{year}{2019}).

\bibitem{buttner_thermal_2020}
\bibinfo{author}{B{\"u}ttner, F.} \emph{et~al.}
\newblock \bibinfo{journal}{\bibinfo{title}{Thermal nucleation and
  high-resolution imaging of submicrometer magnetic bubbles in thin thulium
  iron garnet films with perpendicular anisotropy}}.
\newblock {\emph{\JournalTitle{Physical Review Materials}}}
  \textbf{\bibinfo{volume}{4}}, \bibinfo{pages}{011401},
  \doiprefix\url{10.1103/PhysRevMaterials.4.011401} (\bibinfo{year}{2020}).

\bibitem{caretta_interfacial_2020-1}
\bibinfo{author}{Caretta, L.} \emph{et~al.}
\newblock \bibinfo{journal}{\bibinfo{title}{Interfacial
  {{Dzyaloshinskii}}-{{Moriya}} interaction arising from rare-earth orbital
  magnetism in insulating magnetic oxides}}.
\newblock {\emph{\JournalTitle{Nature Communications}}}
  \textbf{\bibinfo{volume}{11}}, \bibinfo{pages}{1090},
  \doiprefix\url{10.1038/s41467-020-14924-7} (\bibinfo{year}{2020}).

\bibitem{ding_identifying_2020}
\bibinfo{author}{Ding, S.} \emph{et~al.}
\newblock \bibinfo{journal}{\bibinfo{title}{{Identifying the origin of the
  nonmonotonic thickness dependence of spin-orbit torque and interfacial
  Dzyaloshinskii-Moriya interaction in a ferrimagnetic insulator
  heterostructure}}}.
\newblock {\emph{\JournalTitle{Physical Review B}}}
  \textbf{\bibinfo{volume}{102}}, \bibinfo{pages}{054425},
  \doiprefix\url{10.1103/PhysRevB.102.054425} (\bibinfo{year}{2020}).

\bibitem{wang_chiral_2020}
\bibinfo{author}{Wang, H.} \emph{et~al.}
\newblock \bibinfo{journal}{\bibinfo{title}{Chiral {{Spin}}-{{Wave Velocities
  Induced}} by {{All}}-{{Garnet Interfacial Dzyaloshinskii}}-{{Moriya
  Interaction}} in {{Ultrathin Yttrium Iron Garnet Films}}}}.
\newblock {\emph{\JournalTitle{Physical Review Letters}}}
  \textbf{\bibinfo{volume}{124}}, \bibinfo{pages}{027203},
  \doiprefix\url{10.1103/PhysRevLett.124.027203} (\bibinfo{year}{2020}).

\bibitem{gross_real-space_2017}
\bibinfo{author}{Gross, I.} \emph{et~al.}
\newblock \bibinfo{journal}{\bibinfo{title}{{Real-space imaging of
  non-collinear antiferromagnetic order with a single-spin magnetometer}}}.
\newblock {\emph{\JournalTitle{Nature}}} \textbf{\bibinfo{volume}{549}},
  \bibinfo{pages}{252 -- 256}, \doiprefix\url{10.1038/nature23656}
  (\bibinfo{year}{2017}).

\bibitem{lee_investigation_2020}
\bibinfo{author}{Lee, A.~J.} \emph{et~al.}
\newblock \bibinfo{journal}{\bibinfo{title}{Investigation of the {{Role}} of
  {{Rare}}-{{Earth Elements}} in {{Spin}}-{{Hall Topological Hall Effect}} in
  {{Pt}}/{{Ferrimagnetic}}-{{Garnet Bilayers}}}}.
\newblock {\emph{\JournalTitle{Nano Letters}}} \textbf{\bibinfo{volume}{20}},
  \bibinfo{pages}{4667--4672}, \doiprefix\url{10.1021/acs.nanolett.0c01620}
  (\bibinfo{year}{2020}).

\bibitem{li_topological_2020}
\bibinfo{author}{Li, P.} \emph{et~al.}
\newblock \bibinfo{journal}{\bibinfo{title}{{Topological Hall Effect in a
  Topological Insulator Interfaced with a Magnetic Insulator}}}.
\newblock {\emph{\JournalTitle{Nano Letters}}} \textbf{\bibinfo{volume}{21}},
  \bibinfo{pages}{84--90}, \doiprefix\url{10.1021/acs.nanolett.0c03195}
  (\bibinfo{year}{2020}).

\bibitem{jani_antiferromagnetic_2021}
\bibinfo{author}{Jani, H.} \emph{et~al.}
\newblock \bibinfo{journal}{\bibinfo{title}{{Antiferromagnetic half-skyrmions
  and bimerons at room temperature}}}.
\newblock {\emph{\JournalTitle{Nature}}} \textbf{\bibinfo{volume}{590}},
  \bibinfo{pages}{74--79}, \doiprefix\url{10.1038/s41586-021-03219-6}
  (\bibinfo{year}{2021}).

\bibitem{huang_situ_2018}
\bibinfo{author}{Huang, P.} \emph{et~al.}
\newblock \bibinfo{journal}{\bibinfo{title}{In {{Situ Electric Field Skyrmion
  Creation}} in {{Magnetoelectric Cu}}$_2${{OSeO}}$_3$}}.
\newblock {\emph{\JournalTitle{Nano Letters}}} \textbf{\bibinfo{volume}{18}},
  \bibinfo{pages}{5167--5171}, \doiprefix\url{10.1021/acs.nanolett.8b02097}
  (\bibinfo{year}{2018}).

\bibitem{maccariello_electrical_2018}
\bibinfo{author}{Maccariello, D.} \emph{et~al.}
\newblock \bibinfo{journal}{\bibinfo{title}{Electrical detection of single
  magnetic skyrmions in metallic multilayers at room temperature}}.
\newblock {\emph{\JournalTitle{Nature Nanotechnology}}}
  \textbf{\bibinfo{volume}{13}}, \bibinfo{pages}{233--237},
  \doiprefix\url{10.1038/s41565-017-0044-4} (\bibinfo{year}{2018}).

\bibitem{jiang_blowing_2015}
\bibinfo{author}{Jiang, W.} \emph{et~al.}
\newblock \bibinfo{journal}{\bibinfo{title}{Blowing magnetic skyrmion
  bubbles}}.
\newblock {\emph{\JournalTitle{Science}}} \textbf{\bibinfo{volume}{349}},
  \bibinfo{pages}{283--286}, \doiprefix\url{10.1126/science.aaa1442}
  (\bibinfo{year}{2015}).

\bibitem{legrand_room-temperature_2017}
\bibinfo{author}{Legrand, W.} \emph{et~al.}
\newblock \bibinfo{journal}{\bibinfo{title}{Room-{{Temperature
  Current}}-{{Induced Generation}} and {{Motion}} of sub-100 nm
  {{Skyrmions}}}}.
\newblock {\emph{\JournalTitle{Nano Letters}}} \textbf{\bibinfo{volume}{17}},
  \bibinfo{pages}{2703--2712}, \doiprefix\url{10.1021/acs.nanolett.7b00649}
  (\bibinfo{year}{2017}).

\bibitem{schott_skyrmion_2017}
\bibinfo{author}{Schott, M.} \emph{et~al.}
\newblock \bibinfo{journal}{\bibinfo{title}{The {{Skyrmion Switch}}: {{Turning
  Magnetic Skyrmion Bubbles}} on and off with an {{Electric Field}}}}.
\newblock {\emph{\JournalTitle{Nano Letters}}} \textbf{\bibinfo{volume}{17}},
  \bibinfo{pages}{3006--3012}, \doiprefix\url{10.1021/acs.nanolett.7b00328}
  (\bibinfo{year}{2017}).

\bibitem{white_electric-field-induced_2014}
\bibinfo{author}{White, J.~S.} \emph{et~al.}
\newblock \bibinfo{journal}{\bibinfo{title}{Electric-{{Field}}-{{Induced
  Skyrmion Distortion}} and {{Giant Lattice Rotation}} in the {{Magnetoelectric
  Insulator Cu}}$_2${{OSeO}}$_3$}}.
\newblock {\emph{\JournalTitle{Physical Review Letters}}}
  \textbf{\bibinfo{volume}{113}}, \bibinfo{pages}{107203},
  \doiprefix\url{10.1103/PhysRevLett.113.107203} (\bibinfo{year}{2014}).

\bibitem{vazOxideSpinorbitronicsNew2018}
\bibinfo{author}{Vaz, D.~C.}, \bibinfo{author}{Barth{\'e}l{\'e}my, A.} \&
  \bibinfo{author}{Bibes, M.}
\newblock \bibinfo{journal}{\bibinfo{title}{Oxide spin-orbitronics: {{New}}
  routes towards low-power electrical control of magnetization in oxide
  heterostructures}}.
\newblock {\emph{\JournalTitle{Japanese Journal of Applied Physics}}}
  \textbf{\bibinfo{volume}{57}}, \bibinfo{pages}{0902A4},
  \doiprefix\url{10.7567/JJAP.57.0902A4} (\bibinfo{year}{2018}).

\bibitem{ahadiEnhancingSuperconductivitySrTiO2019}
\bibinfo{author}{Ahadi, K.} \emph{et~al.}
\newblock \bibinfo{journal}{\bibinfo{title}{Enhancing superconductivity in
  {{SrTiO}}$_3$ films with strain}}.
\newblock {\emph{\JournalTitle{Science Advances}}}
  \textbf{\bibinfo{volume}{5}}, \bibinfo{pages}{eaaw0120},
  \doiprefix\url{10.1126/sciadv.aaw0120} (\bibinfo{year}{2019}).

\bibitem{barthelemy_quasi-two-dimensional_2021}
\bibinfo{author}{Barthelemy, A.} \emph{et~al.}
\newblock \bibinfo{journal}{\bibinfo{title}{Quasi-two-dimensional electron gas
  at the oxide interfaces for topological quantum physics}}.
\newblock {\emph{\JournalTitle{{EPL} (Europhysics Letters)}}}
  \textbf{\bibinfo{volume}{133}}, \bibinfo{pages}{17001},
  \doiprefix\url{10.1209/0295-5075/133/17001}.
\newblock \bibinfo{note}{Publisher: {IOP} Publishing}.

\end{thebibliography}
\end{document}